\documentclass[twocolumn]{aastex631}
\usepackage{newtxtext,newtxmath}
\usepackage{xcolor}

\def\rasec {\hbox{$\,$\raise 0.6 ex \hbox{\rm s}\kern-.35em
                  \lower 0.0 ex \hbox{.}$\,$}}        
\def\decsec{\hbox{$\,$\raise 0.5 ex \hbox{$\prime\prime$}\kern-.45em
                  \lower 0.0 ex \hbox{.}$\,$}}         
\def\decmin{\hbox{$\,$\raise 0.5 ex \hbox{$\prime$}\kern-.45em
    \lower 0.0 ex \hbox{}$\,$}}

\newcommand{\gtabouteq}{\,\hbox{\raise 0.5 ex \hbox{$>$}\kern-.77em 
                    \lower 0.5 ex \hbox{$\sim$}$\,$}}       
\newcommand{\ltabouteq}{\,\hbox{\raise 0.5 ex \hbox{$<$}\kern-.77em 
                     \lower 0.5 ex \hbox{$\sim$}$\,$}}
\newcommand{\curveshape}{\rotatebox[origin=c]{-40}{\large $\sim$}~}

\begin{document}
\nolinenumbers
\title{CHANG-ES XXXII: Spatially Resolved Thermal/Nonthermal Separation from Radio Data Alone -- New Probes into NGC~3044 and NGC~5775}

\author{Judith Irwin}
\affiliation{Dept. of Physics, Engineering Physics \& Astronomy,
Queen's University, Kingston, K7L 3N6, Canada, irwinja@queensu.ca}

\author{Tanden Cook}
\affiliation{Dept. of Physics, Engineering Physics \& Astronomy
Queen's University, Kingston, K7L 3N6, Canada}

\author{Michael Stein}
\affiliation{Ruhr University Bochum, Faculty of Physics and Astronomy, Astronomical Institute (AIRUB), 44780 Bochum, Germany}

\author{Ralf-Juergen Dettmar}
\affiliation{Ruhr University Bochum, Faculty of Physics and Astronomy, Astronomical Institute (AIRUB), 44780 Bochum, Germany}

\author{Volker Heesen}
\affiliation{
University of Hamburg, Hamburger Sternwarte, Gojenbergsweg 112, 21029 Hamburg, Germany}

\author{Q. Daniel Wang}
\affiliation{Department of Astronomy, University of Massachusetts,
North Pleasant Street\\ Amherst, MA 01003-9305, USA,LGRT-B 619E, 710}

\author{Theresa Wiegert}
\affiliation{Institute of Astrophysics of Andalucia (IAA-CSIC),
Glorieta de la Astronomía s/n, 18008 Granada, Spain}

\author{Yelena Stein}
\affiliation{Ruhr University Bochum, Faculty of Physics and Astronomy, Astronomical Institute (AIRUB), 44780 Bochum, Germany}

\author{Carlos Vargas}
\affiliation{
Dept. of Astronomy \& Steward Observatory, 933 N. Cherry Ave., Tucson, AZ, 85719, USA}



\begin{abstract}
{ We have carried out spatially resolved thermal/nonthermal separation on two edge-on galaxies, NGC~3044 and NGC~5775, using only radio data.  Narrow-band imaging within a frequency band that is almost contiguous from 1.25 to 7.02 GHz (L-band, S-band and C-band) has allowed us to fit spectra  and construct thermal, nonthermal, and nonthermal spectral index maps. This method does not require any ancillary H$\alpha$ and infrared data, or reliance on dust corrections that are challenging in edge-on galaxies. For NGC~3044, at 15 arcsec resolution, we find a median thermal fraction of $\sim\, 13$\% with an estimated uncertainty in this fraction of $\sim\, 50$\%  at 4.13 GHz. This compares well with the H$\alpha$ mixture method results. 
We uncovered evidence for a vertical outflow feature reaching at least $z\,\sim\,3.5$ kpc in projection above the plane, reminiscent of M82's starburst wind.
For the higher SFR galaxy, NGC~5775 at 12 arcsec resolution, we find a median thermal fraction of 44\% at 4.13 GHz with an estimated error on this fraction of 17\%.
Both galaxies show a change of slope (flattening) in L-band. These results suggest that a radio-only method for separating thermal from nonthermal emission is not only feasible, but able to reveal new features that might otherwise be obscured in edge-on disks.}

\end{abstract}



\section{Introduction} \label{sec:intro}

The CHANG-ES (Continuum Halos in Nearby Galaxies -- an EVLA\footnote{The Expanded Very Large Array (EVLA) is now known as the Karl G. Jansky Very Large Array.} Survey) project  \citep{irw2012a,irw12b}
has identified 35 spiral galaxies that are edge-on to the line of sight, in order to investigate the presence and strength of their spatially resolved radio halos and the relationship between this high latitude emission and other galactic properties. The observations used the Karl G. Jansky Very Large array (hereafter the Very Large Array, or VLA) in a variety of array configurations and all four Stokes parameters.  A summary of CHANG-ES goals and some  selected results can be found in \cite{irw19a}.

\begin{figure*}
   \centering
  \includegraphics[width=0.8\textwidth]{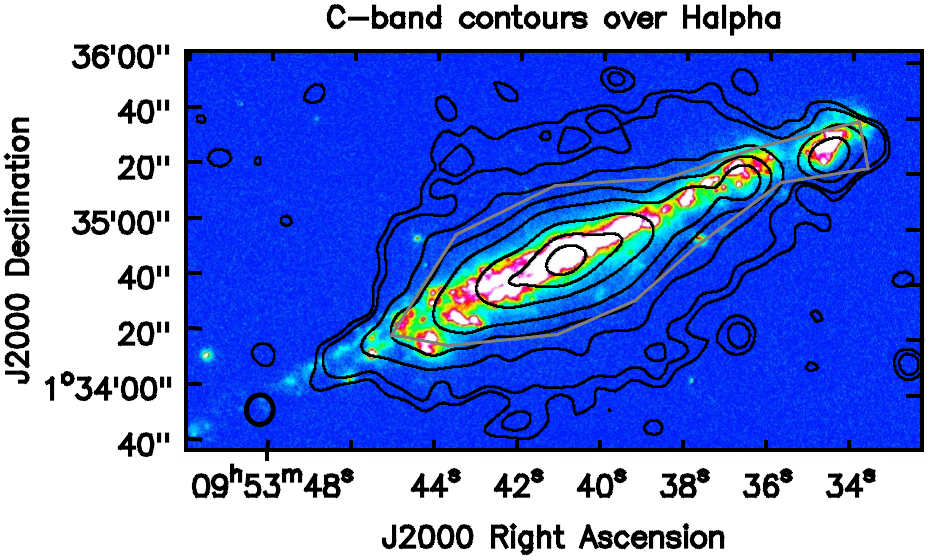}  
   \caption{The edge-on galaxy, NGC~3044, shown in H$\alpha$ (colours), with arbitrary scaling, overlaid with C-band radio contours from D-configuration.  Radio contours are at 15, 25, 70, 150, 300, 800, 2000, and 4000 $\upmu$Jy beam$^{-1}$. The beam size ($10\farcs 7 \times 9\farcs 6$ at a position angle of $-5\fdg 1$) is shown as a thick black circle at lower left. The H$\alpha$ image is from \citet{var19}  and the radio image is from  \citet{wie15}. The grey curve encloses the region that was used for a flux measurement, as described in Sect.~\ref{sec:initial_spectrum}. %Both images in FITS format can be downloaded from the CHANG-ES public release website: {\tt https://projects.canfar.net/changes} 
   }
    \label{fig:N3044_fig1}
\end{figure*}

{ In the CHANG-ES sample, two galaxies with significant and impressive radio halos are NGC~3044 and NGC~5775. In this paper, we will focus on NGC~3044 in some detail, followed by NGC~5775, to be dealt with in a similar, but less detailed fashion. The halo of NGC~3044} is nicely illustrated by
Figure~\ref{fig:N3044_fig1} which shows radio contours from  \citet{wie15} superimposed on a colour image of  H$\alpha$ emission from \citet{var19}. This image nicely shows the contrast between the broad-scale radio halo and the narrower disk. Ionized gas is mostly constrained to the disk within which star-forming regions are found. However,  {\it extra-planar} diffuse ionized gas has been measured in this galaxy as well \citep{col00, ros00, mil03}. Halo emission in the ultraviolet has also been detected and attributed to reflection from dust \citep{hod16}.

The magnetic field in the disk of NGC~3044 is $\sim$ 14 $\upmu$Gauss and the orientation of the magnetic field in the halo is  approximately perpendicular to the major axis \citep{kra20}. The halo radio emission  has been modeled in the $z$ direction (i.e. perpendicular to the galaxy's plane) in terms of synchrotron energy losses with diffusion \citep{kra18}.
This galaxy harbors a historical supernova \citep[SN~1983E, e.g.][]{irw20} as well as an ultraluminous X-ray source, ULX1 which reached $\sim\,10^{40}$ erg s$^{-1}$ at its peak \citep{wal22}. The galaxy is purely `star-forming', that is to say, there is no evidence for an active galactic nucleus (AGN). 
NGC~3044 appears to be isolated, but the HI distribution is quite asymmetric \citep{lee97, pop11} with the north-west advancing side depleted of HI compared to the south-east side, suggesting that some interaction may have occurred in the past. 
Optical light profiles have also recently  been provided by \citet{gil22}.
Basic data for NGC~3044 are in Table~\ref{tab:N3044params}.

{\begin{table*}
\fontsize{7}{11}\selectfont
\begin{center}
\caption{Galaxy Parameters}\label{tab:N3044params}
    \begin{tabular}{lcccccccc}
    \hline
Galaxy & RA$^{\rm a}$ & Dec$^{\rm a}$ & $D^{\rm b}$ & d$^{\rm c}$& i$^{\rm d}$ &
SFR$^{\rm e}$ & $L_{{\rm H}\alpha}$$^{\rm f}$ &$S_{1.6~{\rm GHz}}$$^{\rm g}$\\
& ($^{\rm h}$ $^{\rm m}$ $^{\rm s}$)  & ($\degr$ $\arcmin$ $\arcsec$)& (Mpc) & ($\arcmin$) & ($\degr$) &(M$_\odot$ yr$^{-1}$) & ($10^{41}$ erg s$^{-1}$) & (mJy) \\
\hline
N~3044 & 09 53 40.88   & +01 34 46.7 & 20.3 & 3.07 &85 &1.75 $\pm$ 0.16 & $1.20\,\pm\,0.20$& $104.2\,\pm\,2.1$\\
N~5775 & 14 53 57.60   & +03 32 40.1 & 28.9 & 3.80 & 86 &7.56 $\pm$ 0.65 & $2.70\,\pm\,0.41$
& $255.0\,\pm\,5.1 $
\\
\hline
    \end{tabular}
\end{center}
$^{\rm a}$ Right Ascension and Declination (J2000) of the pointing and image centre.\\
$^{\rm b}$ Galaxy distance \citep{wie15}. \\
$^{\rm c}$ Diameter at 22 $\upmu$m \citep{wie15}.\\
$^{\rm d}$ Inclination \citep{kra18}.\\
$^{\rm e}$ Star formation rate \citep{var19}.\\
$^{\rm f}$ Integrated H$\alpha$ luminosity \citep{var19}.\\
$^{\rm g}$ Flux density at 1.6 GHz \citep{wie15}.
\end{table*}}

CHANG-ES observations were initially in two  wide radio frequency bands (L-band: centered at 1.6 GHz,  and C-band: centered at 6 GHz).  Now, new additional S-band data (3.0 GHz) have been added to the survey complement, giving CHANG-ES almost continuous frequency coverage from 1.2 through to 7 GHz (Table~\ref{tab:observing}) with only a few minor gaps.  This impressive frequency coverage of $\Delta\,\nu\,\sim 4.5$ GHz 
can be exploited in order to recover spatially resolved spectral information about the galaxies and their halos. In this paper, we carry out such an analysis for NGC~3044. 

In Sect.~\ref{sec:separating}, we describe the procedure for thermal/nonthermal separation. Sect.~\ref{sec:observations} describes the radio observations.
Sect.~\ref{sec:results} gives the results, including wide-band imaging, narrow-band imaging and cube-fitting from the narrow-band maps which permits the thermal/nonthermal separation. Sect.~\ref{sec:discussion} discusses sources of error, provides a comparison with H$\alpha$-related results, discusses the vertical feature observed in the thermal fraction map, and the L-band spectral change of shape. Sect.~\ref{sec:conclusions} summarizes the main conclusions.
 
\section{Separating Thermal from Nonthermal Emission}
\label{sec:separating}

In principle, continuum emission at cm wavelengths is relatively simple to analyze because the observed emission results from only two physical processes: thermal Bremsstrahlung radiation (thermal emission) and synchrotron radiation (nonthermal emission). Contributions from other continuum components are negligible. In particular, no significant dust emission is expected at wavelengths longer than $\sim$ 1 mm \citep[$\nu\,\ltabouteq\,300$ GHz, see ][]{sil11}.
 Moreover, dust absorption is also negligible, avoiding long-standing difficulties that are commonly seen at optical wavelengths.  
 Optical depth effects at radio frequencies are also usually negligible unless a compact and/or high-density source is observed, either from synchrotron self-absorption or thermal absorption.  An example of synchrotron self-absorption in the CHANG-ES survey is  the dense active core of NGC~4845, as described in \citet{irw15}. 

In spite of this simplicity,  in practice, it is not a straightforward exercise to separate the two emission components. It is well-known that nonthermal emission dominates thermal over the frequency bands of interest \citep{con92}. This is verified by generally steep  spectral indices (i.e. $\alpha\,\ll\,-0.1$) in all CHANG-ES galaxies, %\footnote{For a list of all CHANG-ES publications, see the public release website at {\tt https://projects.canfar.net/changes}.}, 
where we use the convention, $I_\nu\,\propto\,\nu^\alpha$.  It has also now been established that the spectral index steepens with distance from the galaxy's plane \citep[e.g.][]{wie15,kra20,ste23}.  At any location, however, it has not been clear what fraction of the total emission is due to thermal and nonthermal components, or whether
changes in the slope of the spectrum, if present, are due to curvature in the nonthermal spectral index, or contributions from thermal emission, or both.

The measured emission at any location in the galaxy can be expressed by a frequency-dependent observed specific intensity, $I_\nu$ (usually expressed in
Jy beam$^{-1}$) which is the sum of the nonthermal (NT) and thermal (TH) components. In the optically thin limit,
\begin{eqnarray}
I_\nu&=&I_{\nu_{0{\rm tot}}}\,\left(\frac{\nu}{\nu_0}\right)^{\alpha\,+\,\beta\,\log\left(\frac{\nu}{\nu_{0}}\right)}  \label{eqn:sum1}\\
&=&I_{\nu_{\rm NT}}\,+\, I_{\nu_{\rm TH}}\label{eqn:sum2}\\
&=&I_{\nu_{0_{\rm NT}}}\,\left(\frac{\nu}{\nu_0}\right)^{\alpha_{\rm NT}\,+\,\beta_{\rm NT}\,\log\left(\frac{\nu}{\nu_0}\right)}\,+\, I_{{\nu_0}_{\rm TH}}\,\left(\frac{\nu}{\nu_0}\right)^{-0.1}\label{eqn:sum3}
\end{eqnarray}

Eq.~\eqref{eqn:sum1} represents the total emission, which is described by a total spectral index, $\alpha$, as well as some curvature that is described by the parameter, $\beta$.
The value of $\beta$, should it be required, is not well known. This particular function is fitted in the standard data reduction routine that is used to form and clean maps 
(i.e. {\tt tclean}, Sect.~\ref{sec:radioimaging}). The quantity,  $\nu_0$, is a reference frequency, taken in this work to be the center of all three bands, i.e. $\nu_0\,=\,4.13$ GHz.
Previous CHANG-ES maps, such as those that are available on the public release website\footnote{\tt https://projects/canfar.net/changes} have been made with this recipe, but without $\beta$.
That is, no thermal correction is made, and the spectrum is assumed to {  be represented simply by $I_\nu\,\propto\,\nu^\alpha$.}
This is because the signal-to-noise (S/N) was insufficient to adequately describe any curvature term, if it had been present.  Trials with and without $\beta$ are provided in \citet{irw12b}. {It is worth noting that the mathematical form in which $\beta$ is included in Eqs.~\ref{eqn:sum1} or \ref{eqn:sum3} is simply a way to include curvature, and is not physically motivated. }

Eq.~\eqref{eqn:sum2} reveals that the total emission must be the sum of nonthermal and thermal components. A number of CHANG-ES papers have carried out the thermal/nonthermal separation so that the two contributions can be studied separately, including \citet{var18},\citet{var19}, \citet{mor19}, \citet{mis19}, \citet{stey19}, \citet{sch19}, \citet{stey20}, \cite{hea22}, and \citet{ste23}.  {\it In every case, the thermal component was computed from an H$\alpha$ map, corrected for dust.}

Eq.~\eqref{eqn:sum3} shows the frequency dependence of the thermal and nonthermal terms.  Again, the nonthermal component is represented by a spectral index, $\alpha_{\rm NT}$, and a curvature term, $\beta_{\rm NT}$, and the thermal emission has the well-known $\nu^{-0.1}$ dependence.
The frequency, $\nu$, is normalized by $\nu_0$
so that $I_\nu$, $I_{\nu_{\rm NT}}$, and $I_{\rm TH}$  all have the same units.

Let us now connect the spectrum to the physical properties of the source at any position.  We have  \citep[e.g. see][]{irw21},
\begin{eqnarray}
    I_{{\rm NT}} &\propto& N_{0}\,{B_\perp}^{\frac{\Gamma\,+\,1}{2}} \label{eqn:INB}\\
    I_{\rm TH}&\propto& {{\rm T}_{\rm e}}^{-0.35}\,{\cal EM}\label{eqn:ITEM}
\end{eqnarray}
{where here, and in the following, we drop the $\nu_0$ subscripts for simplicity.}

In Eq.~\eqref{eqn:INB}, $N_0$ and $\Gamma$ are measures of the cosmic ray electron energy spectrum described by $N(E)\,=\,N_0\,E^{-\Gamma}$ and $B_\perp$ is the strength of the magnetic field that is perpendicular to the line of sight.   $\Gamma$ is related to $\alpha_{\rm NT}$ via $\alpha_{\rm NT}\,=\,\frac{\left(1\,-\Gamma\right)}{2}$. Consequently, if thermal and nonthermal emission can be separated and $\alpha_{\rm NT}$ measured, then the slope of the cosmic ray electron energy spectrum, $\Gamma$, can be obtained.

In Eq.~\eqref{eqn:ITEM}, ${\rm T_e}$ is the electron temperature of the thermal gas and $\cal EM$ is the emission measure.  The emission measure is given by
\begin{equation}
{\cal EM}\,=\, \int {n_{\rm e}}^2\,dl
\label{eqn:EM}
\end{equation}
 where $n_{\rm e}$ is the electron density and $l$ is the line of sight distance through the source. We can now see why the thermal emission component can be found from H$\alpha$ emission because an integral over H$\alpha$ line emission frequencies  is also proportional to $\cal EM$, i.e. $\int I_{\rm H\alpha}\,d\nu\,\propto\,{\cal EM}$.  Consequently, an independently measured H$\alpha$ map can be used to find $I_{\rm TH}$ which can then be subtracted from the measured total emission map (Eq.~\eqref{eqn:sum2}) to find $I_{{\rm NT}}$ as has been done in the CHANG-ES papers listed above.

The challenges to using an H$\alpha$ map to find $I_{\rm TH}$ have been pointed out in detail by \citet{var18} and \citet{var19}, the main issue being uncertain dust absorption. H$\alpha$ emission is optical emission that is strongly affected by dust, something that is of particular concern for the edge-on galaxies in the CHANG-ES sample in which line-of-sight distances are large. Consequently, $22~\upmu$m emission maps (or some other appropriate infra-red map) are also required to help correct for the dust.  An assumption of electron temperature is also required, although variations in T$_{\rm{e}}$
may not be so severe because of its weak power in Eq.~\eqref{eqn:ITEM}. A value of ${\rm T_e}\,=\,10^{4}$ K is almost universally adopted. 

We now have a situation that, in order to find $I_{\rm NT}$ and $I_{\rm TH}$ separately, we require two additional maps and therefore two additional observing sessions at different telescopes (optical and infrared, if those maps are not publicly available) beyond the radio observations. Moreover, even though H$\alpha$ data are high resolution (of order an arcsec), IR data are relatively low resolution, in comparison.  For example, \citet{var18} achieved a relatively good 15 arcsec resolution in carrying out their `mixture method' of separating thermal from nonthermal emission in NGC~3044.  In addition to existing CHANG-ES radio data at L-band and C-band, they utilized 22 $\upmu$m as well as H$\alpha$ maps.  

From Eq.~\eqref{eqn:sum3}, we can see that the frequency dependence of the thermal and nonthermal components are different. Although $\alpha_{\rm NT}$ will vary with position,  a typical value is -0.7, indicating that the nonthermal component dominates at low radio frequencies whereas the thermal component, which varies as $\nu^{-0.1}$, becomes more important at high radio frequencies.  The nonthermal dominance at low frequencies has been exploited by telescopes such as the LOw Frequency ARray \citep[LOFAR;][]{van_Haarlem_2013} which observes at frequencies lower than 240 MHz. In that regime, the thermal component can be neglected and $\alpha_{\rm NT}$ can be measured directly from the observations \citep[e.g.][and many others]{pal09}, although the spatial resolution suffers at the low frequencies.

Now that S-band data have been added to the CHANG-ES lexicon, we have such a wide, almost contiguous frequency band ($\sim$ 4.5 GHz) at high spatial resolution that  it should be possible to {\it solve} for the thermal and nonthermal components {\it using only these radio data}. With sufficient signal-to-noise (S/N) and frequency coverage, Eq.~\eqref{eqn:sum3} can be solved for the four unknowns: 
$I_{{\rm NT}},~\alpha_{\rm NT},~\beta_{\rm NT},~{\rm and}~ I_{\rm TH}$. No additional optical or infra-red maps are required, and there is no need to assume a value for ${\rm T}_{\rm e}$ or rely on previous dust-correction calibrations. {The key to the success of this process is accurate, consistent calibration over all bands, as described in Sect.~\ref{sec:initial_spectrum}.}

\section{Radio Observations, Standard Data Reductions and Imaging -- {NGC~3044}}\label{sec:observations}

\subsection{Observations and Calibrations}
\label{sec:radioobs}

A summary of the VLA observations is given in Table~\ref{tab:observing}. 
Altogether, we have 6 independent data sets in three different observing bands and three array configurations (hereafter, 'arrays'). For S-band/C-array and for C-band/C-array, there were also two separate observing sessions, each of which was calibrated separately before the combination.  Each data set used the same primary flux calibrator (3C~286) and one of two possible secondary calibrators. Each band contains a number of `spectral windows' (spws) with each spw containing 64 spectral channels.  Note that there are more, but narrower, spws in L-band than the others\footnote{L-band was designed to have more channels in CHANG-ES observations as an aid for Faraday rotation analysis.}.  

Data reduction was carried out using the Common Astronomy Software Application (CASA) \citep{mcm07, CAS22}\footnote{See casa.nrao.edu.}. 
Reduction of the VLA data has been fully described in previous CHANG-ES papers, for example,  \cite{irw12b}, \cite{irw13}, \cite{irw19b}, and \citet{wie15}.  Calibrations for all data sets except S-band were carried out manually, using the standards described in these papers. 
{Each of these data sets was also self-calibrated \citep[e.g.][]{wie15}, or else it was determined that self-calibration did not produce improvements.}
%and the CASA versions used are indicated in the above-mentioned papers.  
The newer S-band data were calibrated using 
 the VLA pipeline (version 6.2.7.1) and CASA versions 5.7 and 6.4 were used for further processing and imaging of these data.

Prior to imaging, the various data sets had to be consolidated to a common weight scale so that they could be combined.  This was done by measuring the 
scatter in the visibilities as a function of time, antenna and/or baseline, and using this scatter to set the weights and weight errors for all visibilities\footnote{The CASA task, {\tt statwt}, was used.}.

\begin{table*}
\begin{center}
\caption{Observing Data {for NGC~3044}\label{tab:observing}}
\begin{tabular}{lccccccccc}
\hline
Band ($\nu_0$)$^{\rm a}$ & Freq. range$^{\rm b}$ & $\Delta \nu$$^{\rm c}$ &No. Spws$^{\rm d}$& Obs. Date$^{\rm e}$ & TOS$^{\rm f}$&Prim. Calib.$^{\rm g}$&Second. Calib.$^{\rm h}$&PB FWHM$^{\rm i}$ & LAS$^{\rm j}$\\ 
~~~~~~~~~~(GHz) & (GHz) & (GHz)& && (hours)& & &(arcmin) & (arcmin) \\
$\dotfill$ Array & & & & & & & \\
\hline\hline
L-band (1.58) & 1.247 $\to$ 1.503 & 0.512& 32& & &3C~286& &26.6  &  \\ 
                             &  1.647 $\to$ 1.903 & & & & \\
$\dotfill$ B & & &&21-03-2011 & 0.43& &J~1007-0207 (S)& & 2.0 \\  
$\dotfill$  C& & & &25-03-2012  &0.25& &J~0925+0019 (P)& & 16.2 \\                        
$\dotfill$  D& &  &&21-12-2011 &0.16& &J~1007-0207 (S)& & 16.2 \\                                  \hline
S-band (3.00) & 1.987 $\to$ 4.011 & 2.024 & 16 & & & 3C~286&J~0925+0019 (P)& 14.0& 8.2\\
$\dotfill$ C & & & &13-02-2020&0.69& & & & \\
              & & & &14-03-2020 &0.69& & & & \\
\hline
C-band (6.00) & 4.979 $\to$ 7.020 &  2.042& 16 &     &&3C~286&J~0925+0019 (P)&7.0    \\ 
$\dotfill$ C & & & &18-02-2012 & 1.00& & & &4.0\\
              & & & &06-03-2012&0.85& & & &4.0\\
$\dotfill$ D & & & &13-12-2011&0.31& & & &4.0\\
\hline
\end{tabular}
\end{center}
$^{\rm a}$ Observing band and its central frequency. VLA array configurations, corresponding to separate observing sessions, are given on their own lines following. A blank means that the value is the same as indicated immediately above.\\
$^{\rm b}$ Frequency range.  The small gap at L-band was necessary to eliminate radio frequency interference.\\
$^{\rm c}$ Total bandwidth.\\
$^{\rm d}$ Number of spectral windows in the band.\\
$^{\rm e}$ Date(s) of observations (day-month-year).\\
$^{\rm f}$ Time on source.\\
$^{\rm g}$ Primary calibrator. \\
$^{\rm h}$ Secondary calibrators. P = `primary' ( $<~3$\% amplitude close errors);  S = `secondary' ($3$ to $10$\%). \\
$^{\rm i}$ Primary beam full-width at half-maximum at the central frequency, given by $\rm PB ~{FWHM}=42/\nu$, where $\nu$ is in GHz and $\rm PB~{FWHM}$ is in arcmin. \\
$^{\rm j}$ Largest Angular Size detectable at the central frequency, from the {\it VLA Observational Status Summary} ({\tt https://science.nrao.edu/facilities/vla/docs/manuals}).\\ 
%https://science.nrao.edu/facilities/vla/docs/manuals/oss/performance/resolution
\end{table*}
%judith used listobs to get these numbers
%https://science.nrao.edu/facilities/vla/docs/manuals/oss/referencemanual-all-pages
% 

\subsection{Wide-band Radio Imaging}
\label{sec:radioimaging}

Our initial goal is to make use of all available data from all bands as summarized in Table~\ref{tab:observing}. This provides a standard `reference' image that uses the entire band. This was done using 
CASA's routine, {\tt tclean} which fits Taylor terms to Eq.~\eqref{eqn:sum1}. The full algorithm is explained in \cite{rau11}. 
The default {\tt tclean} reference frequency is the center of the entire band, but it can be set by the user.  For our observations, we use the default of   $\nu_0\,=\,4.13$ GHz.

There are several issues associated with this approach, however.  Firstly, the total emission is fit by Eq.~\eqref{eqn:sum1} which does not take into account any thermal component, as outlined in
Sect.~\ref{sec:separating}. Secondly, if $\beta$ is included in the fit, 
then the assumption is that the entire spectrum can be described by a single curvature value.  We will show in Sect.~\ref{sec:initial_spectrum} that this is not the case.
And thirdly, the use of the Taylor expansion assumes that any departure from $\nu_0$ is small so that the Taylor expansion will converge (i.e. $\frac{\nu\,-\,\nu_0}{\nu_0} \,\ll\,1$).  In fact, the CHANG-ES bandwidth is now so large that $\frac{\nu\,-\,\nu_0}{\nu_0}\,\approx\,\frac{2.9~{\rm GHz}}{4.13~{\rm GHz}}\,=\,0.7$ which is approaching 1, suggesting that higher terms in the expansion might have produced a more accurate result, had the S/N been sufficient to accommodate it. If the reference frequency had been shifted away from the center of the band, then the Taylor sequence would have diverged. 
Early work on the CHANG-ES galaxies showed that the best results for typical S/N values was a fit to Eq.~\ref{eqn:sum1} without $\beta$ \citep{irw12b}.

The final adopted inputs can be found in Appendix~\ref{app:fullband} as well as a description of frequency-dependent primary beam (PB) corrections for the total intensity map and spectral index (and error) maps.

\section{Results -- {NGC~3044}}
\label{sec:results}

\subsection{Wide-Band Imaging Results}\label{sec:radioimagingresults}

The best wide-band total intensity image and spectral index map, fit with Eq.~\eqref{eqn:sum1} with {\it no} curvature term, $\beta$,  are shown in Figure~\ref{fig:widebandmap} (see details in Appendix~\ref{app:fullband}). Data from these maps are in Table~\ref{tab:widebandresults}. Note that absence of $\beta$, implies that the spectrum can be fit with a single value of $\alpha$ over all frequencies.  This ignores any contribution from $I_{\rm TH}$ or any curvature in the nonthermal spectral index, $\beta_{\rm NT}$,  should it exist (Eq.~\ref{eqn:sum3}).

\begin{figure*}[hbt!]
    \centering    \includegraphics[width=5truein,height=3.4truein]{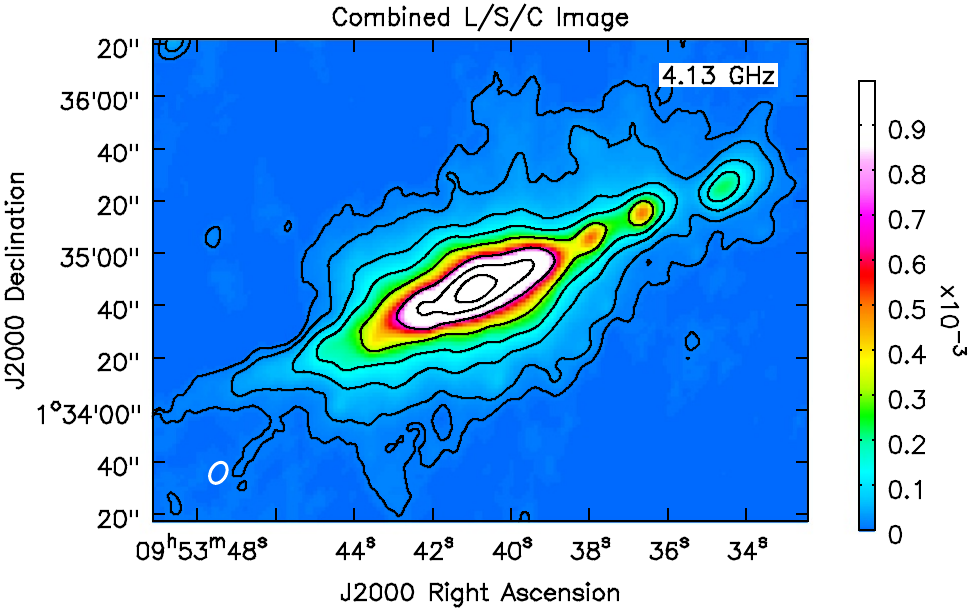}
  \includegraphics[width=5truein,height=3.4truein]{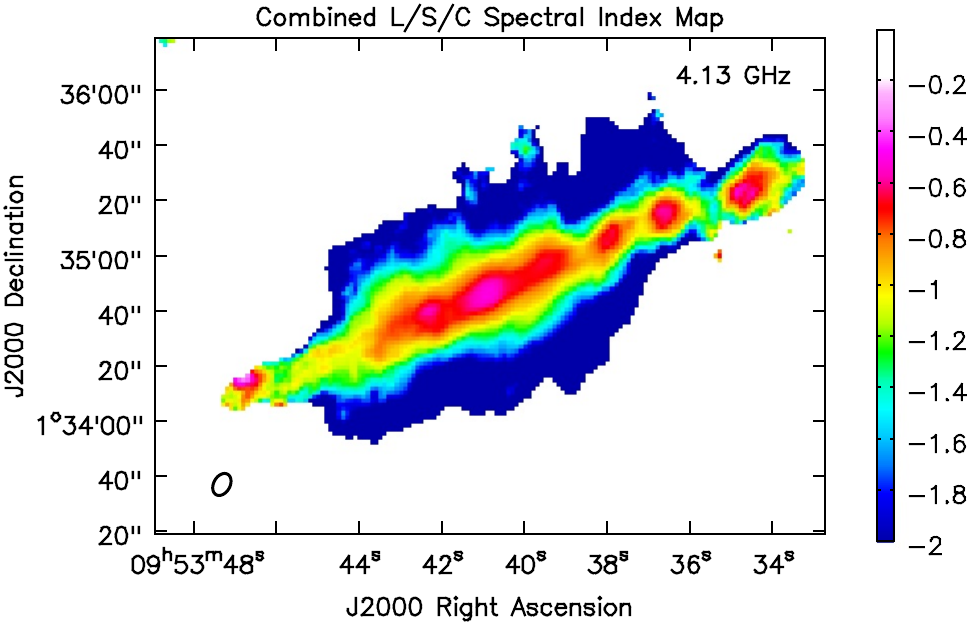}
    \caption{Combined L, S, and C-band images {of NGC~3044} at the central frequency of 4.13 GHz. The beam, shown as as ellipse at lower left, is $8\farcs 5 \times 5\farcs 9$ at a position angle of $-29\fdg 2$. The {\bf Top} image shows the total intensity map, uncorrected for the PB. Contours are at 10 ($2\,\sigma$), 30, 75, 150, 300, 700, 1400, and 2800 $\upmu$Jy beam$^{-1}$ with a  maximum of 4.90 mJy beam$^{-1}$.  The colour wedge is in units of Jy beam$^{-1}$. The {\bf Bottom} image shows the spectral index map, after applying a 5$\sigma$ cutoff and corrected for the PB.
}
    \label{fig:widebandmap}
\end{figure*}

{\begin{table*}[hbt!]
\fontsize{7}{11}\selectfont
\begin{center}
\caption{Wideband (1.25--7.02 GHz) Imaging Results {of NGC~3044}, centered at 4.13 GHz}\label{tab:widebandresults}
    \begin{tabular}{ccccccc}
    \hline
Flux Density$^{\rm a}$ & Map Max$^{\rm b}$& $\alpha$ (max)$^{\rm c}$&$\alpha$ (min)$^{\rm c}$ & $\alpha$ (avg)$^{\rm c}$&$\alpha$ (med)$^{\rm c}$& $\overline{\alpha_w}$ \\
mJy & mJy beam$^{-1}$ & & & & &\\
\hline
$53.4\,\pm\,1.2$  &  4.90 $\pm$ 0.10  & $-0.33\,\pm\,0.11$ &$-4.58\,\pm\,0.86$ &$-1.69\,\pm\,0.21$ &$-1.55\,\pm\,0.12$& $-0.682\,\pm\,0.016$\\
\hline
    \end{tabular}
\end{center}
$^{\rm a}$ Flux density from the PB-corrected map.  The uncertainty is computed from $\sqrt{\rm Nbeams}\,\times\,{\rm rms}\,+\,$ a 2\% calibration error \citep{wie15} added in quadrature.  Nbeams represents the number of beams in the measured area, rms represents the rms map noise. \\
$^{\rm b}$ Maximum value of the map at the center.  The error represents the rms $+$ a 2\% calibration error added in quadrature. \\
$^{\rm c}$ Maximum, minimum, average, and median values from the spectral index map shown in Figure~\ref{fig:widebandmap} (bottom). The errors are from the corresponding spectral index error map (not including calibration error).\\
$^{\rm d}$  Weighted average spectral index over the region shown in Figure~\ref{fig:widebandmap} using Eq.~\eqref{eq:weighted_alpha}. The uncertainty is an estimate, based on Eq.~\eqref{eq:weighted_alpha_error} plus a 2\% error in the weights, added in quadrature.\\
\end{table*}
}

The wideband image using all bands resembles the image of Figure~\ref{fig:N3044_fig1}, as it should, although it shows some additional structure and some extensions to higher $z$ values.  As for the spectral index map, it is clear that the spectral index steepens with distance from the plane.  This is consistent with
\citet{wie15,var18,hea22,ste23} and others.
The average and median spectral indices are very steep (Table~\ref{tab:widebandresults}), but this is because the values are calculated from Figure~\ref{fig:widebandmap} (bottom), as if all points are weighted equally. An observation that detects less of the steeper halo would pick up flatter values of $\alpha$ that are more typical of the disk.  We therefore weigh each value of $\alpha_i$,  by the specific intensity, $I_i$, at any pixel, $i$,  of the total intensity map (Figure~\ref{fig:widebandmap}). That is,
\begin{eqnarray}\label{eq:weighted_alpha}
\overline{\alpha_w}&=&\frac{\sum\, \alpha_i\,w_i}{\sum\,w_i}~~~~~~~~~~
{\rm where}~~~~~~~w_i\,=\,(I_i)^2\\
\overline{\Delta\alpha_w}&=&\frac{\sum\, \Delta\alpha_i\,w_i}{\sum\,w_i}~~~~~~~~~~
{\rm where}~~~~~~~w_i\,=\,(I_i)^2
\label{eq:weighted_alpha_error}
\end{eqnarray}
where Eq.~\eqref{eq:weighted_alpha_error} repeats the process for the spectral index error map.
The weighted average, $\overline{\alpha_w}$ is given in Table~\ref{tab:widebandresults} along with an estimate of its uncertainty. Our result of 
$\overline{\alpha_w}\,\sim\,-0.7$ for is typical of in-disk spectral indices seen in other galaxies 
\citep[e.g.][]{mag15,kle18}.

%\vspace{1truein}

\subsection{Narrow-Band Imaging and the Initial Radio Spectrum}
\label{sec:initial_spectrum}

As a first step towards understanding the spectral shape,  we produce a series of images at each frequency from L-band, through S-band and C-band, each of which covers a small frequency range ($\Delta\,\nu\,=\,0.128$ GHz). {The value of $\Delta\,\nu$ was adopted so that it was the same in all bands, narrow enough to define the spectral shape, and wide enough to ensure sufficient S/N that the individual maps would iterate to a satisfactory image.} Thirty
maps were made over these narrow frequency ranges, as described in Appendix~\ref{app:narrowband}.  Each map was smoothed to the same spatial resolution and corrected for the primary beam (PB).  A list of these maps is given in Table~\ref{tab:im_params}.  The maximum value in the galaxy occurs at the map center, $I_{\rm max}$, and this value is listed, along with the rms noise of the PB-corrected images.  %Note that the rms noise values are high because of the limited frequency ranges used. However, a 2\% 

{We examine the spectrum at the peak position, which corresponds to the center of the galaxy, and we also examine the spectrum of a flux density which incorporates more of the disk rather than a single point. For the latter, we adopt the region enclosed by the grey curve shown in Figure~\ref{fig:N3044_fig1} which encloses 33 beams. These two options should provide a good estimation of galaxy's spectral shape.}

\begin{figure*}[hbt!]
    \centering    \includegraphics[width=5truein,height=3.4truein]{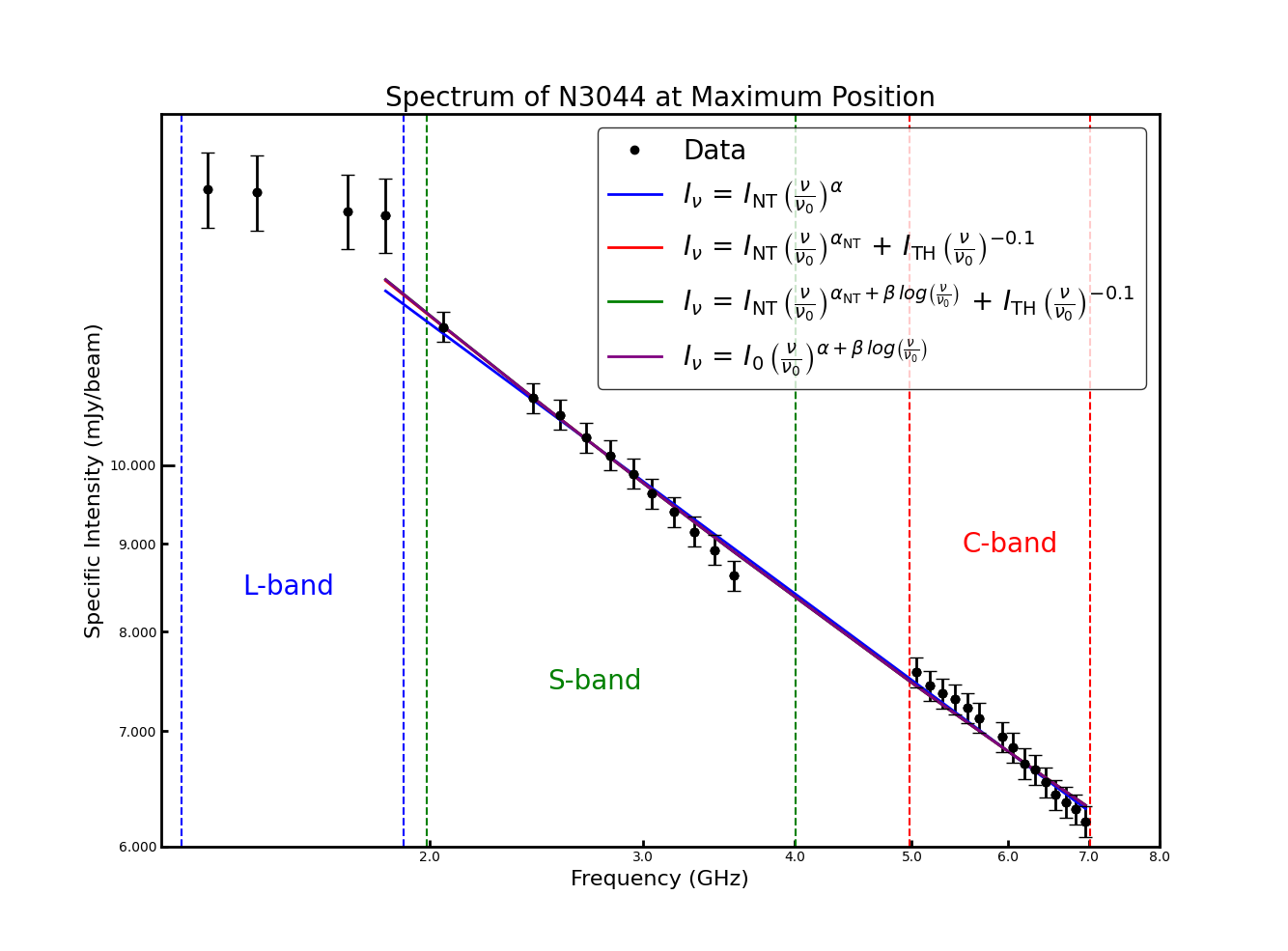}
    \vskip -0.2truein
\includegraphics[width=5truein,height=3.4truein]{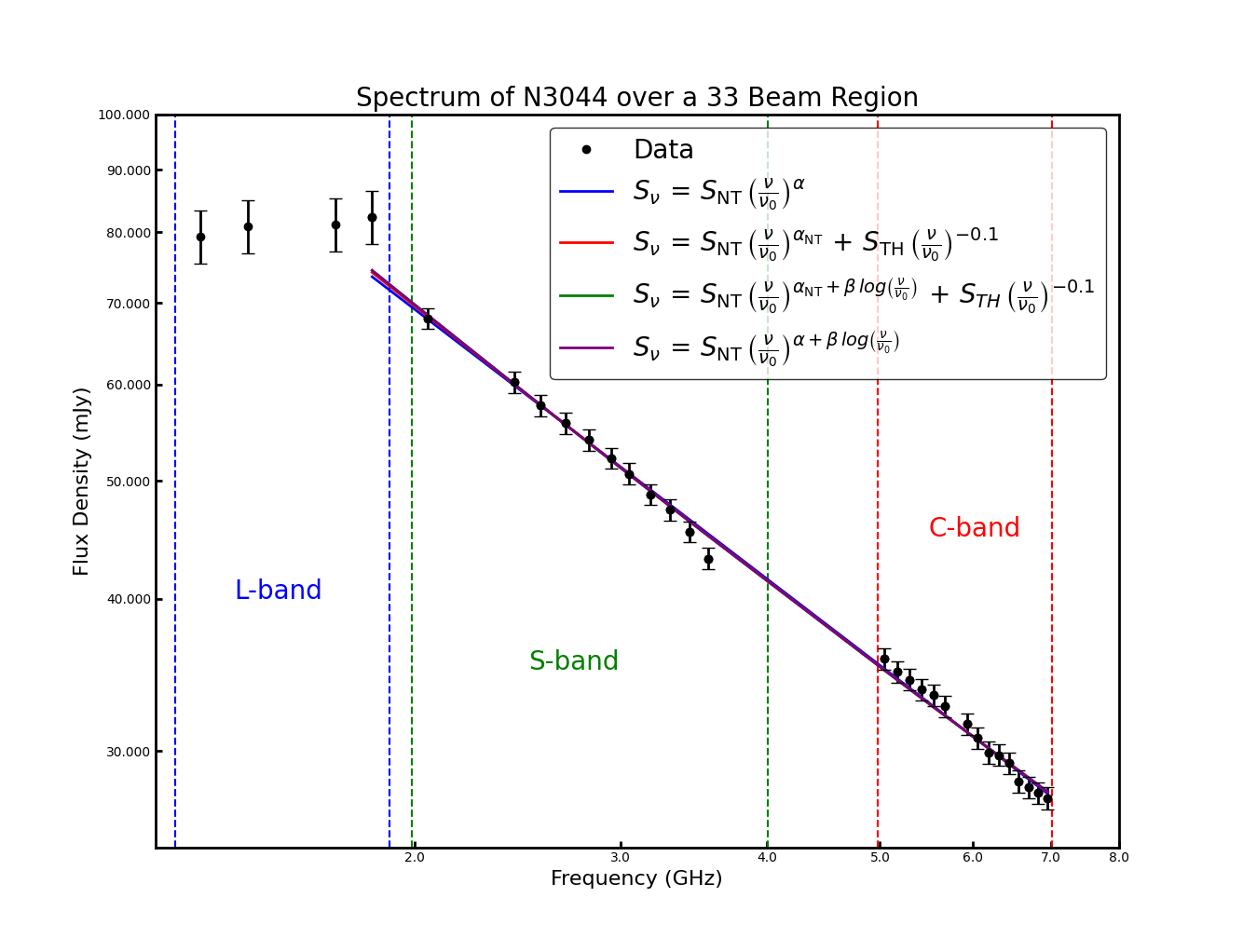}
    \caption{Spectrum of NGC~3044 from narrow-band imaging, omitting the lowest three frequency points.  Top: Values for the central maximum point only. Error bars add in quadrature: a 2\% or 5\% calibration error (Appendix~\ref{app:calibration}) and the rms from
    Table~\ref{tab:im_params}. Bottom: flux density over a 33 beam region that is common to all maps and is encircled by the grey dashed curve in Figure~\ref{fig:N3044_fig1}. Error bars add in quadrature a 2\% or 5\% calibration error (Appendix~\ref{app:calibration}) with the rms from
    Table~\ref{tab:im_params} multiplied by the square root of the number of beams. For each plot, the various frequency bands are marked, and the fitted equations are given in the legend.  Fitted parameters are given in Table~\ref{tab:spectral_fits_27points}.  }
    \label{fig:spectra_27points}
\end{figure*}

The spectrum of the map center, $I_{\rm max}$, is shown in the top plots of 
Figure~\ref{fig:spectra_all} (see Appendix~\ref{app:spectra_allfreqs} for fits to all 30 points) and Figure~\ref{fig:spectra_27points} (fits to 27 points).
{The error bars in these plots include the rms noise of Table~\ref{tab:im_params} along
with a 2\% or 5\% calibration error (described in Appendix~\ref{app:calibration}), added in quadrature.}
The spectrum of the flux, $S_\nu$, that is contained in the 33 beam region enclosed by the grey contour of Figure~\ref{fig:N3044_fig1} is shown in the bottom plots of Figures~\ref{fig:spectra_all} and  \ref{fig:spectra_27points}.
The uncertainty on the flux is taken to be the rms from Table~\ref{tab:im_params} times the square root of the number of beams in the region, with the result added in quadrature to a 2\% or 5\% calibration error.
For all plotted points in these figures, calibration errors dominate over the rms errors.
 A Python code was written to carry out the measurements of rms, map peak, and flux, using the same regions for every map.
% \footnote{This code, called {\it findstats.py}, is available at {\tt https://projects.canfar.net/changes/publications}.}.

Four different functions, shown in the top right box of each plot, were fitted to the data. {These fitted functions are versions of Eqs.~\ref{eqn:sum1} to \ref{eqn:sum3} and are listed as A to D.}
The fitting was carried out via the Python program,
{\tt curve\_fit}, as implemented in the {\tt scipy.optimize} package {(see Appendix~\ref{app:curve_fitting} for details)}. %Each point is weighted by its error, and the quantities, $I_{\rm TH}$ and $I_{\rm NT}$, are restricted to be positive.  %A Python code was written to do the fitting and make the plots.
%\footnote{This code, called {\it fitacurve.py}, is available at {\tt https://projects.canfar.net/changes/publications}.}.
{The fits to all 30 frequencies are shown in Figure~\ref{fig:spectra_all} and corresponding data are listed in Table~\ref{tab:spectral_fits_all}.  %These fits are rather poor, as measured 
%by the rms.  Consequently, we repeated the exercise for just
Fits to 27 frequencies are shown in 
(Figure~\ref{fig:spectra_27points}) with corresponding data in  
Table~\ref{tab:spectral_fits_27points}. %The resulting rms is significantly lower.  %The following discussion, however, refers to both.

{\begin{table*}[ht]
\fontsize{7}{11}\selectfont
\begin{center}
\caption{Map Parameters at 12 arcsec resolution}\label{tab:im_params}
    \begin{tabular}{lccccccccccc}
    \hline
ID Number$^{\rm a}$& Band $^{\rm b}$&  spw$^{\rm c}$& $\nu_0$$^{\rm d}$ & $\Delta\,\nu$$^{\rm e}$
& $I_{\rm max}$$^{\rm f}$ & rms$^{\rm g}$\\ 
& & & (GHz) & (GHz) & (mJy beam$^{-1}$) & ($\upmu$Jy beam$^{-1}$) & & &  &  &\\
\hline
1&L   & 0 $\to$ 7 &  1.31090 & 0.12819 & 14.468  & 87
   & & & & \\
2&L   & 8 $\to$ 15 & 1.43890  & 0.12821 & 14.408& 64
   & & & & \\
3&L   & 16 $\to$ 23 &  1.71090 & 0.12825 & 14.053& 45
   & & & & \\
4&L   & 24 $\to$ 31 & 1.83891  & 0.12826 & 13.980 & 57
   & & & & \\
5&S   & 16 & 2.05109  & 0.12811 & 12.032& 21
   & & & & \\
6&S   & 19 & 2.43511  &  0.12813 &10.940 & 15
   & & & & \\
7&S   & 20 &  2.56311 & 0.12814 & 10.696& 12
   & & & \\
8&S   & 21  &  2.69112 & 0.12814 & 10.374& 11
   & & & \\
9&S   & 22 & 2.81912  & 0.12815 & 10.131& 13
   & & & & \\
10&S   & 23  & 2.94713  &  0.12816& 9.885& 12
   & & & & \\
11&S   & 24 & 3.05113  & 0.12816 & 9.623& 12
   & & & & \\
12& S  & 25&  3.17914 & 0.12817 & 9.389& 12
   & & & & \\
13& S  & 26&  3.30714 & 0.12818 & 9.146& 10
   & & & & \\
14& S  & 27&  3.43515 & 0.12818 & 8.925& 14
   & & & & \\
15& S  & 28&  3.56315 & 0.12819 & 8.624& 15 
    & & & & \\
16& C   & 0  & 5.04199 & 0.12860 & 7.577& 18
   & & & & \\
17&C   & 1 &  5.16999 & 0.12862  & 7.443& 18
   & & & & \\
18&C   & 2 & 5.29799  & 0.12863  & 7.365& 19
   & & & & \\
19&C   & 3 &  5.42599 & 0.12865 & 7.308& 18
   & & & & \\
20&C   & 4 &  5.55399 & 0.12866 & 7.223& 19
   & & & & \\
21&C   & 5 &  5.68199 & 0.12868  & 7.126& 17
   & & & & \\
%22&C   & 6 &  5.80999 & 0.12870  & 6.784& 20
%   & & & & \\
%Judith I've renumbered these next maps -- add one
%to check them on disk
22&C   & 7 &  5.93799 & 0.12871  & 6.946& 21
   & & & & \\
23&C   & 8 &  6.05998 & 0.12873  & 6.849& 25
   & & & & \\
24&C   & 9 &  6.18798 & 0.12874  & 6.705& 24
   & & & & \\
25&C   & 10 &  6.31598 & 0.12876  & 6.650& 24
   & & & & \\
26&C   & 11 &  6.44398 & 0.12877  & 6.539& 20
   & & & & \\
27&C   & 12 &  6.57198 & 0.12879  & 6.430& 27
   & & & & \\
28&C   & 13 &  6.69998 & 0.12880  & 6.366& 22
   & & & & \\
29&C   & 14 &  6.82798 & 0.12882  & 6.307& 16
   & & & & \\
30&C   & 15 &  6.95598 & 0.12883  & 6.206& 28
   & & & & \\
   \hline
    \end{tabular}
\end{center}
$^{\rm a}$ Unique identification number of the map.\\
$^{\rm b}$ Frequency band from which the map was made. \\
$^{\rm c}$ Spectral windows in the band(s) from which the map was made. Ranges for each band were L-band: 0 - 31; S-band: 16 - 31; C-band: 0 - 15. \\
$^{\rm d}$ Central frequency of the map.\\
$^{\rm e}$ Frequency width of the map.\\
$^{\rm f}$ Value of the maximum of the emission at the map center.\\
$^{\rm g}$ Map rms value measured from the primary beam-corrected maps, using 4 different off-source boxes that cover identical regions in each map.\\
\end{table*}
}

%N.B. for total flux spectrum, cut off all .pb corrected maps at 3sigma.  but the sizes are all different, in some the halo is more readily seen than others dependent on s/n etc.  So then cut off all maps to the same area as the smallest map so that total flux is being compared over the same area for all

%The fitted results using all 30 frequencies are given
%in Table~\ref{tab:spectral_fits_all} 
%with the fitted curves listed from A to D. A measure of the goodness of fit is provided by the rms in this table, which measures how closely the fits match the data.
Some important conclusions can be drawn from the simple plots of Figure~\ref{fig:spectra_all} and Figure~\ref{fig:spectra_27points} and their associated tabular values:
{
\begin{itemize}
    \item {\it The spectrum flattens at low frequency (L-band).}  This is an unexpected result which we will discuss in 
    Sect.~\ref{sec:turnover}. The fitted curves cannot reproduce such a turnover without a value of $\beta_{\rm NT}$, i.e. the nonthermal spectrum requires curvature to explain the flattening in L-band.  Even with $\beta_{\rm NT}\,\ne\,0$, however, the fits are poor at low frequencies (Figure~\ref{fig:spectra_all}). 
%    \item {\it Trials (not shown) show a weaker or negligible turnover at high $z$}. It is more difficult to fit a spectrum to the halo because of the lower signal-to-noise in those regions.
    \item {\it Fits that include either curvature and/or a thermal component generally improve over fits that have a simple curve of type A (i.e. $I_\nu\,\propto\,\nu^\alpha$). } This is evident by the lower rms scatter in Table~\ref{tab:spectral_fits_all} and Table~\ref{tab:spectral_fits_27points}, although the improvement in rms tends to be small.
 %   \item {\it Previous measurements of the spectral index, which take a single L-band and a single C-band value and fit a curve of type A, also called band-to-band spectral indices, likely produce values of $\alpha$ that are systematically too flat.}  An example is in \cite{irw19b}. This is a consequence of the flattening in L-band.
    \item{\it Curvature as a result of a thermal contribution only (Type B) cannot be distinguished mathematically from curvature in the nonthermal spectrum only (Type D). } However, we know that $I_{TH}$ must be present; the question is whether its contribution to the curvature dominates over any intrinsic curvature in the nonthermal spectrum.
\end{itemize}
}

{A contribution from a thermal component is  physically motivated.  However, the form of curvature in the nonthermal component, $\beta_{\rm NT}$, is not physically motivated.  It is simply a convenient way to express the possible shape of the curvature of $I_{\rm NT}$. For example, a negative $\beta_{\rm NT}$ flattens the curve at low frequencies and steepens it at high frequencies. One might have adopted more physically motivated curves, such as from radiative or diffusive losses or winds \citep[e.g.][]{hee21,wer21}. However, it is clear from the uncertainties in Tables~\ref{tab:spectral_fits_all} and \ref{tab:spectral_fits_27points} (Type C) that our data cannot handle more free parameters.  
A curve that includes both a flattening at low frequency plus a flattening again at high frequency would have a shape like an inverse integral sign: \curveshape. The low-frequency flattening  
is not well reproduced by any of the fits, 
and the error bars are larger at low frequencies (Appendix~\ref{app:calibration}). 
This argues for omitting the lowest frequency 3 points where it is clear that the flattening is occurring. That is, from high to low frequency, a clear change of spectral shape begins to occur at $\nu\,<\, 1.84$ GHz.
A comparison of Table~\ref{tab:spectral_fits_all} and Table~\ref{tab:spectral_fits_27points} shows that the fitted parameters 
actually agree between the 30-point and 27-point results. But the fits for only 27 frequency points have improved,
as measured by the rms scatter about the curves. For example, the rms scatter about 27 points decreases by $\approx\,75$\% compared to 30 points.

%The simplest curve which includes physically motivated parameters with sufficiently few free parameters that can be handled by the data, would be described by a form of Type B; that is, a non-curving (constant $\alpha$) nonthermal spectrum plus a thermal component. 

Finally, although the various fits are mostly indistinguishable mathematically, the simplest physically-motivated curve has a single value of $\alpha$, with no nonthermal curvature ($\beta_{\rm NT}\,=\,0$).This does not rule out the possibility that some nonthermal curvature could also exist; only that the data cannot support the increase in free parameters that would be required in such a case.}}

%total fit without point 11 gives:S = S_0 nu^alpha_NT + S_th:
%the best fit value of S_0 is 18.8120371886+/- a standard error of 4.05123707321
%the best fit value of alpha is -1.40821803563+/- a standard error of 0.201444773908
%the best fit value of S_th is 25.7481080855+/- a standard error of 3.73989808434
%the thermal fraction is S_th/S_tot =0.577828189902
%the Root Mean Square Error = 1.61735228321

 %we have carried out such a fit and find that it is indistinguishable from our red and green fits in Figure~\ref{fig:spectra_all}. Since this fit is permitted by the CASA task, {\tt tclean}, we can still use it to find the best value of $I_\nu$, as explained in Sect.~\ref{sec:actual_sep}.

{\begin{table*}[!htb]
\fontsize{7}{11}\selectfont
\begin{center}
\caption{Spectral Fits {of NGC~3044}-- 12 arcsec resolution -- 27 points}\label{tab:spectral_fits_27points}
    \begin{tabular}{lcccccc}
    \hline
    {\it Fitted Expression}$^{\rm a}$&&&&&&\\
    \hline
{\bf Maximum Position}$^{\rm b}$&  $I_{\rm NT}$$^{\rm c}$& $\alpha$ or $\alpha_{NT}$ $^{\rm c}$& $\beta$ or  $\beta_{\rm NT}$ $^{\rm c}$&$I_{\rm TH}$$^{\rm c}$ & 
$\frac{I_{\rm TH}}
{I_{\rm tot}}$$^{\rm d}$ 
& rms scatter$^{\rm e}$
\\
 & (mJy beam$^{-1}$) & & & (mJy beam$^{-1}$) & & (mJy beam$^{-1}$)\\
\hline
\hline
{\bf A:} $I_\nu\,=\,I_{\rm NT}\,\left(\frac{\nu}{\nu_0} \right)^\alpha$&
$8.28\,\pm\,0.03 $& $-0.521\,\pm\,0.010$ &  & & & 0.28\\
{\bf B:} $I_\nu\,=\,I_{\rm NT}\,\left(\frac{\nu}{\nu_0} \right)^{\alpha_{\rm NT}}\,+\, I_{\rm TH}\,\left(\frac{\nu}{\nu_0} \right)^{-0.1}$ &
$6.06\,\pm\,1.94$ & $-0.672\,\pm\,0.178$& &$2.19\,\pm\,1.89$ & $0.27\,\pm\,0.18$ & 0.25\\
{\bf C:} $I_\nu\,=\,I_{\rm NT}\,\left(\frac{\nu}{\nu_0} \right)^{\alpha_{\rm NT}\,+\,\beta_{\rm NT}\,log(\frac{\nu}{\nu_0})}\,+\, I_{\rm TH}\,\left(\frac{\nu}{\nu_0} \right)^{-0.1}$ &
$8.00\,\pm\,2396$ & $-0.532\,\pm\,129$ & $0.072\,\pm\,42.0$ &$0.24\,\pm\,2396$ & $0.029\,\pm\,282$ & 0.25\\
{\bf D:} $I_\nu\,=\,I_{\rm NT}\,\left(\frac{\nu}{\nu_0} \right)^{\alpha_{\rm NT}\,+\,\beta\,log(\frac{\nu}{\nu_0})}$ &
$8.24\,\pm\,0.06$& $-0.520\,\pm\,0.010$ &
$0.075\,\pm\,0.085$& & & 0.25\\
\hline
{\bf Region (33 beams)}$^{\rm b}$
& $S_{\rm NT}$$^{\rm c}$& $\alpha$ or $\alpha_{NT}$$^{\rm c}$& $\beta$ or  $\beta_{\rm NT}$ $^{\rm c}$ &$S_{\rm TH}$$^{\rm c}$ & $\frac{S_{\rm TH}}{S_{\rm tot}}$$^{\rm d}$ & rms$^{\rm e}$
\\
& (mJy) & & & (mJy) & & (mJy) \\
\hline\hline
{\bf A:} $S_\nu\,=\,S_{\rm NT}\,\left(\frac{\nu}{\nu_0} \right)^\alpha$&
$40.58\,\pm\,0.16 $& $-0.735\,\pm\,0.010$ &  & & & 1.8\\
{\bf B:} $S_\nu\,=\,S_{\rm NT}\,\left(\frac{\nu}{\nu_0} \right)^{\alpha_{\rm NT}}\,+\, S_{\rm TH}\,\left(\frac{\nu}{\nu_0} \right)^{-0.1}$ &
$36.76\,\pm\,6.61$ & $-0.799\,\pm\,0.121$& &$3.70\,\pm\,6.37$ & $0.091\,\pm\,0.14$ & 1.7\\
{\bf C:} $S_\nu\,=\,S_{\rm NT}\,\left(\frac{\nu}{\nu_0} \right)^{\alpha_{\rm NT}\,+\,\beta_{\rm NT}\,log(\frac{\nu}{\nu_0})}\,+\, S_{\rm TH}\,\left(\frac{\nu}{\nu_0} \right)^{-0.1}$ &
$40.42\,\pm\,149$ & $-0.734\,\pm\,2.36$ & $0.060\,\pm\,1.43$ &$0.00\,\pm\,149$ & $0.00\,\pm\,1.43$ & 1.6\\
{\bf D:} $S_\nu\,=\,S_{\rm NT}\,\left(\frac{\nu}{\nu_0} \right)^{\alpha_{\rm NT}\,+\,\beta\,log(\frac{\nu}{\nu_0})}$ &
$40.42\,\pm\,0.28$ & $-0.734\,\pm\,0.010$& $0.060\,\pm\,0.085$ & & &1.6
\\
\hline
    \end{tabular}
\end{center}
$^{\rm a}$ The type of fit is given by the mathematical expressions shown.  For a single point (Maximum Position), the values are expressed as specific intensities, $I_\nu$, and are in mJy beam$^{-1}$. For a larger region (33 beams), the values are flux densities, $S_\nu$, in mJy. The beam is circular with a FWHM of 12 arcsec. Values are computed at the central frequency, $\nu_0\,=\,4.13$ GHz.\\
$^{\rm b}$ Position at which the fit was carried out. Maximum Position: center of the map at which the emission is a maximum.  Region (33 beams): Spatially integrated region within the areas enclosed by the grey curve in Figure~\ref{fig:N3044_fig1}. \\
$^{\rm c}$ Parameters of the fit with their standard deviations. \\
$^{\rm d}$ Thermal fraction at $\nu_0$.  The denominator is the sum of the thermal and nonthermal emission.\\
$^{\rm e}$ Root-mean-square scatter between the data and fitted curve.\\
\end{table*}
}

\subsection{Narrow-Band Maps and Results from Cube-Fitting}
\label{sec:cube_fits}

From Sect.~\ref{sec:initial_spectrum}, we find that the expression,
\begin{equation}
I_\nu\,=\,I_{{\rm NT}}\,\left(\frac{\nu}{\nu_0} \right)^{\alpha_{\rm NT}}\,+\, I_{\rm TH}\,\left(\frac{\nu}{\nu_0} \right)^{-0.1}
\label{eqn:fittedeqn}
\end{equation}
is the {simplest physically motivated} fit to 27 frequency points (Expression B in Table~\ref{tab:spectral_fits_27points}). %Expressions that include no thermal component are not physical, and the expression that includes both a thermal component, as well as curvature in the nonthermal spectrum (i.e. $\beta\,\ne\,0$), cannot be distinguished as unique by our data.

We now expand upon this approach by forming a cube of all data in order to carry out a fit of Expression B for every real point in the plane. 
The fitting routine is the same as was used for the maximum point and the flux over 33 beams (Sect~\ref{sec:initial_spectrum}).  In addition,  we apply a user-supplied cutoff of 2$\sigma$, where $\sigma$ is the frequency-dependent rms noise given in Table~\ref{tab:spectral_fits_27points}.
Any input data point falling below this threshold is blanked.  Error bars include the rms noise, plus a 2\% or 5\% calibration error (Appendix~\ref{app:calibration}), added in quadrature, and these values enter into the {\tt scipy.optimize curve\_fit} routine, {both as a weighing for each point as well as in calculating the standard deviations, $\sigma_p$, on each parameter, $p$}. As before,
$I_{\rm NT}$ and $I_{\rm TH}$ (the values at $\nu_0\,=\,4.13$ GHz) are restricted to be positive.  

The result provides spatially resolved maps of $I_{\rm NT}$, $\alpha_{\rm NT}$, and $I_{\rm TH}$, along with maps of their standard deviations. The standard deviations for
$I_{\nu_{0_{\rm NT}}}$ and $I_{\rm TH}$  were converted to relative errors { for quick and easy comparison} while the standard deviation for $\alpha_{\rm NT}$ was kept as an absolute error { which is clearer for an exponent }. In addition,
wherever the relative errors exceeded 2.0, the affected points were also blanked.  In practice, this occurred where $I_{\rm TH}$ fell to low levels (i.e. were indistinguishable from zero). 
{The thermal fraction, $Fr$, was calculated according to
\begin{equation}
\label{eqn:thermal_fraction}
Fr\,=\,\frac{I_{\rm TH}}{I_{\rm TH}\,+\,I_{\rm NT}}
\end{equation} 
and its error from the standard expression,
\begin{eqnarray}
\sigma_{Fr}&=&\sqrt{
\left(\frac{\partial Fr}{\partial I_{\rm TH}}\,\sigma_{I_{\rm TH}}\right)^2\,+\,
\left(\frac{\partial Fr}{\partial I_{\rm NT}}\,\sigma_{I_{\rm NT}}\right)^2}\\
&=&
\sqrt{
\frac{
\left(I_{\rm NT}\,\sigma_{I_{\rm TH}}\right)^2
\,+\,
\left(I_{\rm TH}\,\sigma_{I_{\rm NT}}\right)^2
}
{\left({I_{\rm TH} + I_{\rm NT}}\right)^4}
}
\end{eqnarray}
%see https://nicoco007.github.io/Propagation-of-Uncertainty-Calculator/
%or
%https://astro.subhashbose.com/error_propagation_calc
where $\sigma_{I_{\rm TH}}$ and
$\sigma_{I_{\rm NT}}$ are the standard deviations in ${I_{\rm TH}}$ and
${I_{\rm NT}}$, respectively.
The result was then converted to a relative error.}

\begin{figure*}[hbt!]
   \centering
  \includegraphics[width=0.7\textwidth]{ 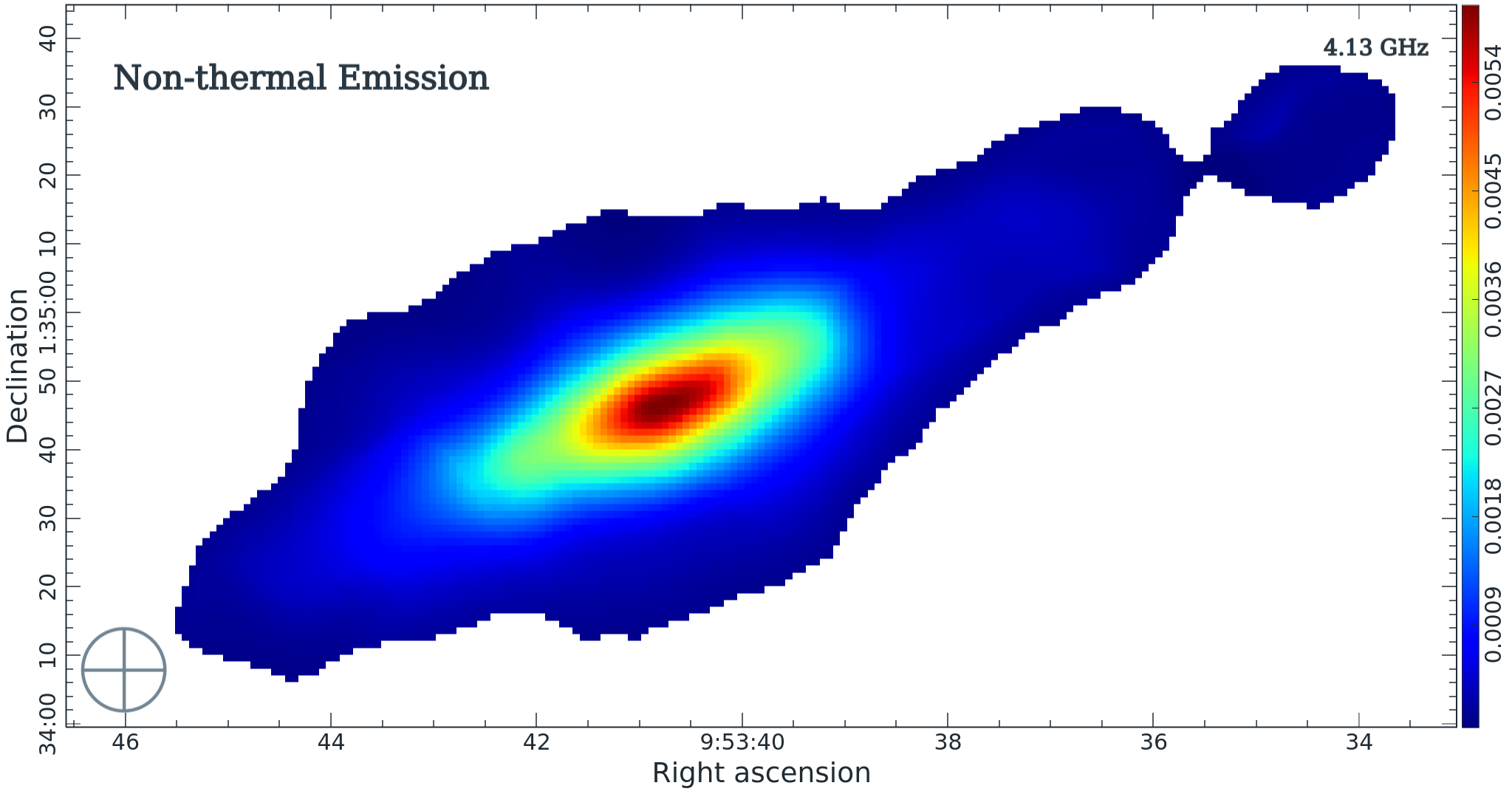}  
  \includegraphics[width=0.70\textwidth]{ 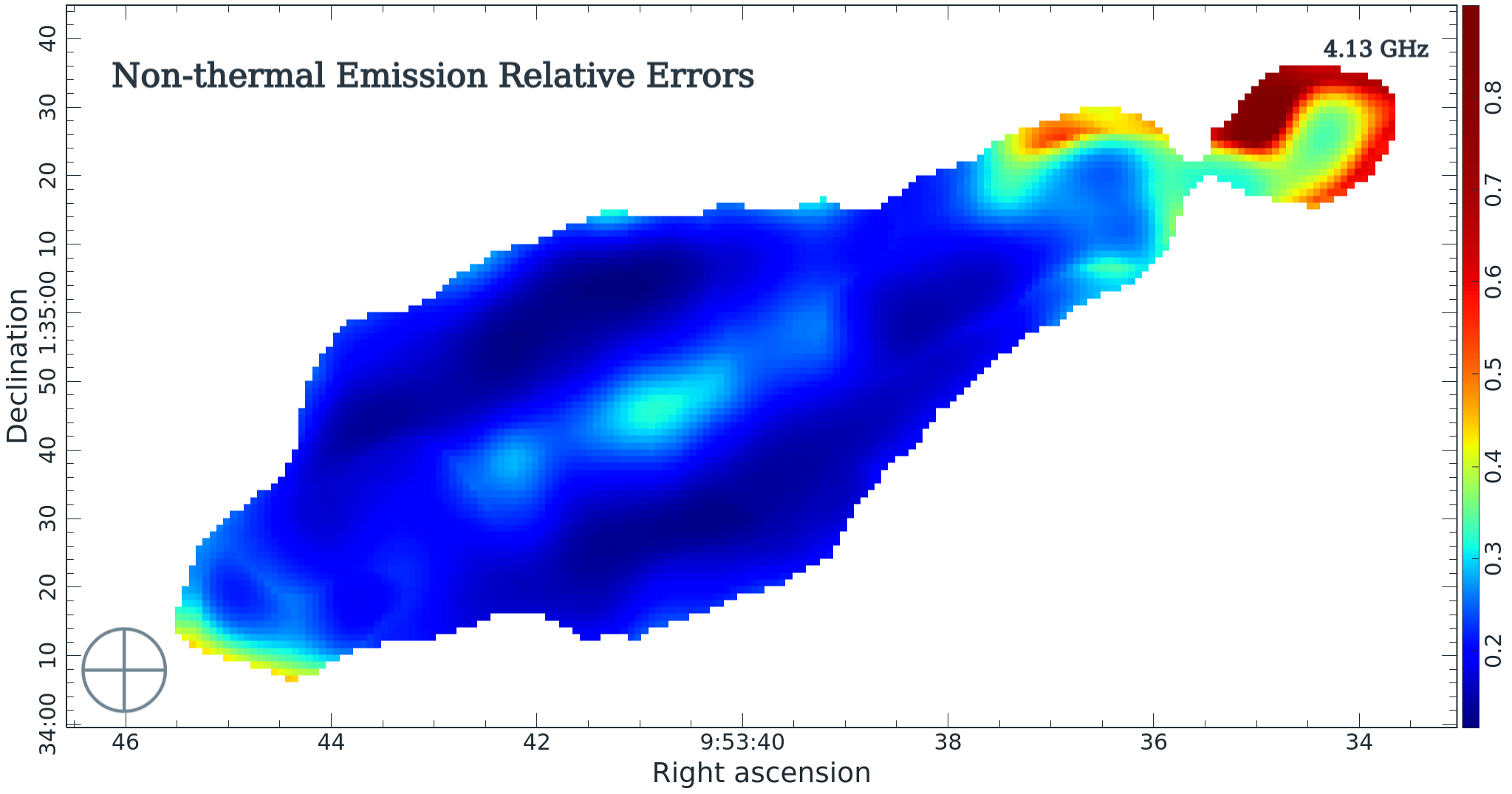}    
  \caption{Nonthermal emission (top) and its relative error (bottom) of NGC3044 resulting from cube-fitting, as described in Sect.~\ref{sec:cube_fits}. Units, for the image, are Jy beam$^{-1}$. The $12\arcsec$ beam is shown at lower left and the frequency is at the upper right.
  Corresponding data are in Table~\ref{tab:imaging_results}. }
  \label{fig:INT_result}
\end{figure*}

\begin{figure*}[hbt!]
   \centering
  \includegraphics[width=0.7\textwidth]{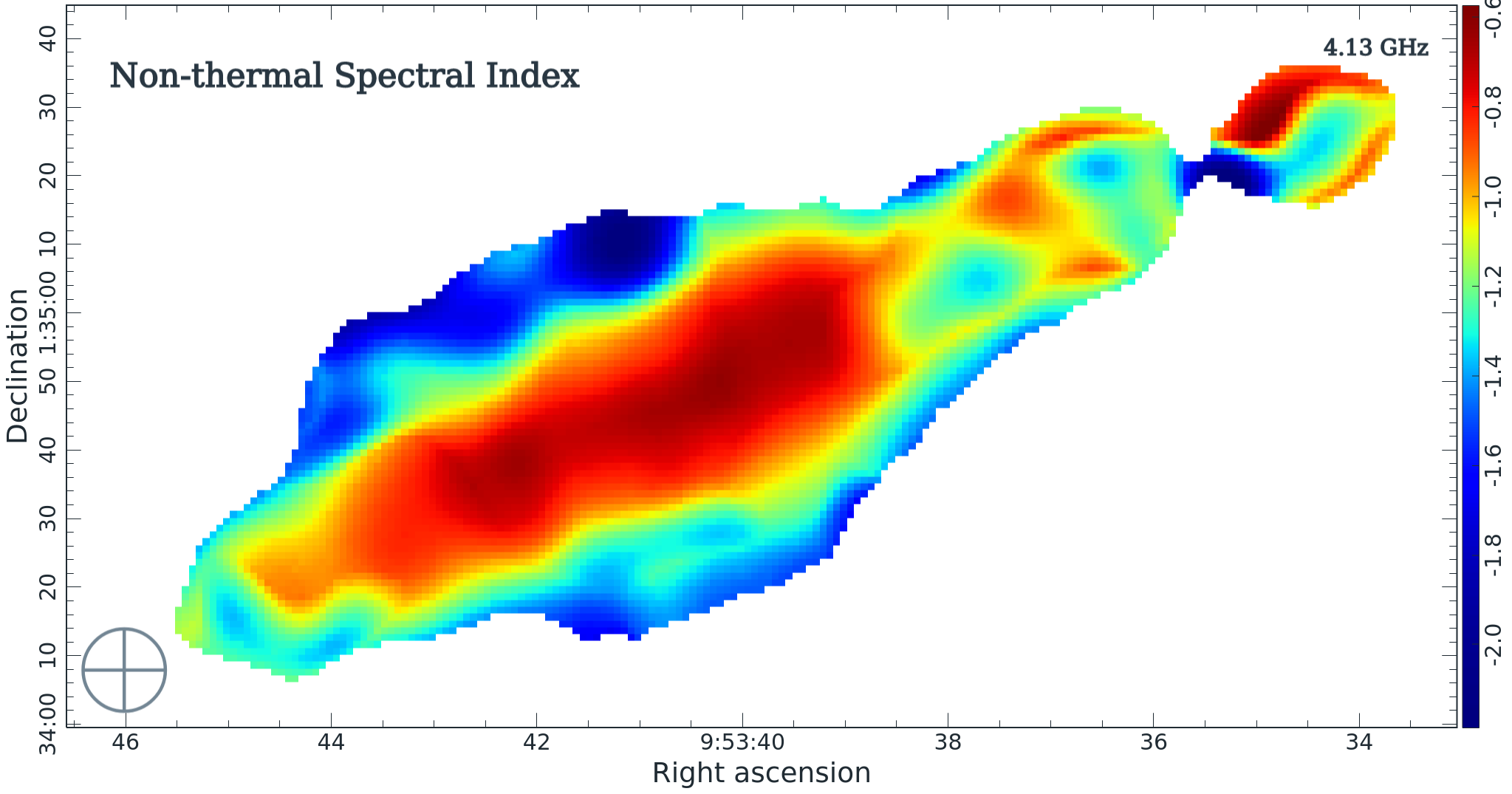}  
  \includegraphics[width=0.70\textwidth]{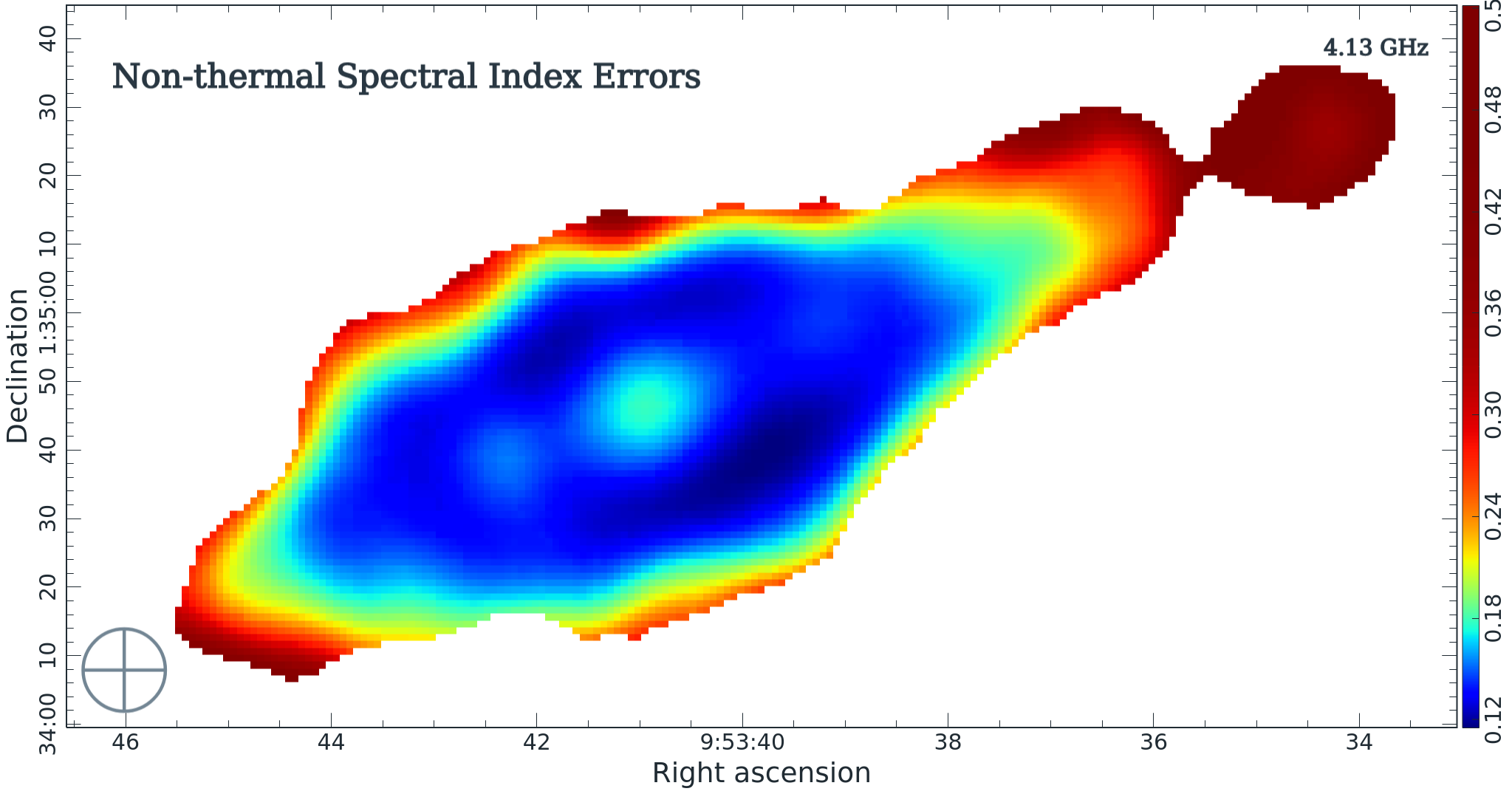}    
  \caption{Nonthermal spectral index (top) and its error (bottom) of NGC3044 resulting from cube-fitting, as described in Sect.~\ref{sec:cube_fits}. The $12\arcsec$ beam is shown at the lower left. Corresponding data are in Table~\ref{tab:imaging_results}.
   }
  \label{fig:alpha_NT_result}
\end{figure*}

\begin{figure*}[hbt!]
   \centering
  \includegraphics[width=0.7\textwidth]{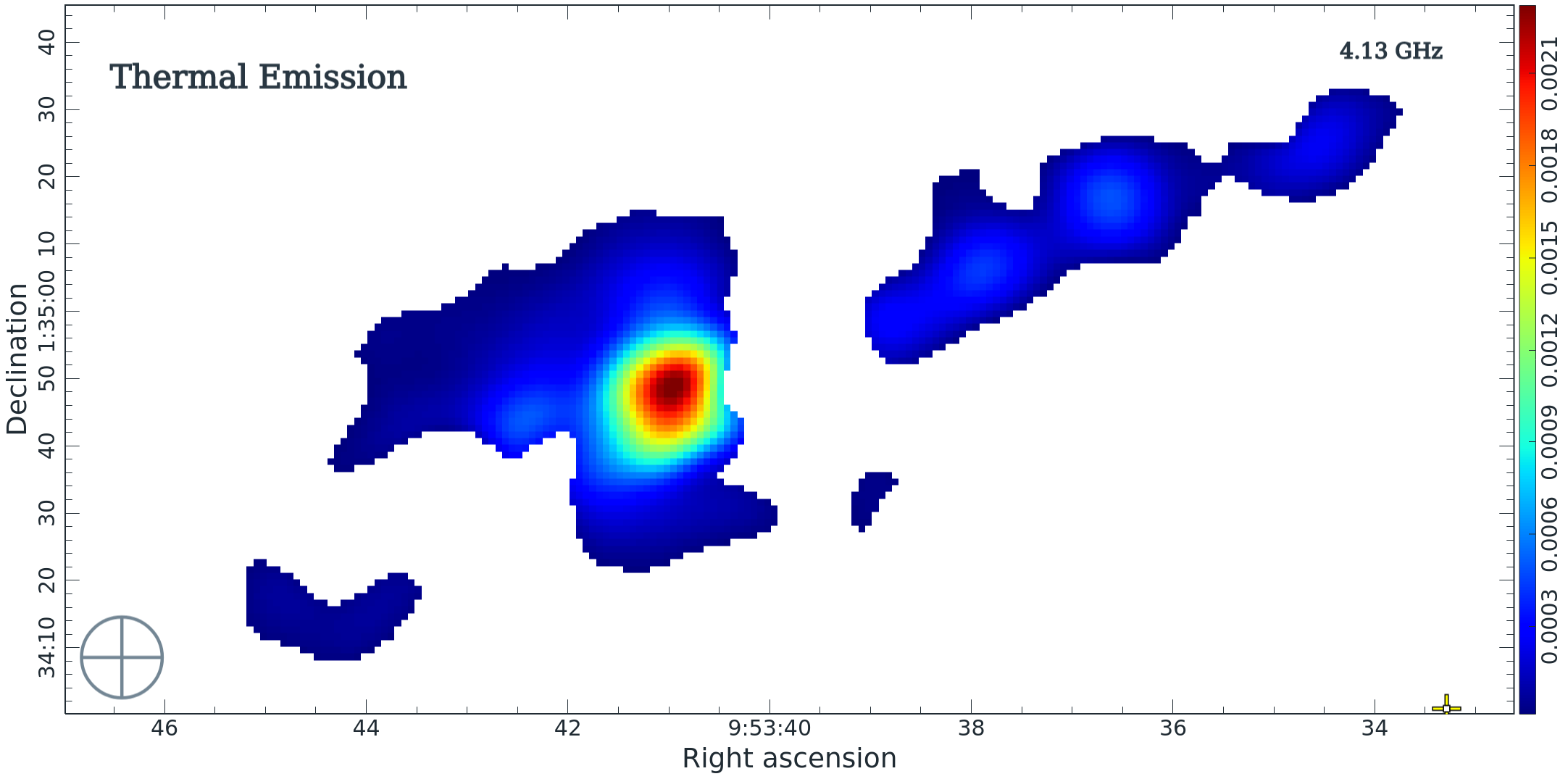}  
  \includegraphics[width=0.70\textwidth]{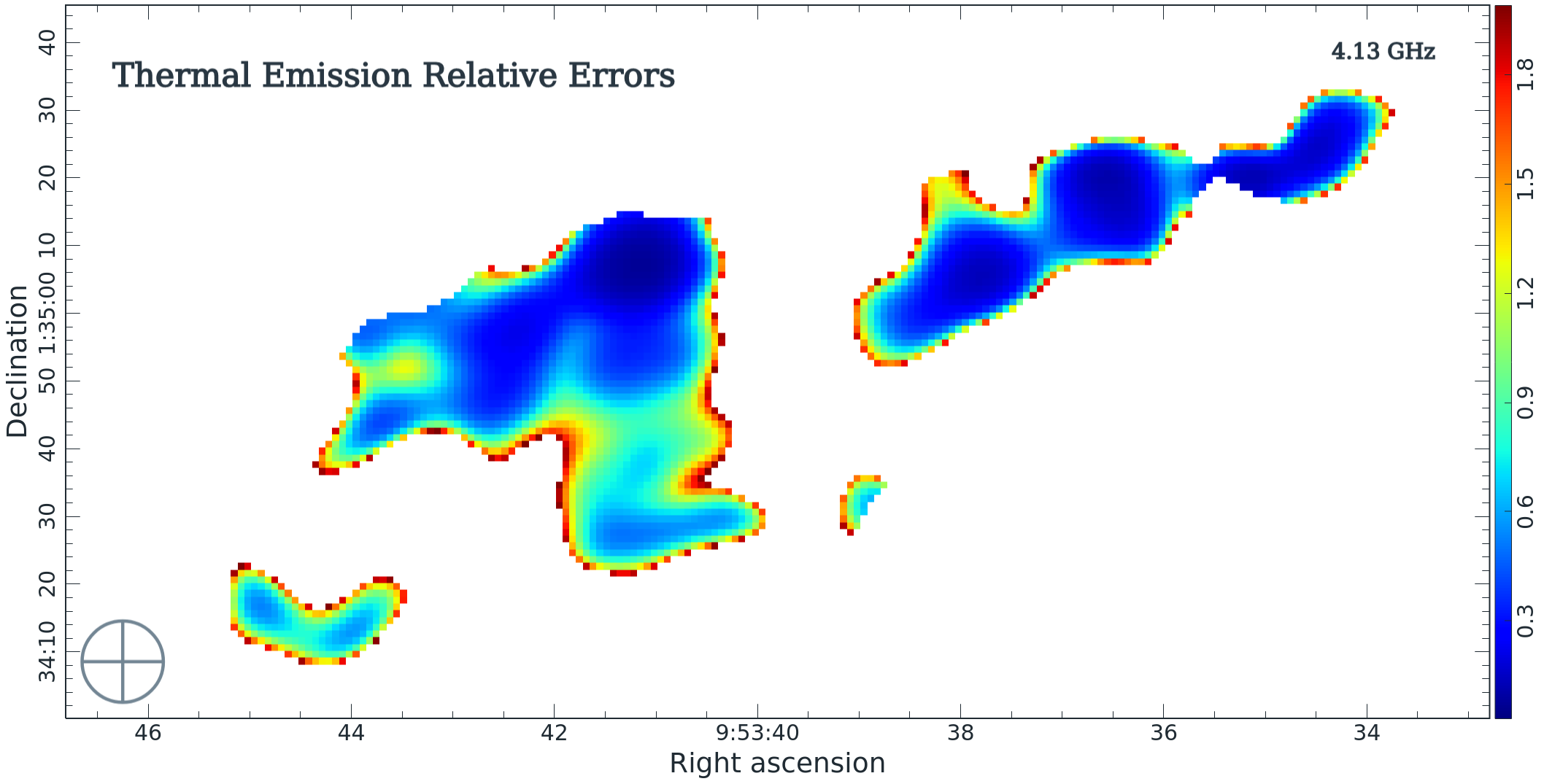}   
    \includegraphics[width=0.70\textwidth]{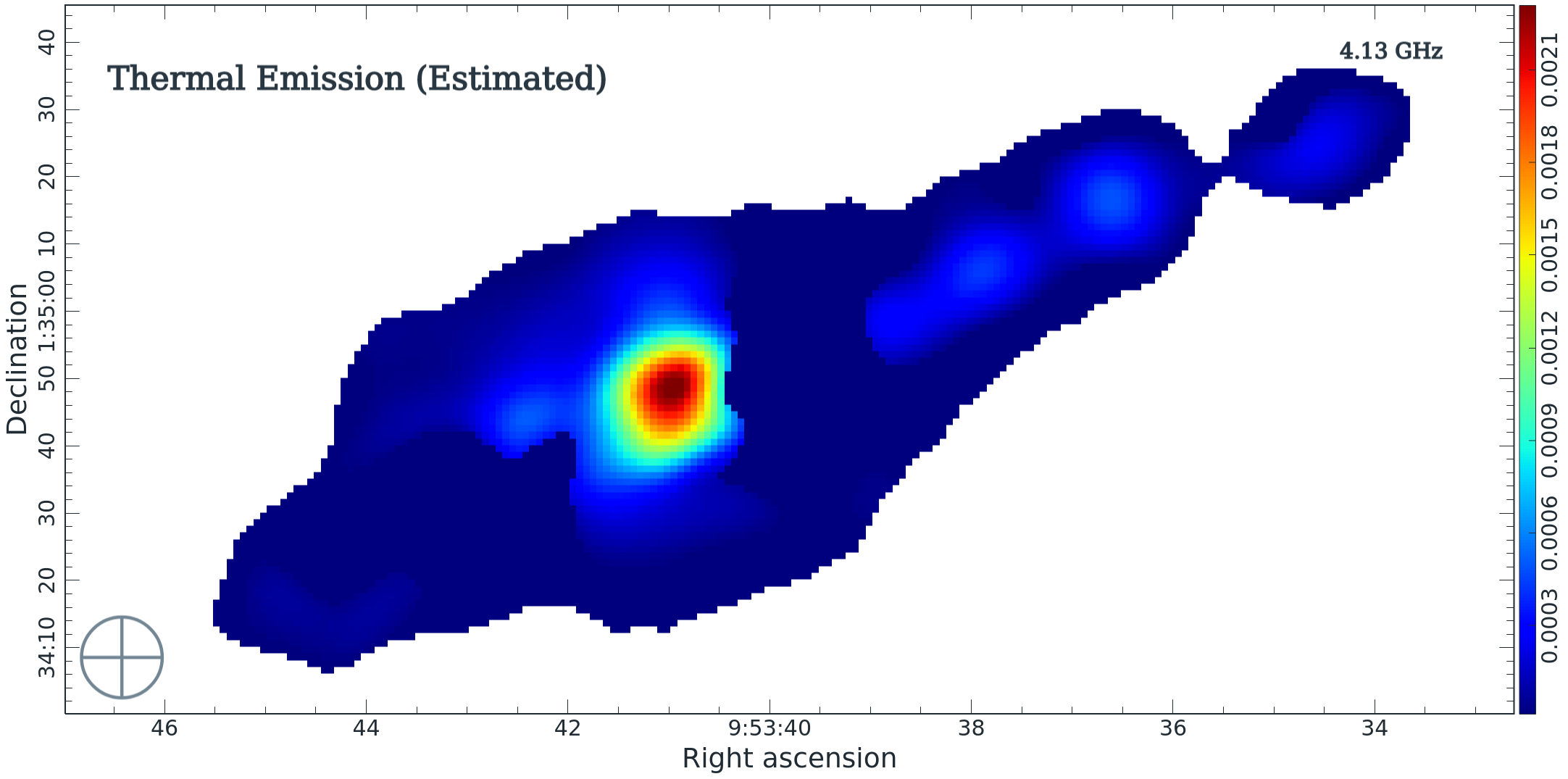 }  
    \caption{
   Thermal emission (top) and its relative error (center) of NGC~3044 resulting from cube-fitting, as described in Sect.~\ref{sec:cube_fits}. Blanked regions that show nonthermal emission (Figure~\ref{fig:INT_result}) are regions within which the thermal emission falls to very low, undetectable levels. The corresponding relative error map is shown at the center.  On the bottom is $I_{\rm TH}$ (max.), i.e. the thermal emission map with blanked regions set to the minimum detectable value of $14.8~\upmu$Jy beam$^{-1}$. The $12\arcsec$ beam is shown at the lower left. Corresponding data are in Table~\ref{tab:imaging_results}.
   }
  \label{fig:ITH_result}
\end{figure*}

\begin{figure*}[!ht]
   \centering
  \includegraphics[width=0.7\textwidth]{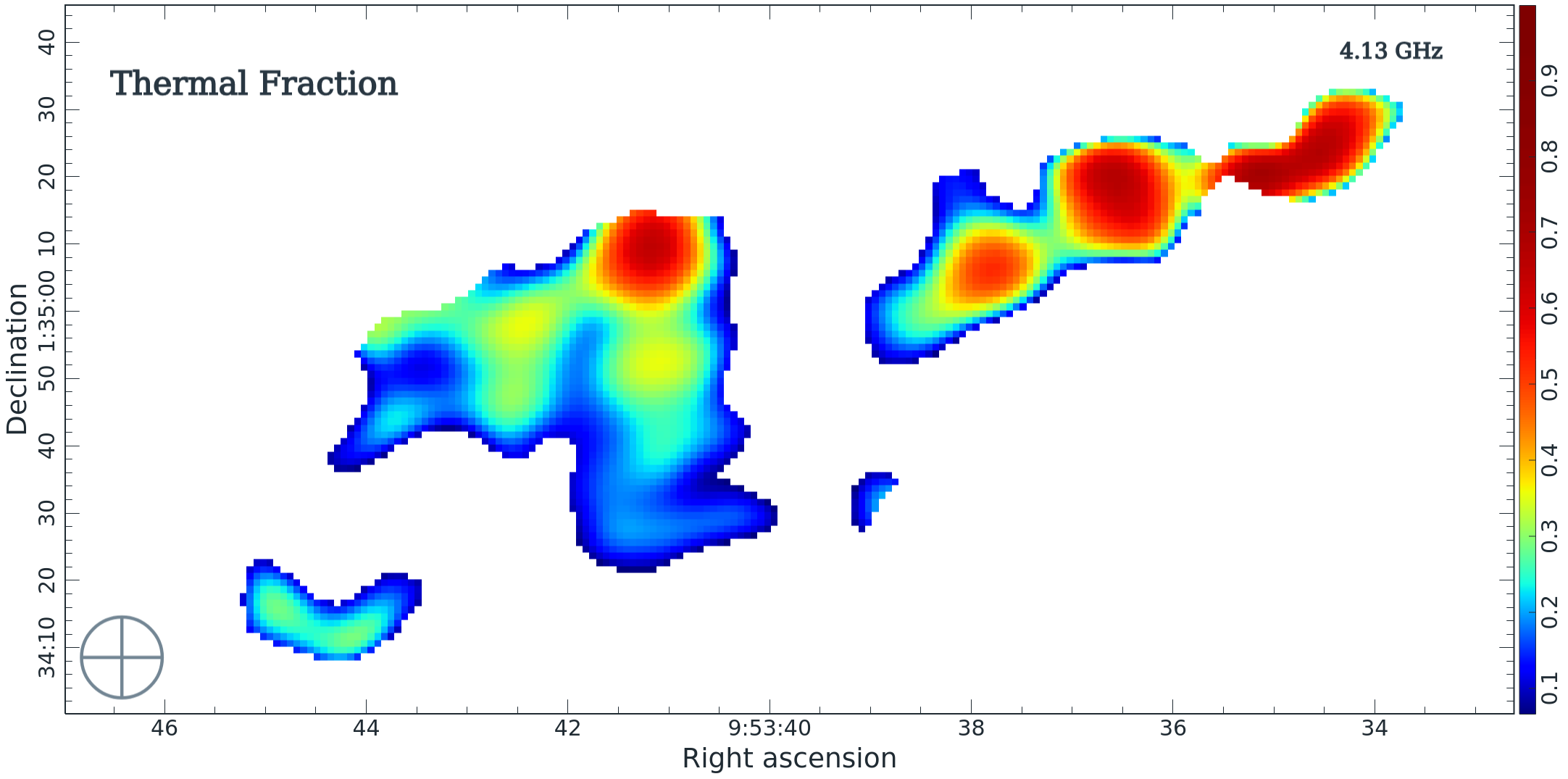}  
  \includegraphics[width=0.70\textwidth]{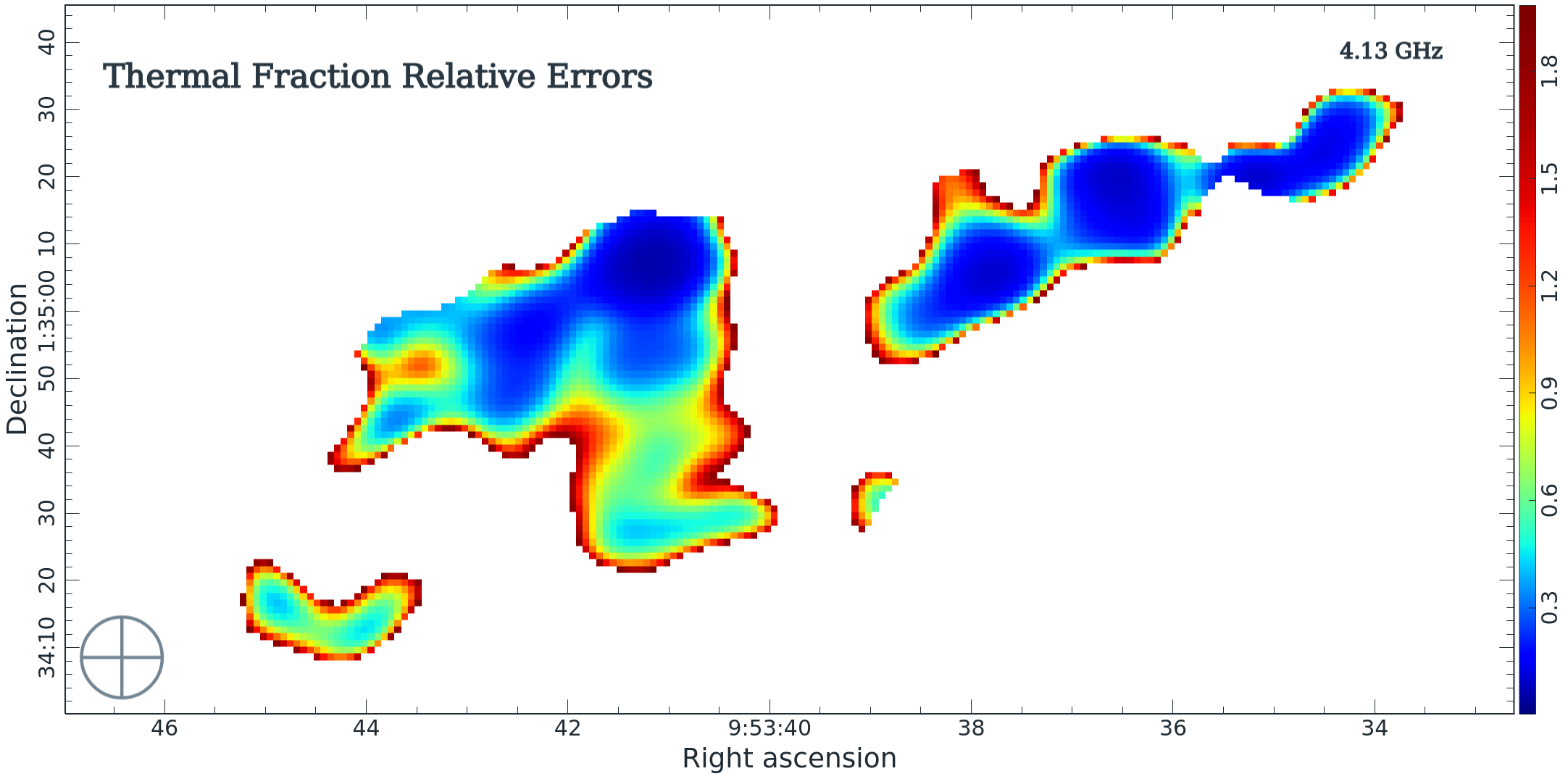}    
    \includegraphics[width=0.70\textwidth]{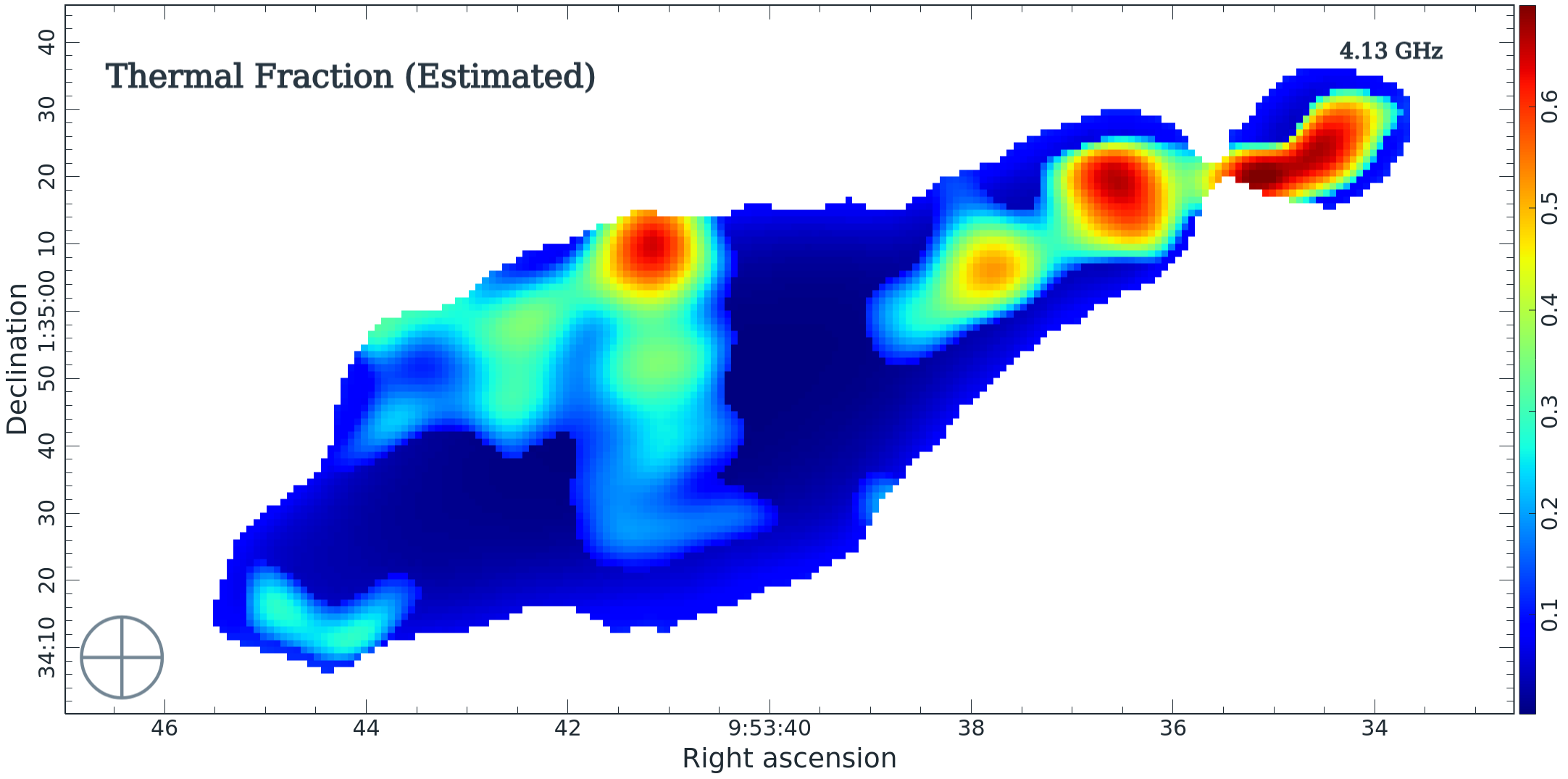 }  \caption{
   Thermal fraction (top) and its relative error (center) of NGC~3044 resulting from cube-fitting, as described in Sect.~\ref{sec:cube_fits}. Blanked regions that have measured values in the nonthermal emission map (Figure~\ref{fig:INT_result}) are regions within which the thermal emission falls to very low, undetectable levels. The bottom image shows $I_{\rm TH}/I_{\rm tot}$ (max.), 
   the result when the low, blanked regions are replaced with the minimum detectable value of $I_{\rm TH}\,=\,14.8~\upmu$Jy beam$^{-1}$. The $12\arcsec$ beam is shown at lower left. Corresponding data are in Table~\ref{tab:imaging_results}.}
  \label{fig:ITH_fraction_result}
\end{figure*}

A Python code was written to make the resulting maps\footnote{This code, called {\tt fitcurves\_CUBE.py}, is available at {\tt https://projects.canfar.net/changes/publications}.} which
are displayed in Figures~\ref{fig:INT_result} through \ref{fig:ITH_fraction_result}. 
Statistics from these maps are given in Table.~\ref{tab:imaging_results}.  We can now examine the spatially resolved images and data from the radio-only fitting method.

%################maps here

\begin{table*}
\begin{center}
\caption{Imaging Results {of NGC~3044} (12 arcsec res.)$^{\rm a}$\label{tab:imaging_results}}
\begin{tabular}{lccccccccc}
\hline
Map & No. Beams$^{\rm b}$ & Flux Density$^{\rm c}$ &Minimum$^{\rm d}$&Maximum$^{\rm d}$ & Mean$^{\rm e}$ & Median$^{\rm f}$ & Median of Error map$^{\rm g}$\\ 
 &  & (mJy) & ($\upmu$Jy beam$^{-1}$) & (mJy beam$^{-1}$) &(mJy beam$^{-1}$)&(mJy beam$^{-1}$)&\\
\hline\hline
$I_{\rm NT}$ & 40.8  &39.4 &39.7 &6.12 &0.866 &0.376 &0.202 &  &  \\ 
$I_{\rm TH}$ &  20.8  & 6.09& 14.8& 2.36 & 0.294& 0.154& 0.592\\ 
$I_{\rm TH}$ (max.)$^{\rm h}$& 40.8  &6.39 & 14.8& 2.36& 0.156& 0.021& ---\\ 
$I_{\rm TH}$ (min.)$^{\rm h}$& 40.8  &6.09 & 0.00& 2.36& 0.149& 0.021& --- \\
\hline
&  && Minimum$^{\rm d}$ & Maximum$^{\rm d}$& Mean$^{\rm e}$& Median$^{\rm f}$ &Median of Error map$^{\rm g}$\\
\hline
$\alpha_{\rm NT}$ & 40.8 & & -2.29 & -0.536 & -1.12 & -1.10& 0.165 \\
$\frac{I_{\rm TH}}{I_{\rm tot}}$& 20.8& & 0.0594 & 0.712 & 0.279 & 0.245 & 0.485 \\
$\frac{I_{\rm TH}}{I_{\rm tot}}$ (max.)$^{\rm h}$ & 40.8 & &
0.0026 & 0.712 & 0.163& 0.102 & --- \\
$\frac{I_{\rm TH}}{I_{\rm tot}}$ (min.)$^{\rm h}$ & 40.8 & &
0.00 & 0.712 & 0.143& 0.084 & --- \\
%$\frac{I_{\rm TH}}{I_{\rm tot}}$ (Var.)$^{\rm i}$ & &\\
\hline\hline
\end{tabular}
\end{center}
$^{\rm a}$ Numerical results from Figures~\ref{fig:INT_result} through \ref{fig:ITH_fraction_result}. The reference frequency is $\nu_0\,=\,4.13$ GHz, the spatial resolution is 12 arcsec, and 27 frequency points were used at each pixel.\\
$^{\rm b}$ Number of beams in the region shown on the maps.\\
$^{\rm c}$ Flux density in the region for $I_{\rm NT}$ and $I_{\rm TH}$. \\
$^{\rm d}$ Minimum and maximum values in the region.\\
$^{\rm e}$ Mean value in the region.\\
$^{\rm f}$ Median value in the region.\\
$^{\rm g}$ Median value of the error maps. For $I_{\rm NT}$,$I_{\rm TH}$, and $\frac{I_{\rm TH}}{I_{\rm tot}}$, it is the relative error, and for $\alpha_{\rm NT}$, it is the absolute error. See Sect.~\ref{sec:cube_fits} for more information.\\
$^{\rm h}$ Values for thermal emission and the thermal fraction  
for a region corresponding to the same 40.8 beam region in the nonthermal maps.  Blanked regions of $I_{\rm TH}$ have been replaced with the minimum detectable value of $I_{\rm TH}$ or with zero (see text) to obtain a maximum and minimum, respectively. Error maps are not calculated, given the assumptions involved.\\
\end{table*}
%judith used listobs to get these numbers
%https://science.nrao.edu/facilities/vla/docs/manuals/oss/referencemanual-all-pages
% 

Nonthermal emission (Fig.~\ref{fig:INT_result}) is smooth with a median relative error of 20\% (Table~\ref{tab:imaging_results}). The nonthermal emission, being stronger than thermal, can be seen over $\approx$ 41, 12-arcsec beams.

The nonthermal spectral index (Fig.~\ref{fig:alpha_NT_result}) is generally steep, typically $\alpha_{\rm NT}\,\approx\, -1$ (median), but ranges from $\approx$~-2 to -0.5 with flattest values in the disk.
%as was already found for the maximum and regional values of Table~\ref{tab:spectral_fits_27points}. 
A typical error is 0.17.
We also see a steepening of the spectral index with distance from the plane, consistent with \citet{var18} who find nonthermal spectral indices of  $\alpha_{\rm NT}\,\ltabouteq\,-1$ away from the plane.
As was noted in Sect.~\ref{sec:radioimagingresults}, the mean and median values are influenced by the steeper values of $\alpha_{NT}$ away from the plane.   
%Even so, our in-disk values are steeper than found by \cite{var18} because they have measured L-band and C-bands only, without S-band, and have included all L-band points which we now know (Sect.~\ref{sec:initial_spectrum}) to be flattened. 
By comparison, a steep 
global nonthermal spectral index ($\alpha_{\rm NT}\,\sim\,-0.9$) has similarly been found for NGC~891 \citep{kle84} and for a larger sample of nearby galaxies \citep{tab17}.

By contrast to the nonthermal emission, the thermal map (Fig.~\ref{fig:ITH_result}) is detected in a region of only $\approx$ 21 beams, 51\% of the area covered by $I_{\rm NT}$. This weaker emission has a larger median relative error (59\%) than the nonthermal map. The $I_{\rm TH}$ map looks patchy because the thermal emission falls to values that are too weak to be distinguished from the nonthermal emission  in the 
fitting process in some regions, so those regions have been blanked. 

Nevertheless, wherever the blanking occurs as a result of such a weak thermal flux, we can still put limits on the thermal emission, either by setting the value to the lowest detectable value in Table~\ref{tab:imaging_results}, i.e. 14.8 $\upmu$Jy beam$^{-1}$, or by setting the value to zero.  We have called these results $I_{\rm TH}$ (max.) and $I_{\rm TH}$ (min.), respectively and are a way of extrapolating the thermal emission into all regions that are also occupied by $I_{\rm NT}$. The $I_{\rm TH}$ (max.) map is shown at the bottom of Figure~\ref{fig:ITH_result} and statistics are given in Table~\ref{tab:imaging_results}. Evidently, the variation in flux density between these two limits is minor and less than the uncertainties. The mean and media, though, are reduced due to the weak estimated extrapolated emission. 

%but relative errors in low regions that have been set to zero or 12.4 $\upmu$Jy beam$^{-1}$ are much higher, of order 100\% at any given point.

The thermal fraction map (Fig.~\ref{fig:ITH_fraction_result}, Top) has a median value of $I_{\rm TH}/I_{\rm tot}$ of $\sim$ 25\%  
with some peaks as high as 71\%. However, the thermal fraction map also
covers only $\approx$ 21 beams because this was the measured region of the thermal emission map. The  
 median relative error is $\sim\, 49$\% in this restricted region. 
If we instead use $I_{\rm TH}$ (max.) and $I_{\rm TH}$ (min.) so as to estimate the thermal fraction over $\approx$ 41 beams, equivalent in size to $I_{\rm NT}$, then we can recalculate $I_{\rm TH}/I_{\rm tot}$ (max.) and $I_{\rm TH}/I_{\rm tot}$ (min.).
The former map is shown at the bottom of Fig.~\ref{fig:ITH_fraction_result} and statistics are given in Table~\ref{tab:imaging_results}.
 The median thermal fraction has now fallen to $\sim$ 10\% because we are including regions within which the thermal emission has dropped below detectable values.  
 %The median uncertainty in measured regions is still $\sim\,28$\% but is higher in regions in which  $I_{\rm TH}$ has been replaced with zero or low values.
  In summary, the median thermal fraction in measured regions is $\sim$ 25\%, but reduces to an estimate of $\sim$ 10\% when we extrapolate to all regions (41 beams) over the target area considered.

\subsection{Sources of Error}
\label{sec:uncertainties}

As evident from Figures~\ref{fig:spectra_all} and \ref{fig:spectra_27points}, the effect of a thermal contribution to the total spectrum is not readily visible by eye and only emerges with a fit to the spectrum.  This requires careful calibration over all bands from L to S to C.  As evident by the error bars, it is the calibration error which is the dominant error for high S/N points. For example, if the rms map noise is 15 $\upmu$Jy beam$^{-1}$ (Table~\ref{tab:im_params}), then any point greater than 0.75 mJy beam$^{-1}$ will be dominated by calibration error. There are two important consequences related to this fact.

The first is that some careful thought should go into calibrations, especially bandpass calibrations, and calibrations that span frequencies from L-band through C-band. For NGC~3044, we used the same primary flux calibrator, 3C~286, for all bands. Most secondary calibrators were also the same and were `primary' in the sense that they are known to have $<~3$\% amplitude closure errors (Table~\ref{tab:observing}). Even with these considerations, however, several spws at the upper frequency end of the S-band had to be discarded (Appendix~\ref{app:narrowband}) and there are also some gaps in the spectrum either because of severe RFI or because part of the spectrum (4 to 5 GHz, Figure~\ref{fig:spectra_all}, or \ref{fig:spectra_27points}) was unobserved.  
Having the 4--5 GHz gap filled in may have improved the results.  Moreover, the C-band can extend to 8 GHz, so an addition to this high frequency might also improve the results.  Going to an even higher frequency (8--12 GHz: X-band) is {\it not} recommended because of the possibility of an additional contribution from `anomalous microwave emission' (AME) as has been seen in the galaxy, M~31 \citep{bat19}. {The largest angular size detectable by the VLA at X-band is also only 2.4 arcmin} which makes comparison with the lower-frequency bands less accurate. 

The second is that, if a calibration uncertainty of $\sim~2$\% is adopted, then 
a thermal component must exceed approximately 6\%, in order to be distinguished in its contribution to the spectrum.
 In the current case, this was not an issue over most of the regions covered, but if the fit were carried out with a lower reference frequency than 4.13 GHz, this may not be the case.  For example, early work by \citet{con92} suggests a global thermal fraction for galaxies of $\sim$ 13\% at 2.0 GHz if $\alpha\,=\,-0.7$, which suggests local values both above and below 13\%.  \citet{par18} find global thermal fractions between 6 and 14\% for star-forming galaxies at 1.4 GHz and \citet{tab17} find a thermal fraction of $10\,\pm\,9$\% at the same frequency. For NGC~3044, \citet{var18} find a global thermal fraction of 5.9\% at 1.5 GHz. Consequently, the reference frequency should be higher than approximately 4 GHz, but still central to the total bandwidth, in order to have a sufficient thermal fraction to be distinguished by the fit. 
 
 Given the change of slope observed in the L-band (low frequencies) and the possibility of AME at high frequency, complete contiguous measurements from $\sim$ 1.3 through $\sim$ 8.0 GHz (L-band, S-band and C-band) appear to be the `sweet spot' for carrying out a thermal/nonthermal separation using radio data only. One would also want to target galaxies with total flux densities that are greater than NGC~3044  {(cf. Sect.~\ref{sec:N5775})}.

 %Related to contiguous coverage is the fact that we have fitted 27 frequency points rather than all 30 because of the L-band spectral flattening.  This was necessary in order to capture the steepening spectrum above $\sim$ 1.7 GHz  
% The full cube analysis was repeated by including all L-band data points but combining them into a single map whose central frequency was $\nu_L\,=\,1.57$ GHz. This map was then used with the remaining 26 frequency points from S-band and C-band.  Maps similar to Figure~\ref{fig:INT_result} to \ref{fig:ITH_fraction_result} were then remade. Because we are now including all L-band data, the nonthermal spectral index is somewhat flatter than before, for example, the median is now $\alpha_{\rm NT}\,=\,-0.98\,\pm\,0.12$ as opposed to  $\alpha_{\rm NT}\,=\,-1.16\,\pm\,0.16$ (Table~\ref{tab:imaging_results}), although individual points could exceed these error bars. The main conclusion from this repetition is that the vertical feature is still present in the thermal fraction map.  This will be discussed in more detail in Sect.~\ref{sec:vertical_feature}.

At the highest frequencies (C-band only) we must also consider the possibility of missing flux (see LAS in Table~\ref{tab:observing}). In C-band, the LAS ranges 
from $4\farcm 7$ at the low-frequency end of C-band to $3\farcm 6$ at the high-frequency end.  The infrared diameter of NGC~3044 is only $3\farcm 07$ (Table~\ref{tab:N3044params}), the modeled region (e.g. ~\ref{fig:INT_result}) extends to a diameter of $3\farcm 2$, and the region over which thermal emission was detected is smaller still  (Figure~\ref{fig:ITH_result}). These regions are smaller than the C-band LAS.
\citet{gre91} find a single-dish, Green Bank Telescope (GBT) flux density at 4.85 GHz of $S_{\rm 4.85~GHz}\,=\,43\,\pm\,10$ mJy. We recomputed our C-band flux density for the same reference frequency and find 
$S_{\rm 4.85~GHz}\,=\,40.4\,\pm\,0.9$ mJy. These values agree within uncertainties, although the GBT error bar is generous.
Although we can't rule out a small amount of missing flux at high frequencies
if the broad scale halo were included, 
 it is unlikely that missing flux has influenced the models presented in Sect.~\ref{sec:cube_fits} which apply to a smaller, restricted spatial region.

\subsection{Comparison with H-alpha-Related Results}
\label{sec:comparison}

\begin{figure*}[hbt!]
    \centering    \includegraphics[width=0.7\textwidth]{
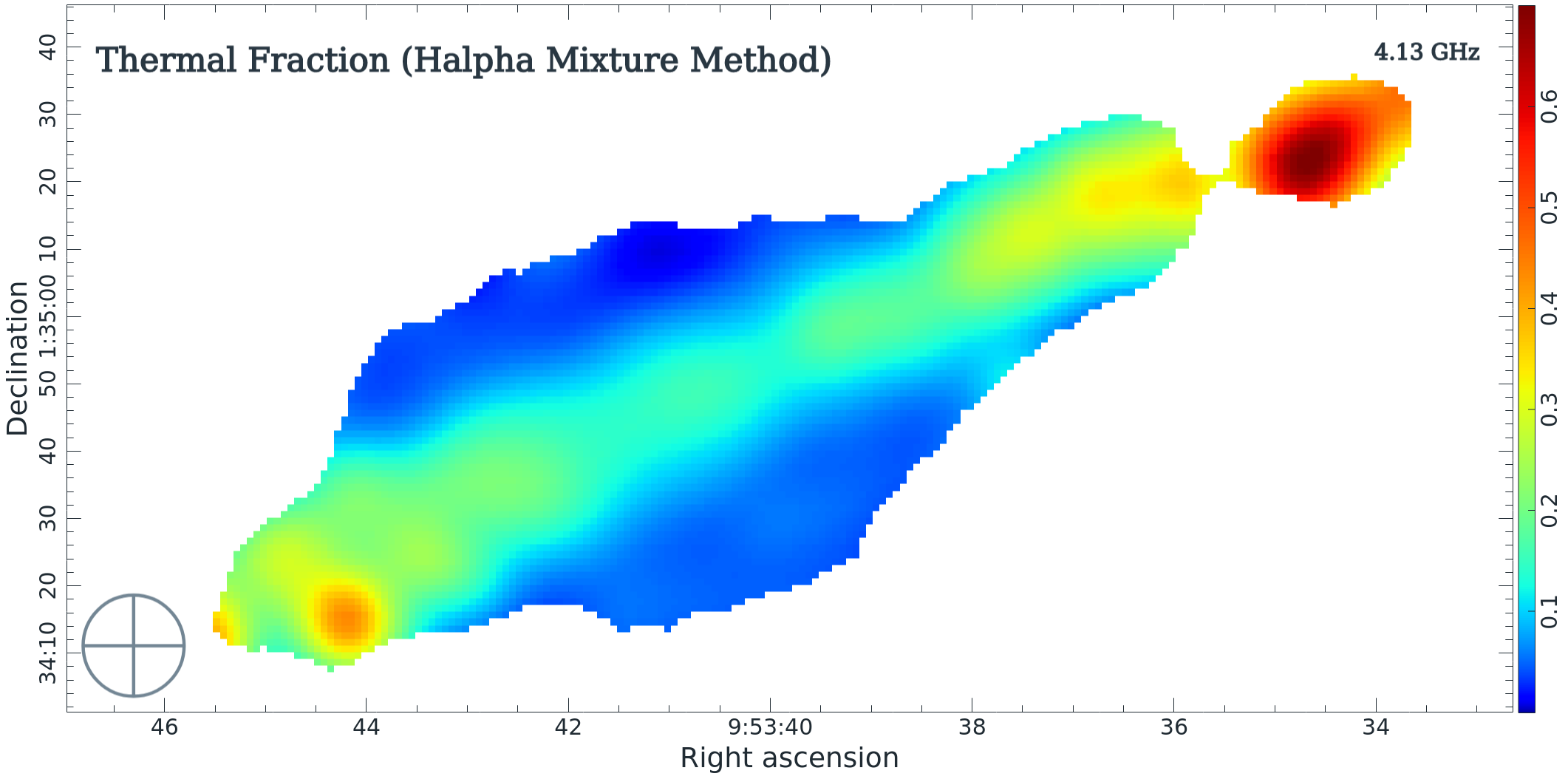}
  \includegraphics[width=0.7\textwidth]{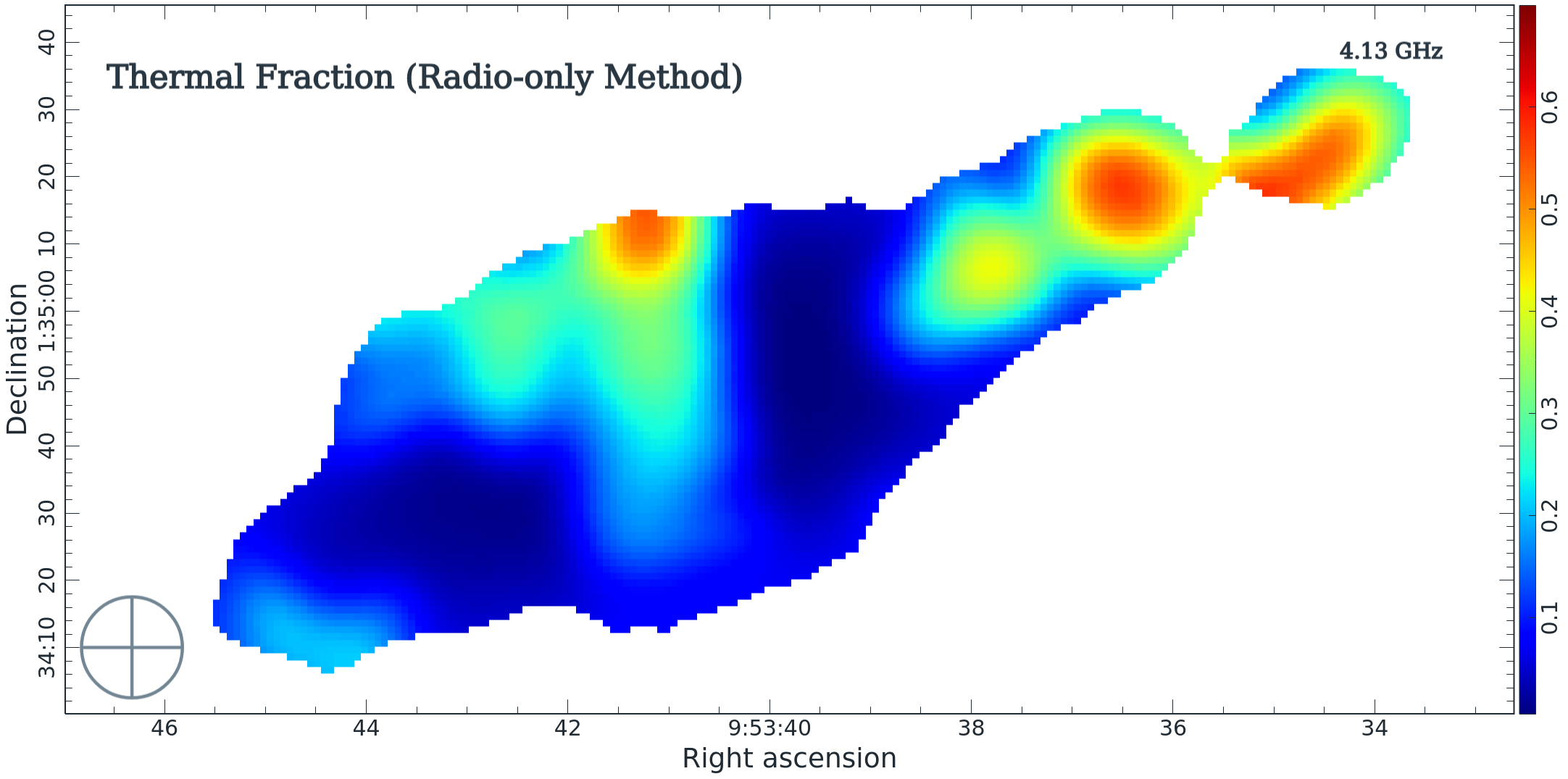}
    \caption{Thermal fractions of NGC~3044 at 15 arcsec resolution. Top: Mixture method using H$\alpha$ and IR data from \citet{var18}. The data have been interpolated to 4.13 GHz and given the same blanking as the radio-only data. Bottom: Radio-only result smoothed to 15 arcsec resolution. $I_{\rm TH}$ (max.) has been used. Notice the vertical feature near the center.  Corresponding data are given in Table~\ref{tab:imaging_results_15arcsec}. The beam is shown at the lower left.}
    \label{fig:ITH_fractions_15arcsec}
\end{figure*}

\begin{figure*}[hbt!]
   \centering
  \includegraphics[width=0.7\textwidth]{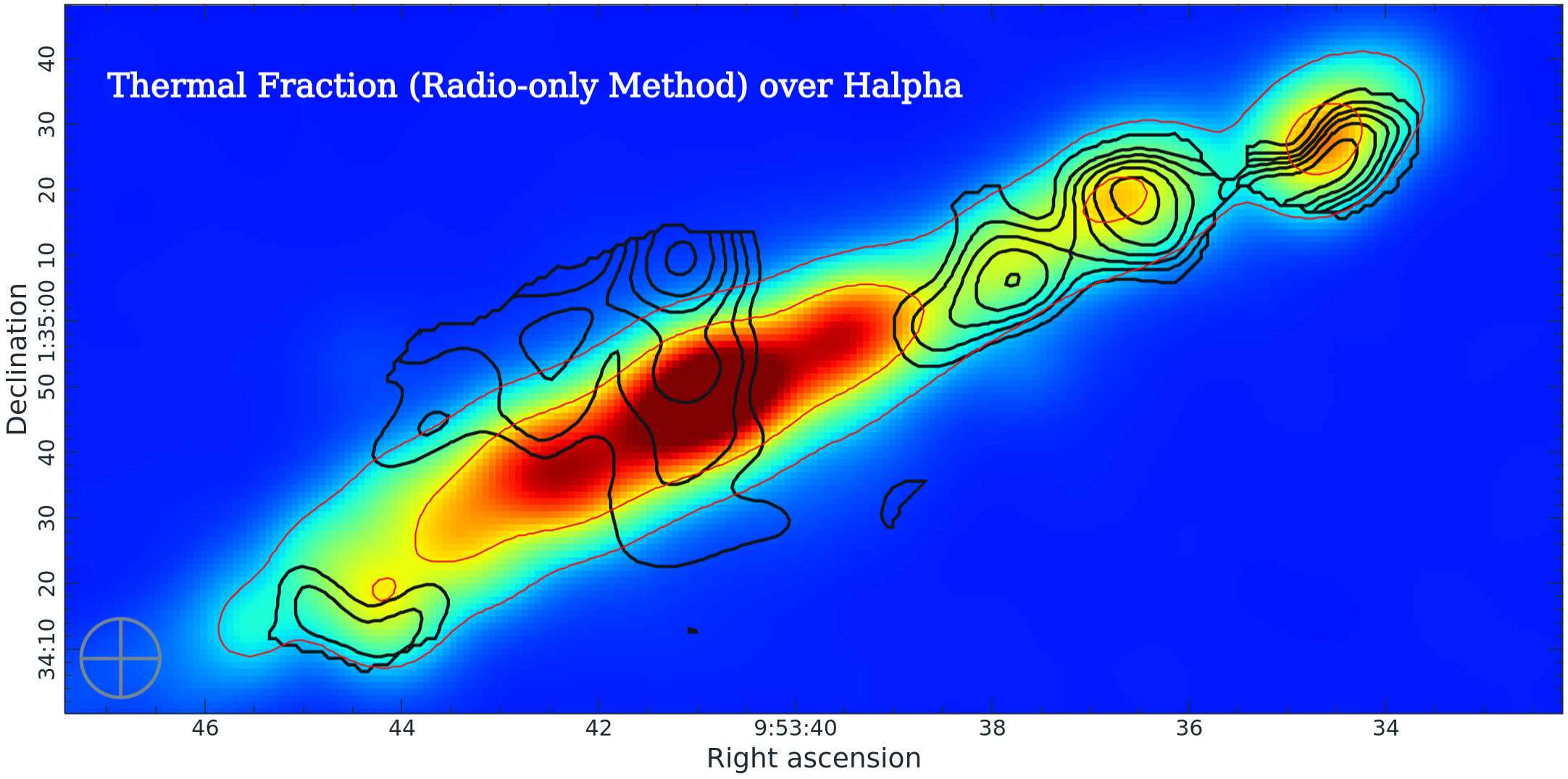}   
  \includegraphics[width=0.7\textwidth]{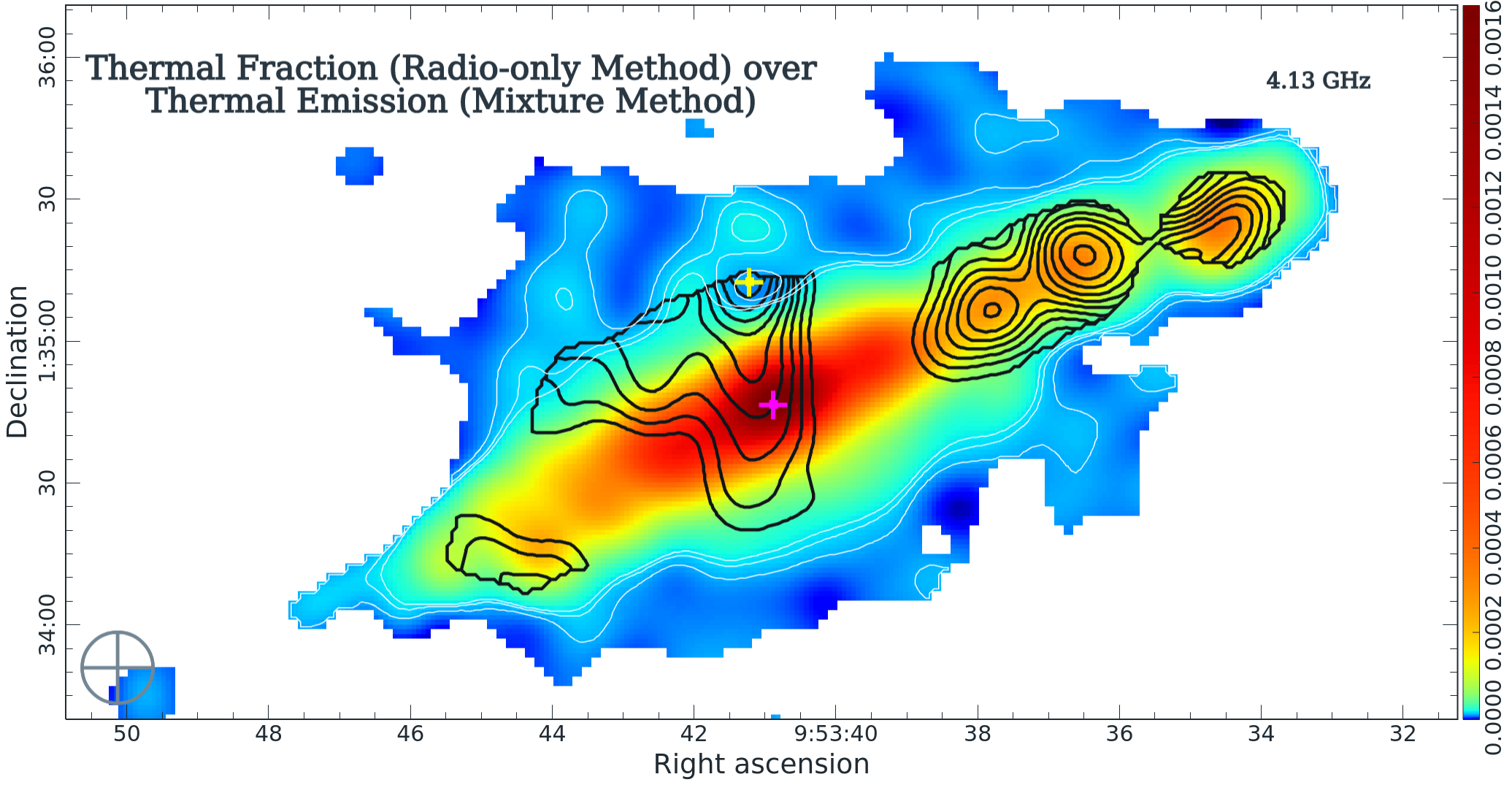}
  \caption{Thermal comparisons for NGC~3044. Top: Colour image and two thin red contours represent the H$\alpha$ image from Figure~\ref{fig:N3044_fig1} smoothed to 12 arcsec resolution, arbitrarily scaled.  Thick black contours represent the thermal fraction image of Figure~\ref{fig:ITH_fraction_result}, with contours at 0.1, 0.2, 0.3, 0.4, 0.5, and 0.6.  The 12 arcsec beam is shown at  lower left.
  Bottom: Radio-only thermal fraction map in black contours at 15 arcsec resolution. Contours are at 0.1, 0.15, 0.2, 0.25, 0.3, 0.35, 0.4, 0.45, 0.5, and 0.55. The colour image is the \citet{var19} $I_{\rm TH}$ map from the mixture method, at the same resolution.  Several white contours delineate the lower level thermal emission at 8, 12.5, and 15 $\upmu$Jy beam$^{-1}$. The frequency is given at top right, beam size at lower left, the galaxy center (from Table~\ref{tab:N3044params}) is marked with a magenta `+' and the peak of the vertical feature in the radio thermal fraction map is marked with a yellow `+'.}
    \label{fig:Vargas_radio_comparison}
\end{figure*}

\cite{var18} have determined C-band and L-band thermal maps and thermal fraction maps using the `mixture method' (Sect.~\ref{sec:separating}), which uses the H$\alpha$ map to estimate the thermal emission and the $22~\upmu$m map to correct for dust.
%For a comparison with our results at $\nu_0\,=\,4.13$ GHz, we have interpolated the \cite{var18} values (their Table~6) using their total spectral index of $\alpha\,=\,-0.737\,\pm\,0.052$, their thermal and total flux densities, and the known thermal spectral index of $\alpha_{\rm TH}\,=\,-0.1$, to give a value at our central frequency of $\nu_0\,=\,4.13$ GHz. The result is $\frac{I_{\rm TH}}{I_{\rm tot}}\,=\,0.154\,\pm\,0.029$ at 4.13 GHz.   We note that the \citet{var18} results cover a larger region than shown in 
%Figure~\ref{fig:INT_result}.  If their results are restricted to an equivalent region, their thermal fraction increases to 
%$\frac{I_{\rm TH}}{I_{\rm tot}}\,=\,0.185\,\pm\,0.035$. The value increases because the thermal fraction is higher closer to the disk than in the halo. These 
%results compare favourably with our estimated median value of $\frac{I_{\rm TH}}{I_{\rm tot}}\,=\,0.17\,\pm\,30$\% at the same frequency.  Our thermal fraction does not change beyond the quoted uncertainty if we smooth our
 %12 arcsec resolution maps to 15 arcsec in order to match the \citet{var18} resolution.
 These values are not at frequencies that correspond to our central $\nu_0\,=\,4.13$ GHz.  Consequently, we have taken their thermal map, $I_{\rm TH}$, and adjusted to 4.13 GHz using $I_\nu\,\propto\,\nu^{-0.1}$.  We have also used their nonthermal map, $I_{\rm NT}$, and adjusted to 4.13 GHz using 
 $I_{\rm NT}\,\propto\, \nu^{\alpha_{\rm NT}}$, where $I_{\rm TH}$, $I_{\rm NT}$, and $\alpha_{NT}$ maps are all taken from  Figure 7 of \citet{var18}.  Their thermal fraction map, $I_{\rm TH}/I_{\rm tot}$, was then recalculated at our value of $\nu_0$. Each map was then regridded to match the sampling of our radio-only maps, converted into Jy beam$^{-1}$ units, and blanked to match the same coverage of our radio-only maps (e.g. as in Figure~\ref{fig:INT_result}). The resulting thermal fraction map is shown at the top of Figure~\ref{fig:ITH_fractions_15arcsec}.  It is not surprising that this image resembles the H$\alpha$ map (Figure~\ref{fig:N3044_fig1} and Figure~\ref{fig:Vargas_radio_comparison}, Top) since it was generated from H$\alpha$ data.

Our results for $I_{\rm NT}$ (Figure~\ref{fig:INT_result}) $I_{\rm TH}$ (max.) (Fig.~\ref{fig:ITH_result}) and $I_{\rm TH}$ (min.) were then smoothed to $15\arcsec$ resolution to match the \citet{var18} resolution. The thermal fraction map was then recalculated to form $\frac{I_{\rm TH}}{I_{\rm tot}}$ (max.) and $\frac{I_{\rm TH}}{I_{\rm tot}}$ (min.).
Numerical results from these $15\arcsec$ resolution maps are given in Table~\ref{tab:imaging_results_15arcsec} and the thermal fraction maps are shown in Figure~\ref{fig:ITH_fractions_15arcsec}. 
 
{Median values of $I_{\rm NT}$ differ between the mixture-method maps and the radio-only derived maps by 19\%.  A similar comparson of $I_{\rm TH}$ gives a difference of 12\%.  These values agree within the 20\% uncertainty quoted by \citet{var18} for the mixture method alone.}
 As for the thermal fraction, the mixture method gives a 
 thermal fraction of 12.5\%, again  with estimated uncertainties {on this fraction} of order 20\%. The median thermal fraction from the radio-only method is also $\sim\,13$\% with an estimated uncertainty {on this fraction} of $\sim\,50$\% (Table~\ref{tab:imaging_results}). 
 {As can be seen, the radio-only uncertainties are larger than those of the estimated mixture method, although the latter estimate has not factored in the assumptions and calibration issues associated with optical and infrared data. Clearly, the values agree within uncertainties,} but 
 the morphologies of the two maps are quite different towards the center.  We will address this further in Sect.~\ref{sec:vertical_feature}.
\cite{var18, var19} also note that some very high thermal fractions are observed in certain discrete regions of the disk, for example, they find
C-band thermal fractions of $\sim\,40$\% increasing to $\sim\,80$\% in certain discrete regions of the disk.  We also see high thermal fractions in some regions of the galaxy at 4.13 GHz, for example, $\sim\,60$\%. %Higher values towards the ends of the disk (far west and far east) are similar between the two maps.

We have also compared our radio-only thermal fraction map to the H$\alpha$ map directly. In Figure~\ref{fig:Vargas_radio_comparison} (Top), we show our $12\arcsec$ radio-only 
$I_{\rm TH}/I_{\rm tot}$ (max.) map (from Figure~\ref{fig:ITH_fraction_result} top) superimposed on the H$\alpha$ map from  Figure~\ref{fig:N3044_fig1} which has been smoothed to the same $12\arcsec$ spatial resolution.
It is important to note that the two maps are completely independent. The thermal emission and thermal fraction are derived from radio data only, so a correlation of peaks (contours) with H\,{\sc ii} regions (H$\alpha$ peaks) offers some validity regarding the effectiveness of the radio technique. {For example, two HII region complexes on the far western end of the disk are circled in light red contours and the radio-only method (black contours) also have peaks at those locations.}

{The comparison can also been seen in Figure~\ref{fig:Vargas_radio_comparison} (Bottom), where we have extended the $I_{\rm TH}$ map from the mixture method (colours) to lower emission levels. In the outer disk towards the north-west, we see the good alignment between several peaks in both $I_{\rm TH}$ and the radio-only thermal fraction (contours).}
The main difference is towards the center of the map, where we see an apparent vertical feaure in the radio-only thermal fraction map but not the mixture-method map. We we will address this issue in more detail in Sect.~\ref{sec:vertical_feature}.

We have also failed to detect thermal emission in a gap along the disk to the west of the center where there is clearly H$\alpha$ emission. In this region, although thermal emission should be present, the nonthermal component is also high and the thermal fraction has fallen below 6\%.  As pointed out in Sect.~\ref{sec:uncertainties}, when the thermal fraction drops below this value, it cannot be distinguished in the fitting.

%15 arcsec beam constitutes 254.945 pixels out of 6664 pixels so 26.1 beams
\begin{table*}
\begin{center}
\caption{Imaging Results {of NGC~3044} (15 arcsec res.)$^{\rm a}$\label{tab:imaging_results_15arcsec}}
\begin{tabular}{lcccccccc}
\hline
Map & No. Beams$^{\rm b}$ & Flux Density$^{\rm c}$ &Minimum$^{\rm d}$&Maximum$^{\rm d}$ & Mean$^{\rm e}$ & Median$^{\rm f}$ \\ 
 &  & (mJy) & ($\upmu$Jy beam$^{-1}$) & (mJy beam$^{-1}$) &(mJy beam$^{-1}$)&(mJy beam$^{-1}$)&\\
\hline\hline
\color{blue}{Radio only:} & & & & & & & & \\
$I_{\rm NT}$ & 26.1  &28.3 &58.1 &5.26 &1.08 &0.575 & &  \\ 
$I_{\rm TH}$ (max.) &  26.1 & 12.0& 8.11&5.13 & 0.460& 0.115\\
$I_{\rm TH}$ (min.) &  26.1 & 11.8& 0.00&5.13 & 0.453& 0.107\\
\hline
\color{blue}{H$\alpha$ mixture method:}\\
$I_{\rm NT}$  & 26.1  &37.8 &110.2 &9.49 &1.45 &0.696 & &  \\
$I_{\rm TH}$  & 26.1  &5.95 &4.58 &1.76 &0.227 &0.125 & &   \\
\hline
\hline
& No. Beams$^{\rm b}$ &  &Minimum$^{\rm d}$&Maximum$^{\rm d}$ & Mean$^{\rm e}$ & Median$^{\rm f}$ \\ 
\hline
\color{blue}{Radio only:} & & & & & & & & \\
$\frac{I_{\rm TH}}{I_{\rm tot}}$ (max.) & 26.1 &--- & 0.005 & 0.578 & 0.173 & 0.141\\
$\frac{I_{\rm TH}}{I_{\rm tot}}$ (min.) & 26.1 &--- & 0.000 & 0.577 & 0.159 & 0.128\\
\hline
\color{blue}{H$\alpha$ mixture method:}\\
$\frac{I_{\rm TH}}{I_{\rm tot}}$ &26.1& -- & 0.0077 & 0.712 & 0.154 & 0.125\\
%$\frac{I_{\rm TH}}{I_{\rm tot}}$ (Var.)$^{\rm i}$ & &\\
\hline\hline
\end{tabular}
\end{center}
$^{\rm a}$ Comparative numerical results from this work,  smoothed to 15 arcsec resolution. The \cite{var18} values have been interpolated to a central frequency of 4.13 GHz and blanked identically to Figures~\ref{fig:INT_result} through \ref{fig:ITH_fraction_result}. Uncertainties are similar to those of Table~\ref{tab:imaging_results}.\\
$^{\rm b}$ Number of beams in the region shown on the maps.\\
$^{\rm c}$ Flux density in the region for $I_{\rm NT}$ and $I_{\rm TH}$. \\
$^{\rm d}$ Minimum and maximum values in the region.\\
$^{\rm e}$ Mean value in the region.\\
$^{\rm f}$ Median value in the region.\\
\end{table*}

\clearpage

{ 
\section{NGC~5775}
\label{sec:N5775}

NGC~5775 is a well-known and well-studied galaxy in the CHANG-ES sample.  Examples of recent papers include \citet{hea22}, \citet{boe19}, \cite{gho09}, \cite{li08}, and \citet{hea06}, among others. Some basic data for this galaxy are in Table~\ref{tab:N3044params}. This galaxy is similar to NGC~3044 in that it has a significant radio halo and there is no evidence of an active galactic nucleus (AGN).  It was adopted as a second test of the radio-only method of thermal/nonthermal separation because of its similarity to NGC~3044 and also because of its small angular size, which makes LAS issues minor.  However, NGC~5775 has a higher star formation rate than NGC~3044 by a factor of 4.3 and it also has a higher radio flux and H-alpha luminosity by a factor of $\sim$ 2.2. Figure~\ref{fig:N5775_fig10} shows C-band radio contours over an H$\alpha$ image.

\begin{figure*}
   \centering
  \includegraphics[width=0.5\textwidth]{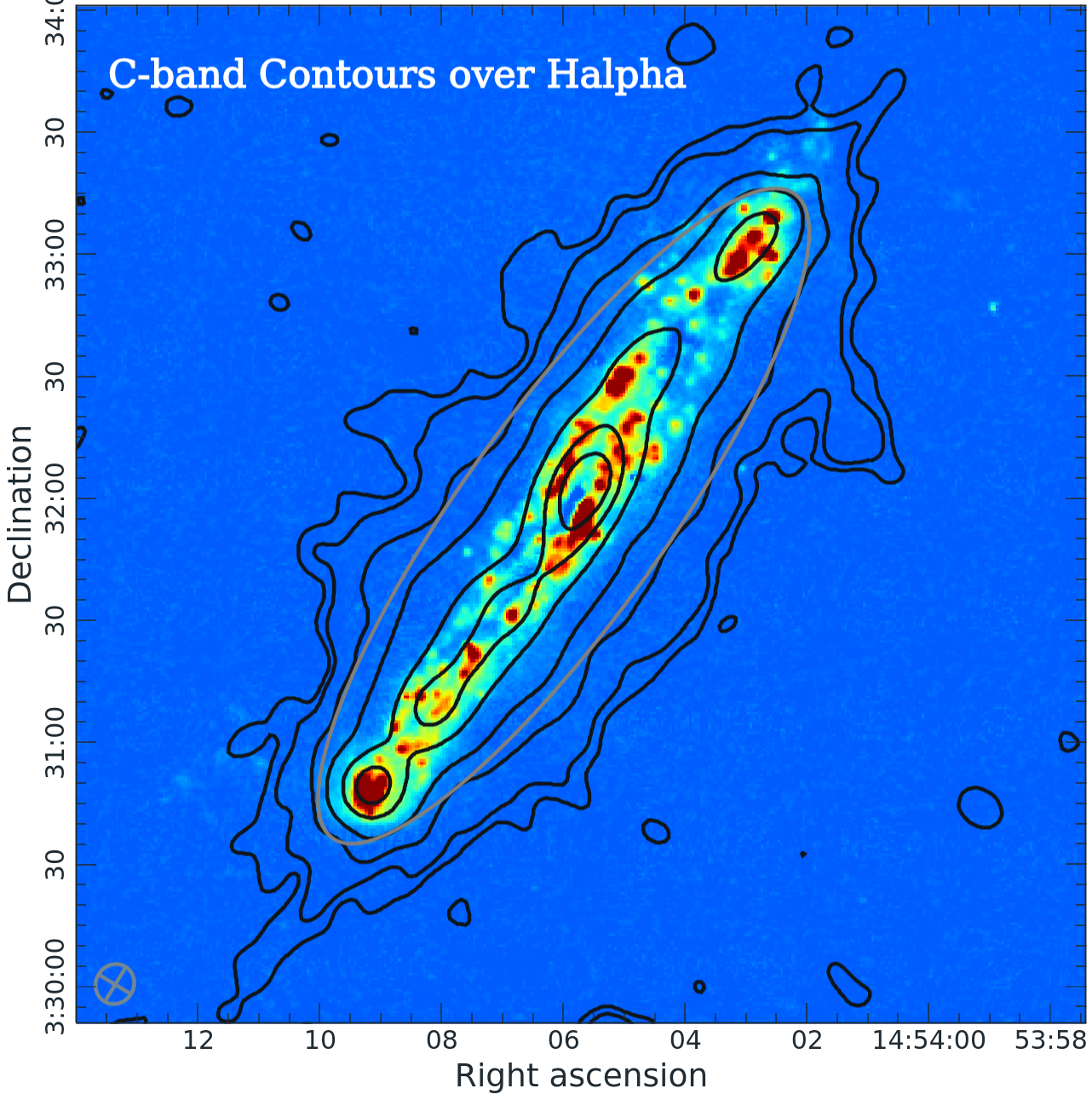}  
   \caption{
   NGC~5775, shown in H$\alpha$ (colours) with arbitrary scaling, overlaid with C-band radio contours from D-configuration.  Radio contours are at 14, 30, 100, 300, 1000, 2000, and 3000 $\upmu$Jy beam$^{-1}$. The beam size ($10\farcs 0 \times 9\farcs 3$ at a position angle of $-31\fdg 2$) is shown as a grey circle with a cross at lower left. The H$\alpha$ 
   and radio data are from  \citet{col00}  and \citet{wie15}, respectively. %Both images in FITS format can be downloaded from the CHANG-ES public release website: {\tt https://projects.canfar.net/changes} 
   The grey curve encloses the region that was used for a flux measurement, as described in Sect.~\ref{sec:narrowband_N5775}. }
  \label{fig:N5775_fig10}
\end{figure*}

Observing data for NGC~5775 are given in Table~\ref{tab:N5775observing}. Narrow-band imaging was carried out in the same fashion as for NGC~3044 with the same adopted spectral width (cf. Table~\ref{tab:im_params}). Typical rms values were 30 $\mu$Jy beam$^{-1}$ at L-band and C-band and 20 $\mu$Jy beam$^{-1}$ at S-band for a common beam of 12 arcsec FWHM.

\begin{table*}
\begin{center}
\caption{Observing Data {for NGC~5775}\label{tab:N5775observing}}
\begin{tabular}{lccccccccc}
\hline
Band ($\nu_0$)$^{\rm a}$ & Freq. range$^{\rm b}$ & $\Delta \nu$$^{\rm c}$ &No. Spws$^{\rm d}$& Obs. Date$^{\rm e}$ & TOS$^{\rm f}$&Prim. Calib.$^{\rm g}$&Second. Calib.$^{\rm h}$&PB FWHM$^{\rm i}$ & LAS$^{\rm j}$\\ 
~~~~~~~~~~(GHz) & (GHz) & (GHz)& && (hours)& & &(arcmin) & (arcmin) \\
$\dotfill$ Array & & & & & & & \\
\hline\hline
L-band (1.58) & 1.247 $\to$ 1.503 & 0.512& 32& & &3C~286& &26.6  &  \\ 
                             &  1.647 $\to$ 1.903 & & & & \\
$\dotfill$ B & & &&05-04-2011 & 1.90& &J~1445+0958 (P)& & 2.0 \\  
$\dotfill$  C& & & &30-03-2012  &0.67& &J~1445+0958 (P)& & 16.2 \\                        
$\dotfill$  D& &  &&30-12-2011 &0.30& &J~1445+0958 (P)& & 16.2 \\                                  \hline
S-band (3.00) & 1.987 $\to$ 4.011 & 2.024 & 16 & & & 3C~286&& 14.0& 8.2\\
$\dotfill$ C & & & &13-03-2020&3.20& & J~1445+0958 (P)& & \\
\hline
C-band (6.00) & 4.979 $\to$ 7.020 &  2.042& 16 &     &&3C~286&&7.0    \\ 
$\dotfill$ C & & & &04-02-2012 & 1.50& & J~1505+0326 (P)& &4.0\\
              & & & &21-02-2013&1.60& & J~1505+0326 (P)& &4.0\\
$\dotfill$ D & & & &10-12-2011&0.67& & J~1445+0958 (P)& &4.0\\
\hline
\end{tabular}
\end{center}
$^{\rm a}$ Observing band and its central frequency. VLA array configurations, corresponding to separate observing sessions, are given on their own lines following. A blank means that the value is the same as indicated immediately above.\\
$^{\rm b}$ Frequency range.  The small gap at L-band was necessary to eliminate radio frequency interference.\\
$^{\rm c}$ Total bandwidth.\\
$^{\rm d}$ Number of spectral windows in the band.\\
$^{\rm e}$ Date(s) of observations (day-month-year).\\
$^{\rm f}$ Time on source.\\
$^{\rm g}$ Primary calibrator. \\
$^{\rm h}$ Secondary calibrators. P = `primary' ( $<~3$\% amplitude close errors);  S = `secondary' ($3$ to $10$\%). \\
$^{\rm i}$ Primary beam full-width at half-maximum at the central frequency, given by $\rm PB ~{FWHM}=42/\nu$, where $\nu$ is in GHz and $\rm PB~{FWHM}$ is in arcmin. \\
$^{\rm j}$ Largest Angular Size detectable at the central frequency, from the {\it VLA Observational Status Summary} ({\tt https://science.nrao.edu/facilities/vla/docs/manuals}).\\ 
%https://science.nrao.edu/facilities/vla/docs/manuals/oss/performance/resolution
\end{table*}
%judith used listobs to get these numbers
%https://science.nrao.edu/facilities/vla/docs/manuals/oss/referencemanual-all-pages
% 

\subsection{Narrow-Band Spectra of NGC~5775}
\label{sec:narrowband_N5775}

Narrow-band spectra of NGC~5775 and corresponding tabular values are given in Figure~\ref{fig:spectra_27points_N5775} and Table~\ref{tab:spectral_fits_27points_N5775}, respectively. We use the same narrow-bands as for NGC~3044 (Table~\ref{tab:im_params}), except that we have restored one point at $\nu\,=\,5.81$ GHz which had been unreliable for NGC~3044. Fits were carried out for the central maximum point as well as a representative 47 beam region that is shown in Figure~\ref{fig:N5775_fig10}.

Our first conclusion is that there is still a flattening of the spectrum at low frequencies, though it is much more subtle and gradual than for NGC~3044. For consistency with NGC~3044, we omit the lowest frequency points for the fitting (see arguments in Sect.~\ref{sec:initial_spectrum}). However, also following NGC~3044, we repeat the fitting
procedure using all 31 points. The resulting parameters for the two cases agree within uncertainties but the fits improve for the 28-point case.  

Our second main conclusion is that the fit of type C, in which both nonthermal curvature as well as a thermal component are included, is demonstrably a better fit than any of the others. Adding a thermal component alone (fit of type B) is indistinguishable from no thermal component at all (type A). That is, $I_{\rm TH}$ is indistinguishable from zero without the inclusion of nonthermal curvature.  The same conclusions are reached for the 31-point fits. This is unlike NGC~3044 where, mathematically, the curves are indistinguishable (Sect.~\ref{sec:initial_spectrum}). 

Note that, on average, $\beta$ is negative.  In the absence of a thermal contribution, negative $\beta$ flattens the curve at low frequencies and steepens it at high frequencies.

\subsection{Results from Cube-Fitting}
\label{sec:cube_fitting_N5775}

As we did for NGC~3044 (Sect.~\ref{sec:cube_fits}), we fit a cube, point-by-point to NGC~5775.  However, instead of using Eq.~\ref{eqn:fittedeqn}, the results of Sect.~\ref{sec:narrowband_N5775} indicate that the fitted equation should now be,
\begin{equation}
I_\nu\,=\,I_{{\rm NT}}\,\left(\frac{\nu}{\nu_0} \right)^{\alpha_{\rm NT}\,+\,\beta_{\rm NT}\,{\rm log}{\frac{\nu}{\nu_0}}}
\,+\, I_{\rm TH}\,\left(\frac{\nu}{\nu_0} \right)^{-0.1}.
\label{eqn:fittedeqnbeta}
\end{equation}
Further tests generating images from cube-fitting showed that Eq.~\ref{eqn:fittedeqnbeta} improves over Eq.~\ref{eqn:fittedeqn} for this galaxy.  For these fits, we add the restrictions $-3.0\,\leq\,\alpha_{\rm NT}\,\leq\,1.0$, $-5.0\,\leq\,\beta_{\rm NT}\,\leq\,5.0$ and, as before, $I_{\rm NT}$ and $I_{\rm TH}$ must be positive.  As with NGC~3044, we blank regions in which $I_{\rm TH}$ goes to zero.  However, to obtain smoother images, we do not blank regions where the relative error exceeds 2.0.  Tests with and without this relative error cutoff show that the medians, as given in Table~\ref{tab:imaging_results_N5775}, differ by $\ltabouteq~10$\%.
Further information about the fitting can be found in Appendix~\ref{app:curve_fitting}.

\begin{figure*}[hbt!]
    \centering    \includegraphics[width=5truein,height=3.4truein]{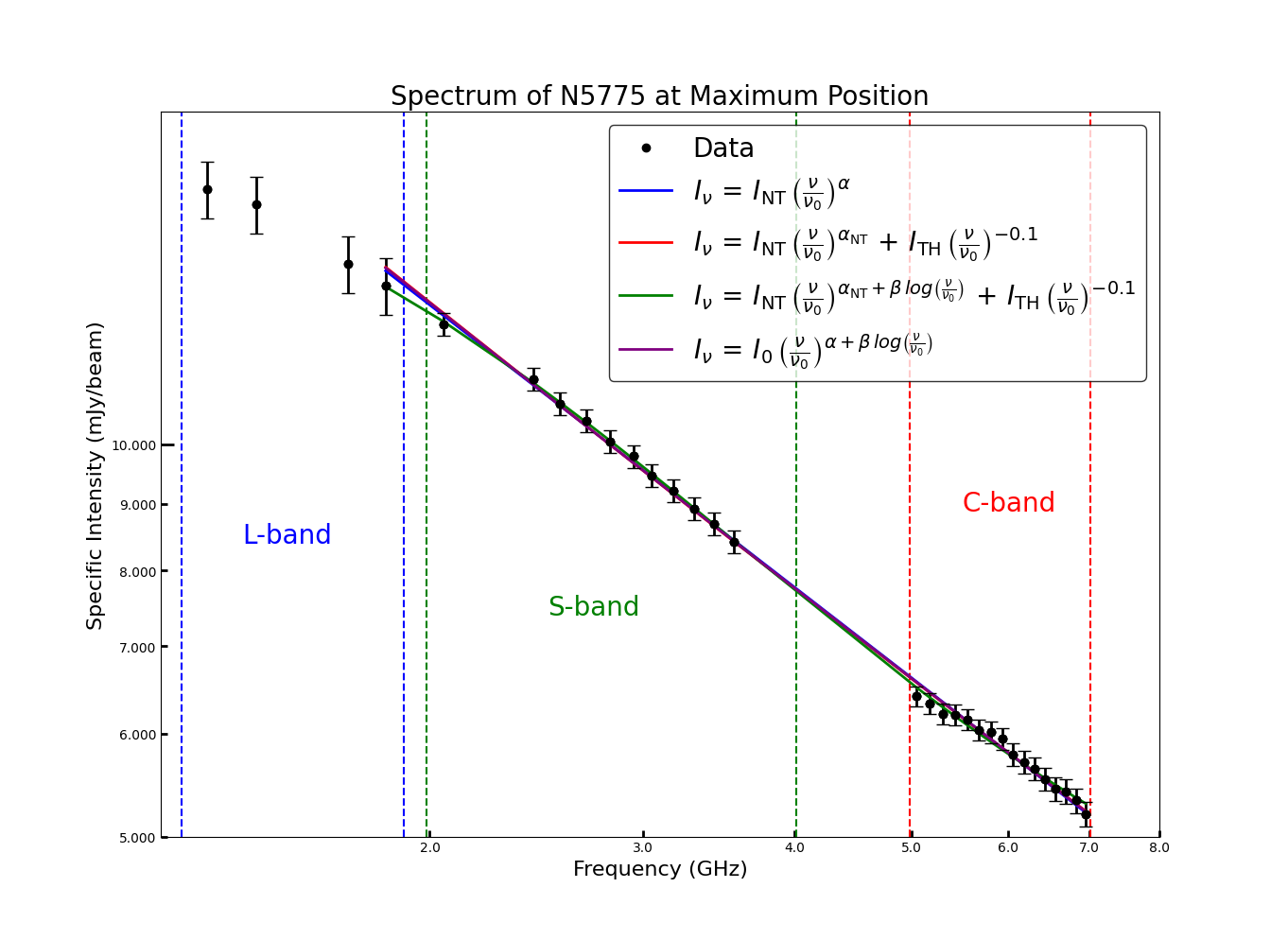}
    \vskip -0.2truein
\includegraphics[width=5truein,height=3.4truein]{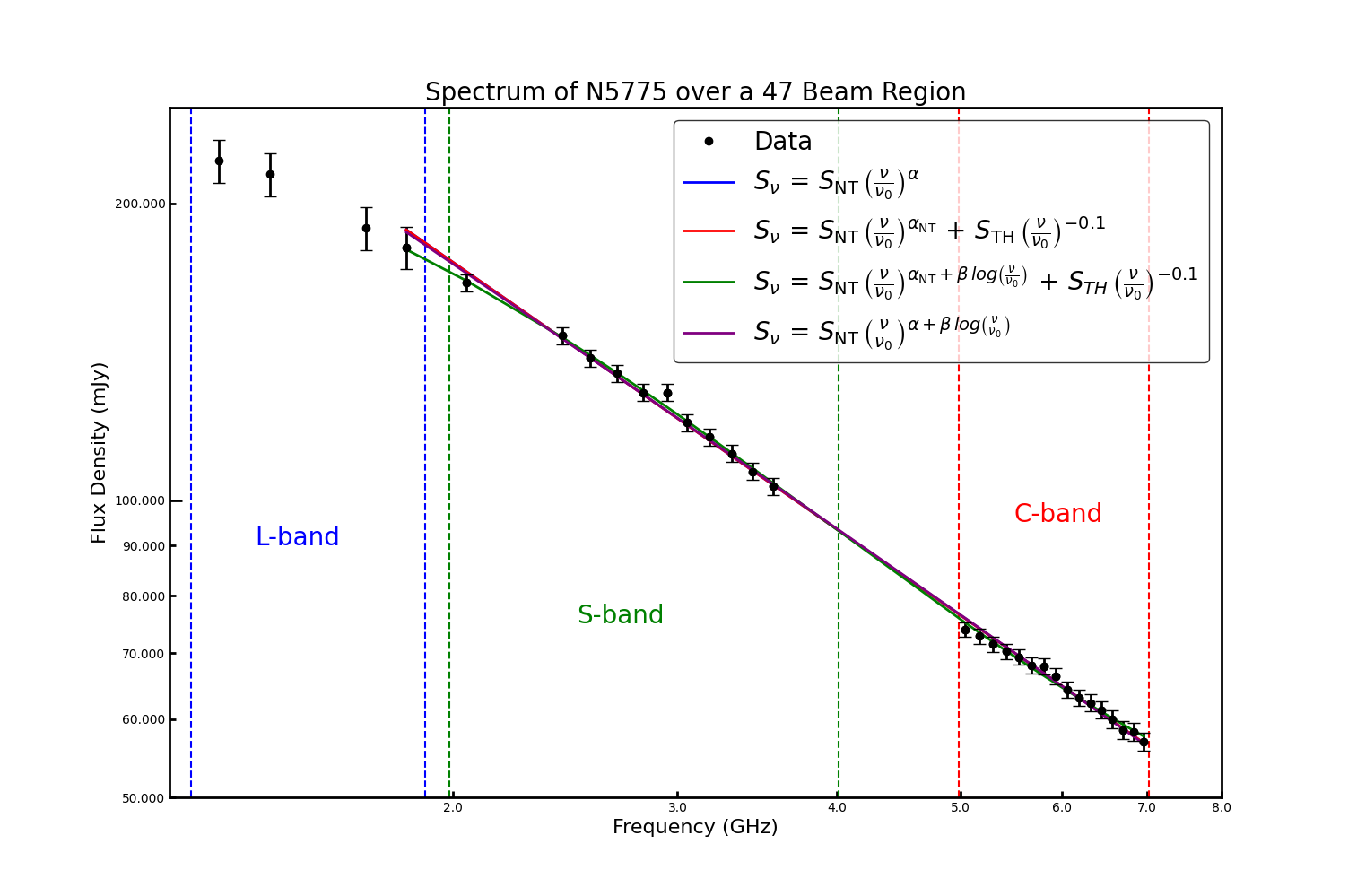}
    \caption{Spectrum of NGC~5775 from narrow-band imaging, omitting the lowest three frequency points.  Top: Values for the central maximum point only. Error bars add in quadrature: a 2\% or 5\% calibration error (Appendix~\ref{app:calibration}) and the rms of each map. Bottom: flux density over a 47 beam region that is common to all maps and is encircled by the grey dashed curve in Figure~\ref{fig:N5775_fig10}. Error bars add in quadrature a 2\% or 5\% calibration error (Appendix~\ref{app:calibration}) with the rms from each map
    multiplied by the square root of the number of beams. For each plot, the various frequency bands are marked, and the fitted equations are given in the legend.  Fitted parameters are given in Table~\ref{tab:spectral_fits_27points_N5775}.  }
    \label{fig:spectra_27points_N5775}
\end{figure*}

{\begin{table*}[!htb]
\fontsize{7}{11}\selectfont
\begin{center}
\caption{Spectral Fits {of NGC~5775}-- 12 arcsec resolution -- 28 points}\label{tab:spectral_fits_27points_N5775}
    \begin{tabular}{lcccccc}
    \hline
    {\it Fitted Expression}$^{\rm a}$&&&&&&\\
    \hline
{\bf Maximum Position}$^{\rm b}$&  $I_{\rm NT}$$^{\rm c}$& $\alpha$ or $\alpha_{\rm NT}$ $^{\rm c}$& $\beta$ or $\beta_{\rm NT}$ $^{\rm c}$&$I_{\rm TH}$$^{\rm c}$ & 
$\frac{I_{\rm TH}}
{I_{\rm tot}}$$^{\rm d}$ 
& rms scatter$^{\rm e}$
\\
 & (mJy beam$^{-1}$) & & & (mJy beam$^{-1}$) & & (mJy beam$^{-1}$)\\
\hline
\hline
{\bf A:} $I_\nu\,=\,I_{\rm NT}\,\left(\frac{\nu}{\nu_0} \right)^\alpha$&
$7.59\,\pm\,0.03 $& $-0.720\,\pm\,0.010$ &  & & & 0.10\\
{\bf B:} $I_\nu\,=\,I_{\rm NT}\,\left(\frac{\nu}{\nu_0} \right)^{\alpha_{\rm NT}}\,+\, I_{\rm TH}\,\left(\frac{\nu}{\nu_0} \right)^{-0.1}$ &
$6.96\,\pm\,1.31$ & $-0.774\,\pm\,0.122$& &$0.61\,\pm\,1.26$ & $0.08\,\pm\,0.15$ & 0.12\\
{\bf C:} $I_\nu\,=\,I_{\rm NT}\,\left(\frac{\nu}{\nu_0} \right)^{\alpha_{\rm NT}\,+\,\beta_{\rm NT}\,log(\frac{\nu}{\nu_0})}\,+\, I_{\rm TH}\,\left(\frac{\nu}{\nu_0} \right)^{-0.1}$ &
$3.13\,\pm\,0.60$ & $-1.68\,\pm\,0.33$ & $-1.30\,\pm\,0.63$ &$4.41\,\pm\,0.56$ & $0.58\,\pm\,0.06$ & 0.05\\
{\bf D:} $I_\nu\,=\,I_{\rm NT}\,\left(\frac{\nu}{\nu_0} \right)^{\alpha\,+\,\beta\,log(\frac{\nu}{\nu_0})}$ &
$7.58\,\pm\,0.05$& $-0.719\,\pm\,0.010$ &
$0.026\,\pm\,0.084$& & & 0.11\\
\hline
{\bf Region (47 beams)}$^{\rm b}$
& $S_{\rm NT}$$^{\rm c}$& $\alpha$ or $\alpha_{NT}$$^{\rm c}$& $\beta$ or $\beta_{\rm NT}$ $^{\rm c}$ &$S_{\rm TH}$$^{\rm c}$ & $\frac{S_{\rm TH}}{S_{\rm tot}}$$^{\rm d}$ & rms$^{\rm e}$
\\
& (mJy) & & & (mJy) & & (mJy) \\
\hline\hline
{\bf A:} $S_\nu\,=\,S_{\rm NT}\,\left(\frac{\nu}{\nu_0} \right)^\alpha$&
$90.78\,\pm\,0.34 $& $-0.900\,\pm\,0.010$ &  & & & 2.1\\
{\bf B:} $S_\nu\,=\,S_{\rm NT}\,\left(\frac{\nu}{\nu_0} \right)^{\alpha_{\rm NT}}\,+\, S_{\rm TH}\,\left(\frac{\nu}{\nu_0} \right)^{-0.1}$ &
$90.78\,\pm\,11.1$ & $-0.900\,\pm\,0.093$& &$1.3\,\pm\,10.6$ & $0.00\,\pm\,0.12$ & 2.1\\
{\bf C:} $S_\nu\,=\,S_{\rm NT}\,\left(\frac{\nu}{\nu_0} \right)^{\alpha_{\rm NT}\,+\,\beta_{\rm NT}\,log(\frac{\nu}{\nu_0})}\,+\, S_{\rm TH}\,\left(\frac{\nu}{\nu_0} \right)^{-0.1}$ &
$46.7\,\pm\,6.2$ & $-1.75\,\pm\,0.25$ & $-1.31\,\pm\,0.49$ &$43.8\,\pm\,5.9$ & $0.48\,\pm\,0.05$ & 1.1\\
{\bf D:} $S_\nu\,=\,S_{\rm NT}\,\left(\frac{\nu}{\nu_0} \right)^{\alpha\,+\,\beta\,log(\frac{\nu}{\nu_0})}$ &
$90.96\,\pm\,0.60$ & $-0.901\,\pm\,0.010$& $-0.030\,\pm\,0.082$ & & &1.9
\\
\hline
    \end{tabular}
\end{center}
$^{\rm a}$ The type of fit is given by the mathematical expressions shown.  For a single point (Maximum Position), the values are expressed as specific intensities, $I_\nu$, and are in mJy beam$^{-1}$. For a larger region (47 beams), the values are flux densities, $S_\nu$, in mJy. The beam is circular with a FWHM of 12 arcsec. Values are computed at the central frequency, $\nu_0\,=\,4.13$ GHz.\\
$^{\rm b}$ Position at which the fit was carried out. Maximum Position: center of the map at which the emission is a maximum.  Region (47 beams): Spatially integrated region within the areas enclosed by the grey curve in Figure~\ref{fig:N5775_fig10}. \\
$^{\rm c}$ Parameters of the fit with their standard deviations. \\
$^{\rm d}$ Thermal fraction at $\nu_0$.  The denominator is the sum of the thermal and nonthermal emission.\\
$^{\rm e}$ Root-mean-square scatter between the data and fitted curve.\\
\end{table*}
}

\begin{figure*}[hbt!]
   \centering
  \includegraphics[width=0.45\textwidth]{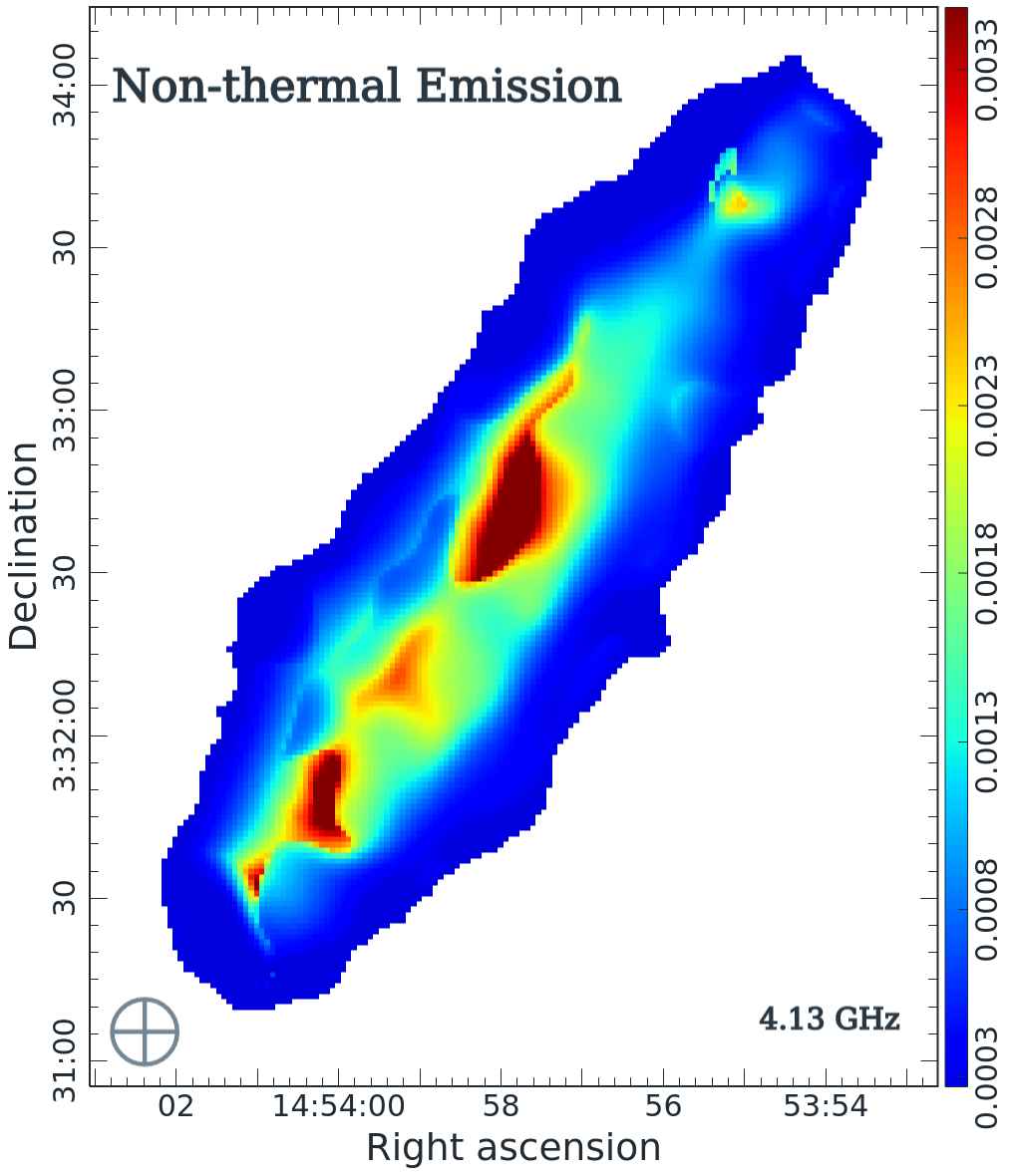}  
  \includegraphics[width=0.45\textwidth]{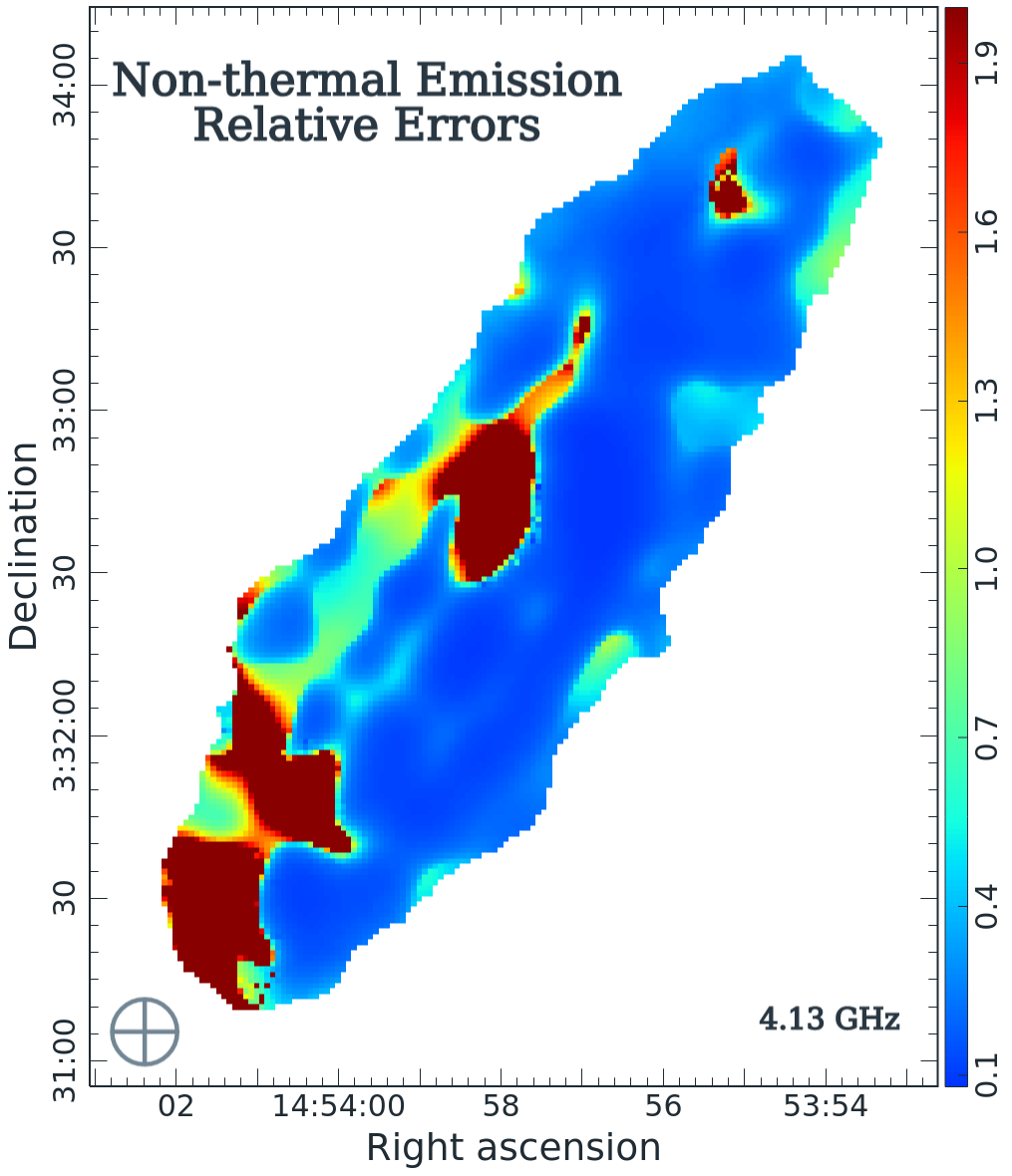}    
  \caption{Nonthermal emission (left) and its relative error (right) of NGC~5775 resulting from cube-fitting, as described in Sect.~\ref{sec:cube_fitting_N5775}. Units, for the image, are Jy beam$^{-1}$. The $12\arcsec$ beam is shown at lower left and the frequency is at lower right. %Open spaces are regions that have been blanked. 
  Corresponding data are in Table~\ref{tab:imaging_results_N5775}.
   }
  \label{fig:INT_result_N5775}
\end{figure*}

\begin{figure*}[hbt!]
   \centering
  \includegraphics[width=0.45\textwidth]{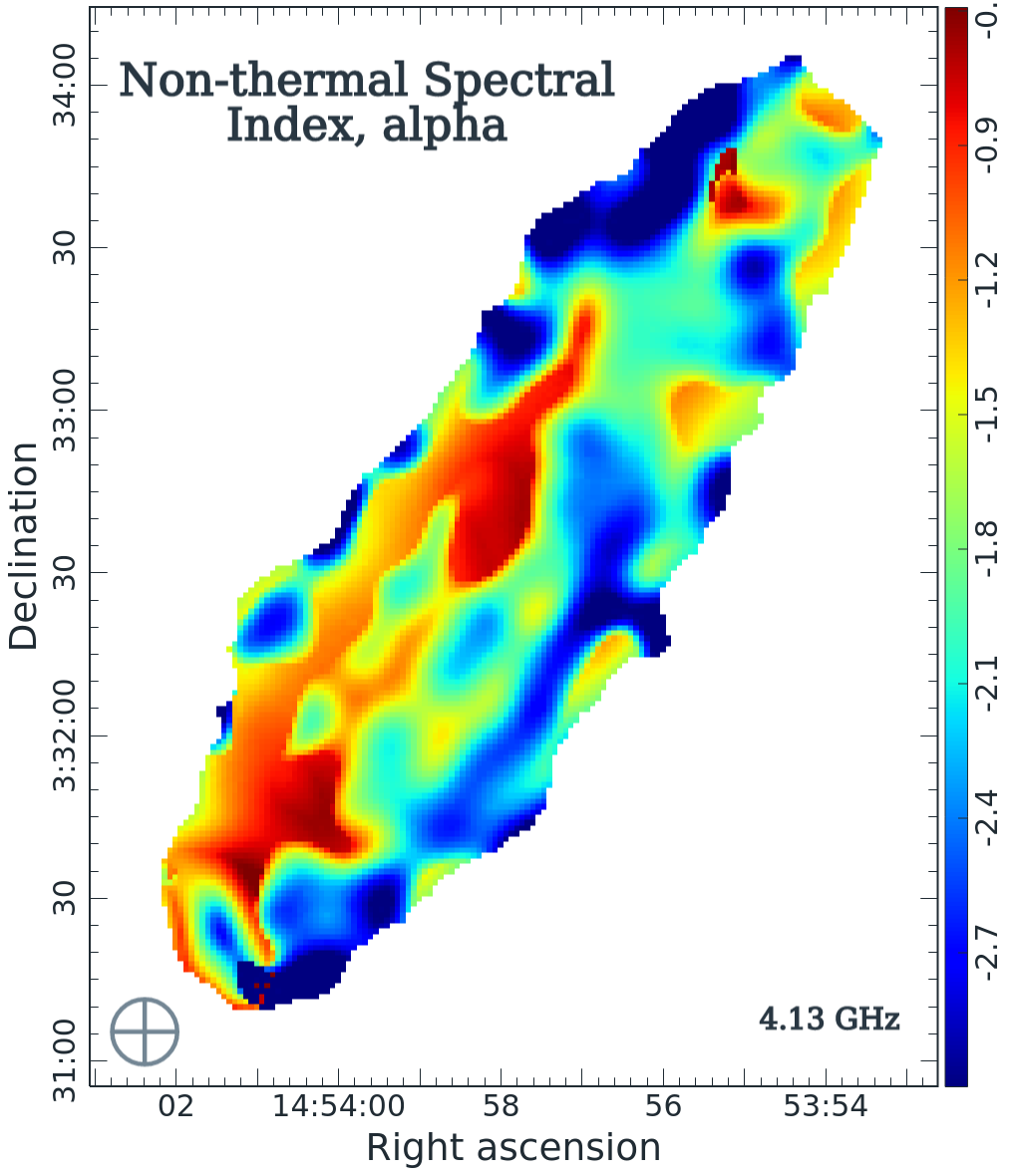}  
  \includegraphics[width=0.45\textwidth]{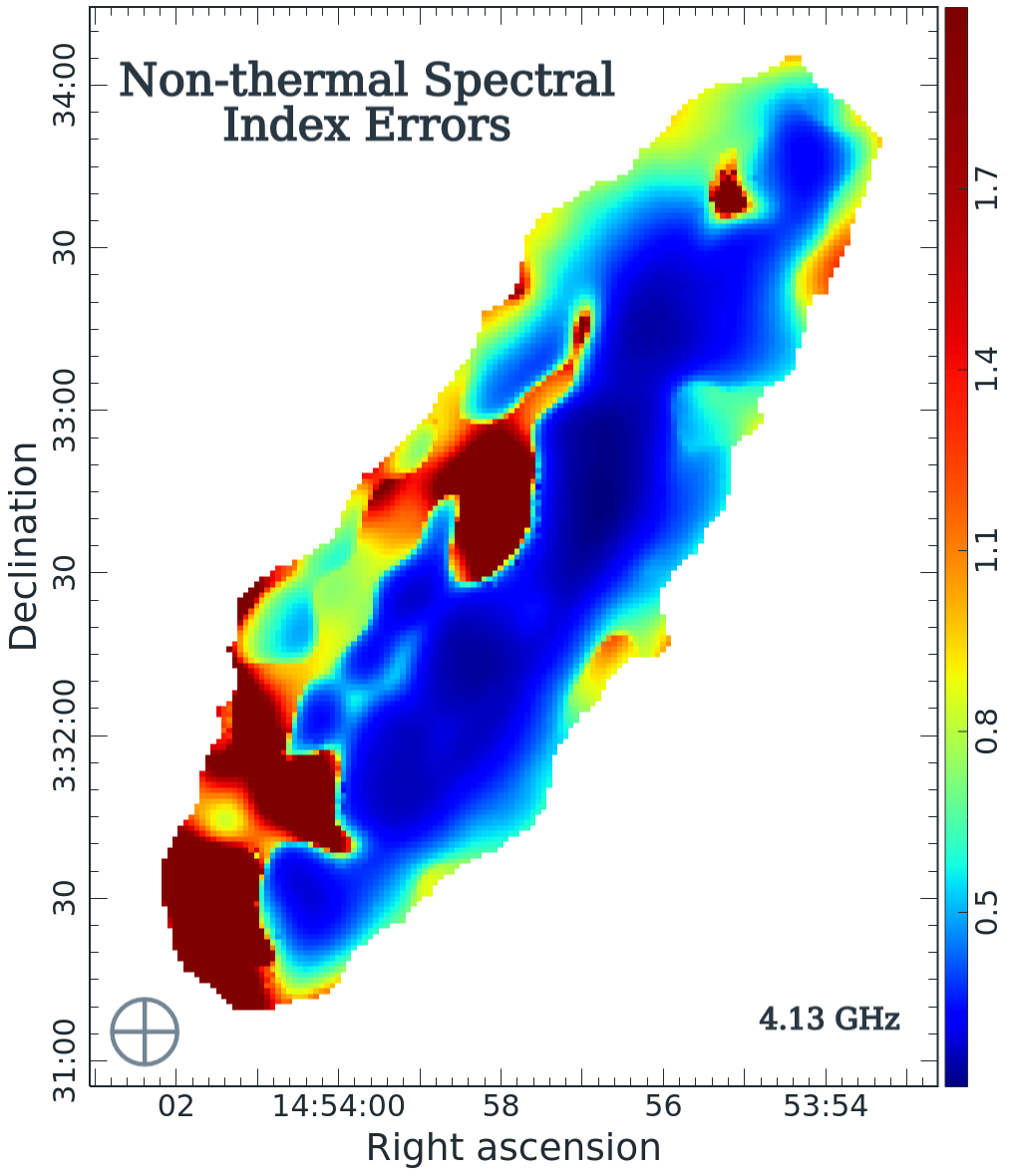}    
  \caption{Nonthermal spectral index, $\alpha_{\rm NT}$ (left) and its error (right) of NGC~5775 resulting from cube-fitting, as described in Sect.~\ref{sec:cube_fitting_N5775}.  The $12\arcsec$ beam is shown at lower left and the frequency is at lower right. Open spaces are regions that have been blanked.
   Corresponding data are in Table~\ref{tab:imaging_results_N5775}.}
  \label{fig:alpha_result_N5775}
\end{figure*}

\begin{figure*}[hbt!]
   \centering
  \includegraphics[width=0.45\textwidth]{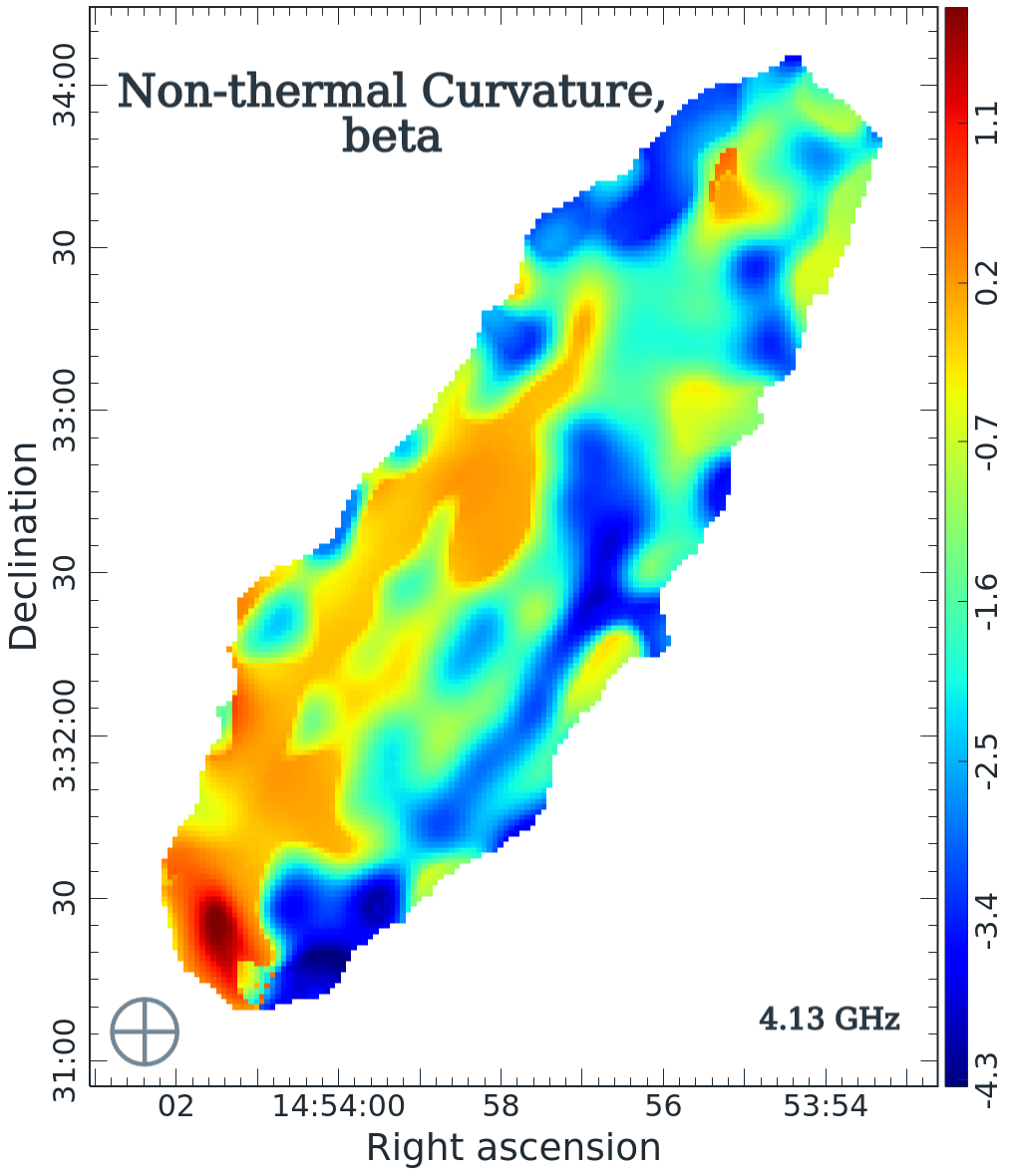 }  
  \includegraphics[width=0.45\textwidth]{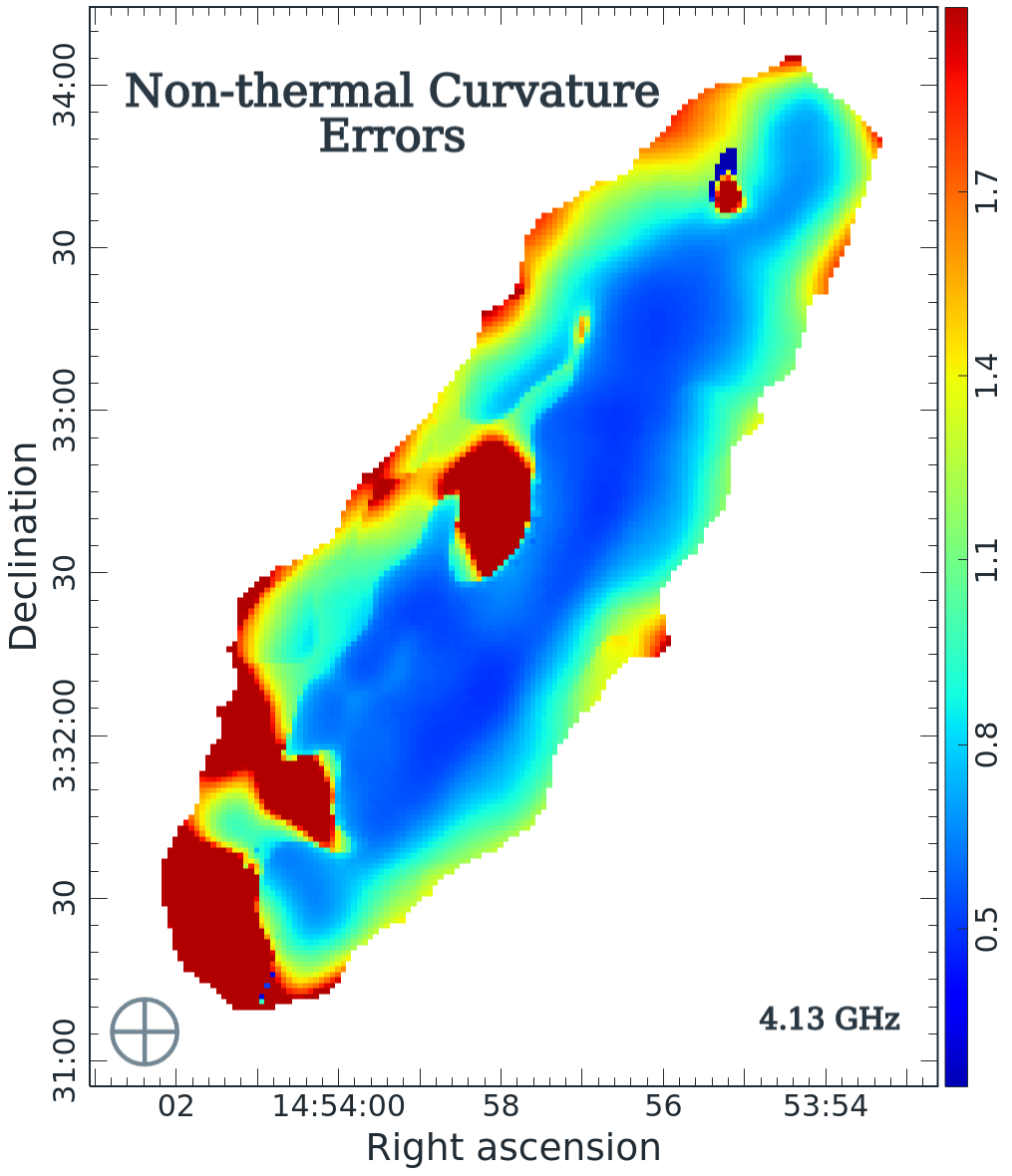}    
  \caption{Nonthermal curvature term, $\beta_{\rm NT}$ (left) and its error (right) of NGC~5775 resulting from cube-fitting, as described in Sect.~\ref{sec:cube_fitting_N5775}.  The $12\arcsec$ beam is shown at lower left and the frequency is at lower right. Open spaces are regions that have been blanked.
   Corresponding data are in Table~\ref{tab:imaging_results_N5775}.}
  \label{fig:beta_result_N5775}
\end{figure*}

\begin{figure*}[hbt!]
   \centering
  \includegraphics[width=0.45\textwidth]{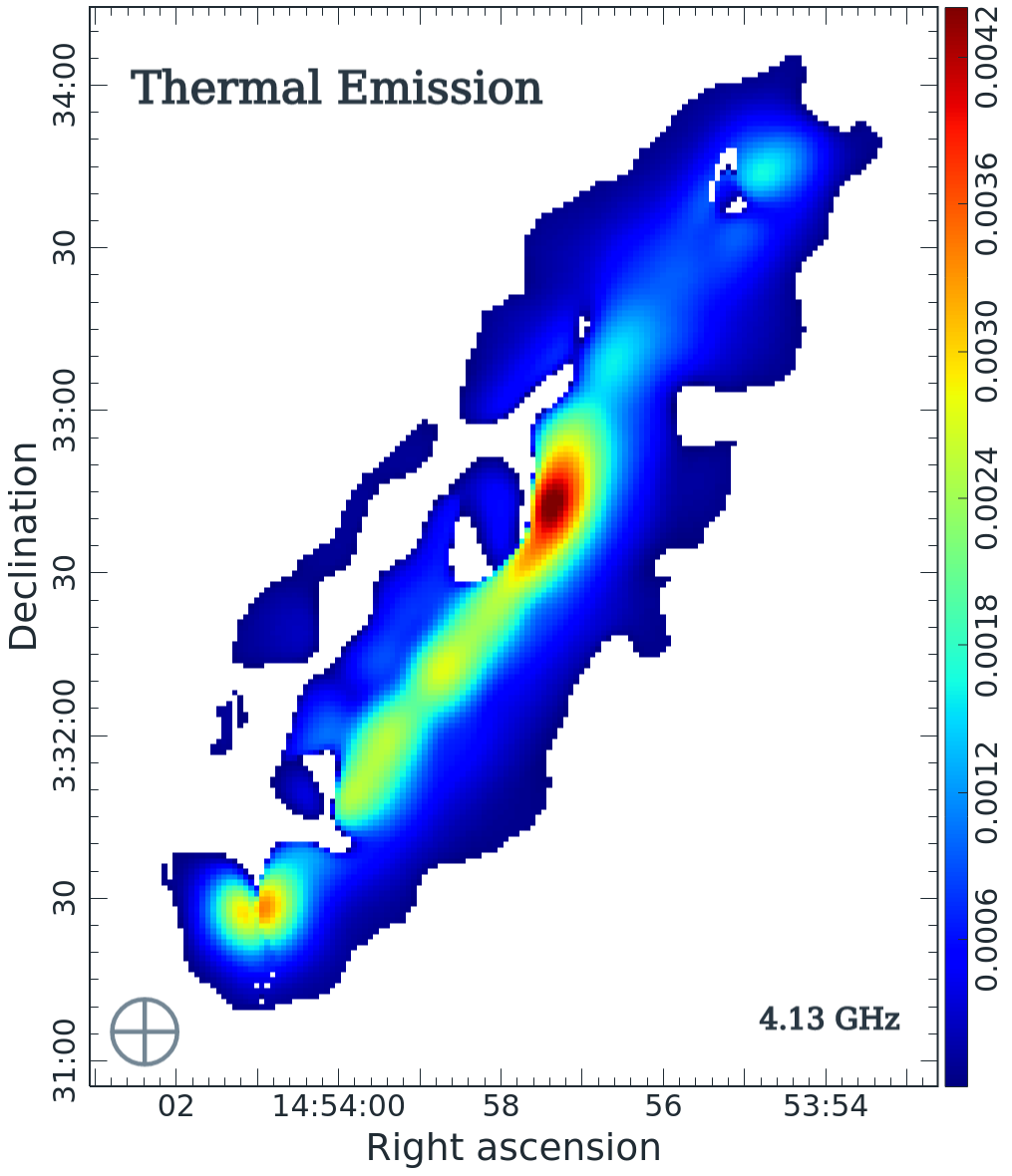}  
  \includegraphics[width=0.45\textwidth]{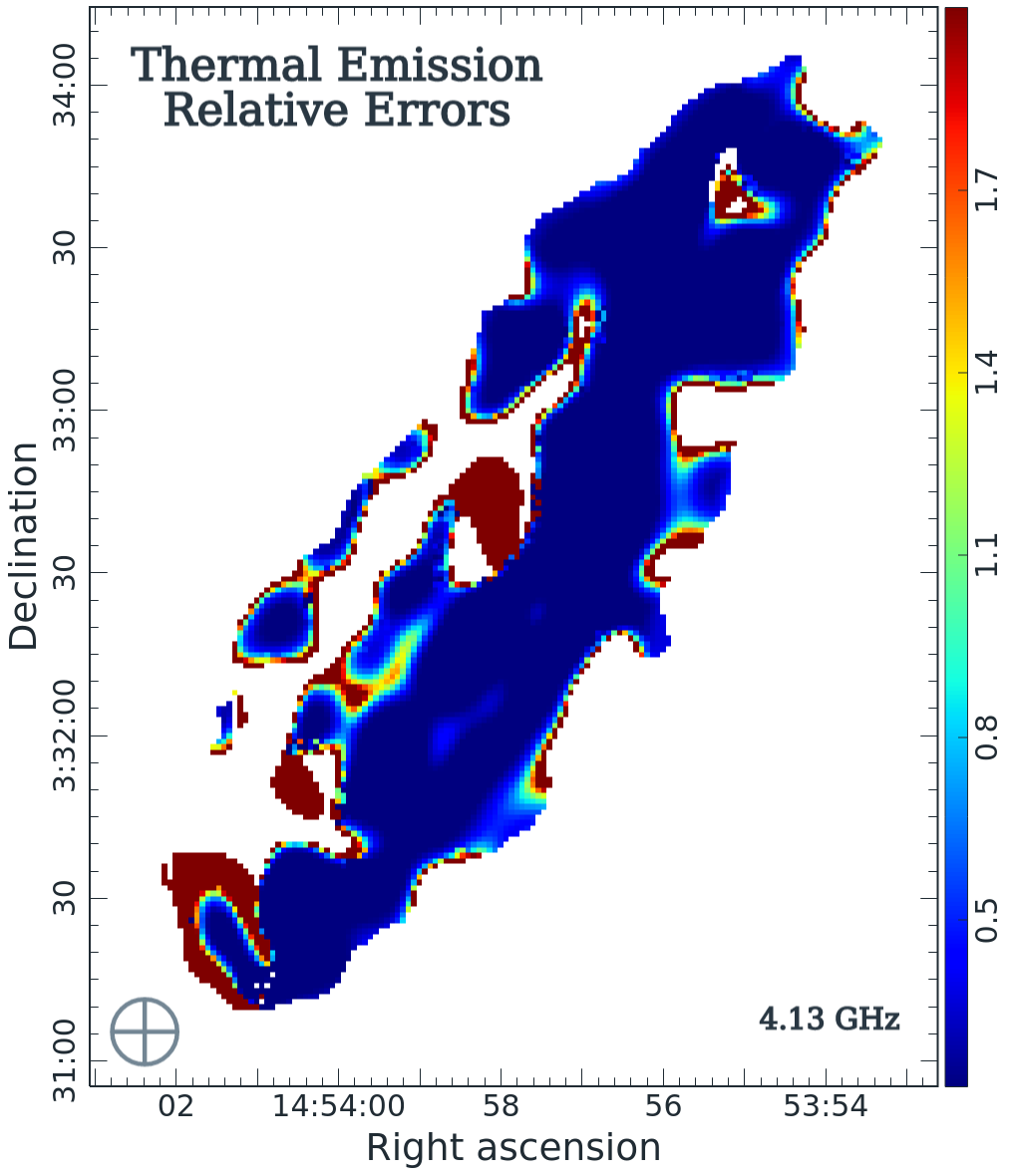}    
  \caption{Thermal emission (left) and its relative error (right) of NGC~5775 resulting from cube-fitting, as described in Sect.~\ref{sec:cube_fitting_N5775}.  The $12\arcsec$ beam is shown at lower left and the frequency is at lower right. Open spaces are regions that have been blanked.
   Corresponding data are in Table~\ref{tab:imaging_results_N5775}.}
  \label{fig:ITH_result_N5775}
\end{figure*}

\begin{figure*}[hbt!]
   \centering
  \includegraphics[width=0.45\textwidth]{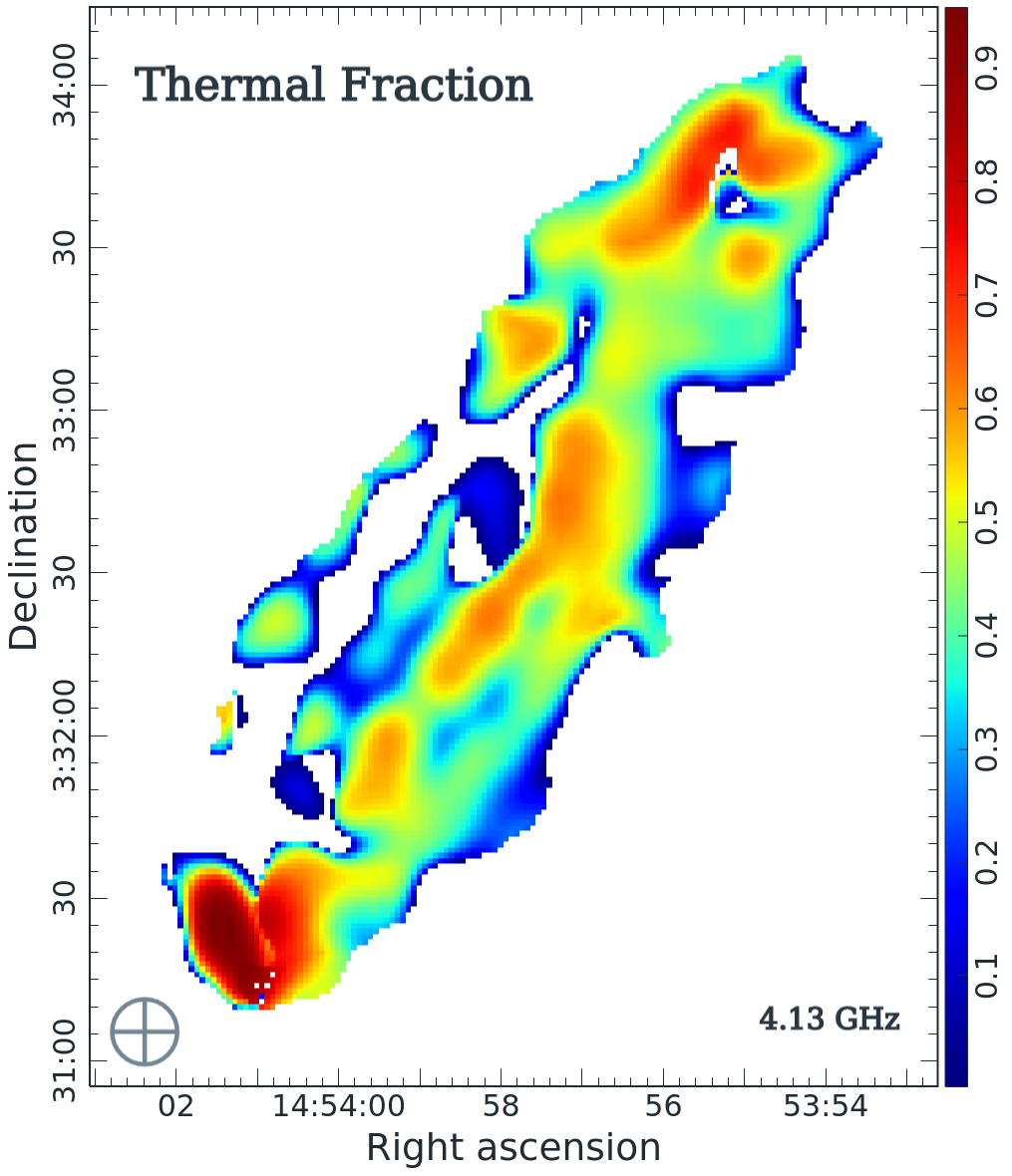}  
  \includegraphics[width=0.45\textwidth]{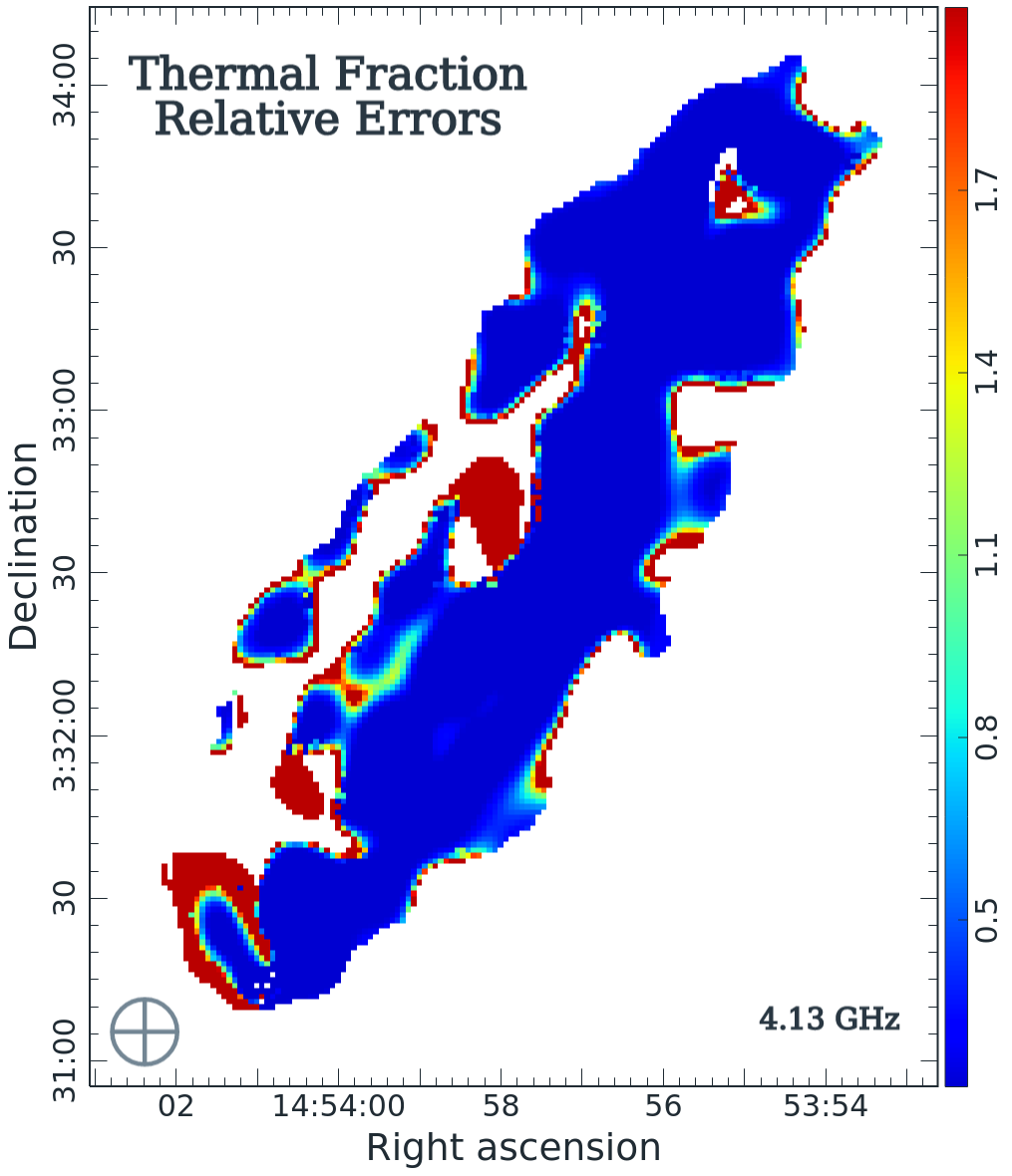}    
  \caption{Thermal fraction (left) and its relative error (right) of NGC~5775 resulting from cube-fitting, as described in Sect.~\ref{sec:cube_fitting_N5775}.  The $12\arcsec$ beam is shown at lower left and the frequency is at lower right. Open spaces are regions that have been blanked. Corresponding data are in Table~\ref{tab:imaging_results_N5775}.
   }
  \label{fig:ITHfrac_result_N5775}
\end{figure*}

\begin{table*}
\begin{center}
\caption{Imaging Results {of NGC~5775} (12 arcsec res.)$^{\rm a}$\label{tab:imaging_results_N5775}}
\begin{tabular}{lccccccccc}
\hline
Map & No. Beams$^{\rm b}$ & Flux Density$^{\rm c}$ &Minimum$^{\rm d}$&Maximum$^{\rm d}$ & Mean$^{\rm e}$ & Median$^{\rm f}$ & Median of Error map$^{\rm g}$\\ 
&  & (mJy) & ($\upmu$Jy beam$^{-1}$) & (mJy beam$^{-1}$) &(mJy beam$^{-1}$)&(mJy beam$^{-1}$)&\\
\hline\hline
$I_{\rm NT}$ & 60.5  &57.3 & 0.00& 7.10 &0.947 &0.628 &0.243 &  &  \\ 
$I_{\rm TH}$ &  49.2  & 37.1& 0.00& 4.54 & 0.754& 0.481& 0.216\\ 
\hline
Map & No. Beams$^{\rm b}$ &  &Minimum$^{\rm d}$&Maximum$^{\rm d}$ & Mean$^{\rm e}$ & Median$^{\rm f}$ & Median of Error map$^{\rm g}$\\
\hline\hline
$\alpha_{\rm NT}$ & 60.5 & & -3.0 & -0.522 & -1.88 & -1.89& 0.496\\
$\beta_{\rm NT}$ & 60.5 & & -4.48 & 1.89 & -1.47 & -1.52& 0.876\\
$\frac{I_{\rm TH}}{I_{\rm tot}}$& 49.2& & 0.669 & 0.959 & 0.423 & 0.439 & 0.167 \\
%$\frac{I_{\rm TH}}{I_{\rm tot}}$ (Var.)$^{\rm i}$ & &\\
\hline\hline
\end{tabular}
\end{center}
$^{\rm a}$ Numerical results from Figures~\ref{fig:INT_result_N5775} through \ref{fig:ITHfrac_result_N5775}. The reference frequency is $\nu_0\,=\,4.13$ GHz, the spatial resolution is 12 arcsec, and 28 frequency points were used at each pixel.\\
$^{\rm b}$ Number of beams in the region shown on the maps.\\
$^{\rm c}$ Flux density in the region for $I_{\rm NT}$ and $I_{\rm TH}$. \\
$^{\rm d}$ Minimum and maximum values in the region.\\
$^{\rm e}$ Mean value in the region.\\
$^{\rm f}$ Median value in the region.\\
$^{\rm g}$ Median value of the error maps. For $I_{\rm NT}$, $I_{\rm TH}$, and $\frac{I_{\rm TH}}{I_{\rm tot}}$, it is the relative error, and for $\alpha_{\rm NT}$ and $\beta_{\rm NT}$, it is the absolute error. \\
%See Sect.~\ref{sec:cube_fits} for more information.\\
\end{table*}
%judith used listobs to get these numbers
%https://science.nrao.edu/facilities/vla/docs/manuals/oss/referencemanual-all-pages
% 
Imaging results for NGC~5775 show complex structure, as well as some regions with high relative errors. The thermal emission (Figure~\ref{fig:ITH_result_N5775}), ignoring blanking, shows a fairly smooth increase towards the center with some additional peaks along the major axis, especially at the far north-western and south-eastern tips of the major axis.
The thermal fraction (Figure~\ref{fig:ITHfrac_result_N5775}) maps shows a particularly high value at the south-eastern location.

\subsection{Comparison with H$\alpha$}
\label{sec:halpha_comparison_N5775}

For NGC~5775, we do not have a spatially resolved map of $I_{\rm TH}$ from the mixture method, as we did for NGC~3044.  However, we can compare our results to an H$\alpha$ map and can also compare some global results.

Figure~\ref{fig:Halpha_ITH_N5775} shows an overlay of H$\alpha$ emission (contours), smoothed to 12 arcsec resolution.  The colour image is the same as Figure~\ref{fig:ITH_result_N5775} showing the thermal emission obtained using the radio-only method.  This is a good example of how well the radio-only thermal emission matches the completely independent H$\alpha$ map.  The H$\alpha$ contours follow the radio thermal emission along the disk, and both peaks due to the two HII regions, one on the far north-west tip and one on the far south-east tip, are also matched. The radio-only thermal/nonthermal separation has recovered these HII regions very well.

\begin{figure*}[hbt!]
   \centering
  \includegraphics[width=0.45\textwidth]{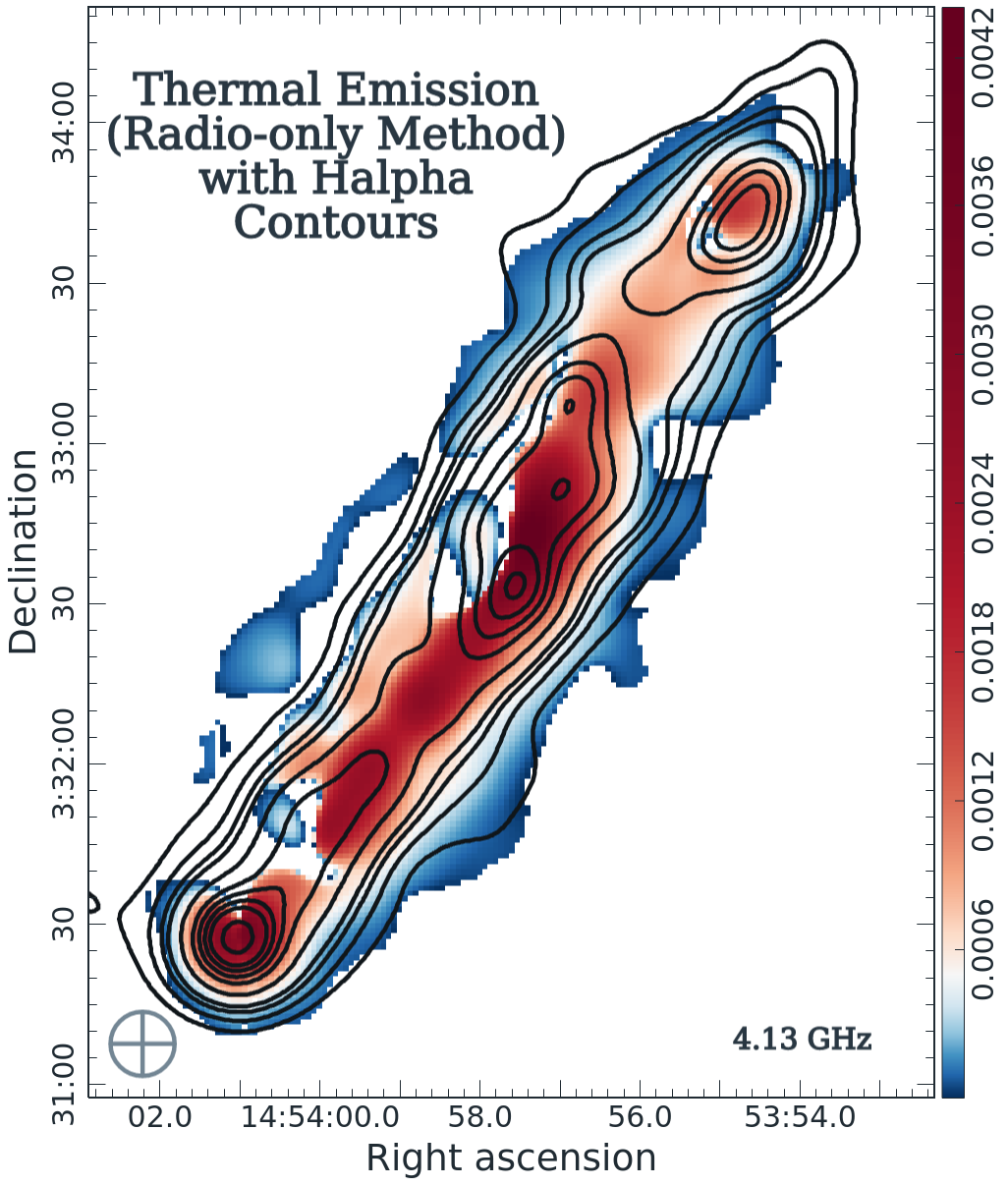}     
  \caption{Colour image of thermal emission from Fig.~\ref{fig:ITH_result_N5775} with contours (at arbitrary scaling) of H$\alpha$ emission superimposed.  The colour bar is in units of Jy. The H$\alpha$ map is taken from Figure~\ref{fig:N5775_fig10} smoothed to the same 12 arcsec resolution of the thermal map. The beam is shown at lower left.
   }
  \label{fig:Halpha_ITH_N5775}
\end{figure*}

As for the thermal fraction map, the radio-only method gives a value of 44\% with an estimated error {on this fraction} of 17\% as listed in Table~\ref{tab:imaging_results_N5775}, which is applicable to the non-blanked regions of Figures~\ref{fig:INT_result_N5775} through \ref{fig:ITHfrac_result_N5775}. This can also be compared with the thermal fraction that has been estimated from H$\alpha$ data.
From \citet{hea22}, the thermal fraction using H$\alpha$ data at $\nu\,=\,1500$ MHz is $15\,\pm\,4$\% with higher values up to $57\,\pm\,23$\% found in locations of star-forming regions in the disk. If their thermal emission is increased by a factor of 1.36 following \cite{var18} for edge-on galaxies, then this value increases to 19\%. Adjusting to $\nu\,=\,4.13$ GHz adopting $\alpha_{NT}\,=\,-0.8$, gives an H$\alpha$-generated thermal fraction of 32\% at 4.13 GHz. If we adopt a steeper value of $\alpha_{NT}$, this fraction would increase. Therefore, our radio-only thermal fraction, taken at face value, is a factor of 1.4 higher than the Heald et al. value. Therefore, it is possible that some thermal emission has been missed from the H$\alpha$ Heald et al. method, but our spatial region is somewhat smaller than theirs.  Given the assumptions, the two values are in rough agreement.

}

\clearpage

\section{Discussion}
\label{sec:discussion}
{Our radio-only method of separating nonthermal and thermal emission in NGC~3044 and NGC~5775 gives reasonable results at 4.13 GHz, both quantitatively and qualitatively, for extracting the thermal emission over the map regions investigated. 

For NGC~3044, we cannot distinguish between the best fit (Table~\ref{tab:spectral_fits_27points}) in terms of the fit rms.  It is clear that a thermal component must be present, but a nonthermal curvature component (non-zero $\beta_{\rm NT}$) could also be present without being  distinguished in our data. By insisting on a fit without curvature (Type B), our solution still allows for the possibility that some nonthermal emission could be represented in the thermal map. For example, positive $\beta$ can mimic the flattening at high frequency where thermal emission is also flattening the curve. A comparison between the thermal fraction determined from the radio-only method and the H$\alpha$ mixture method (Figure~\ref{fig:Vargas_radio_comparison}), however, shows good agreement in the disk.  In regions away from the disk, such as where the thermal fraction shows a peak (yellow cross), the thermal fraction may be overestimated (see Sect.~\ref{sec:vertical_feature}). 

At 12 arcsec resolution for NGC~3044, the thermal fraction in the measured region only, is 25\% with an estimated uncertainty {on this fraction} of 50\% (Table~\ref{tab:imaging_results}). However, if we extrapolate to a larger 41 beam region, the thermal fraction reduces to $\sim\,10$\%.  To match the mixture method H$\alpha$ map at 15 arcsec resolution, the estimated thermal fraction, now over $\sim$ 26 beams (Table~\ref{tab:imaging_results_15arcsec}), 
 we find a thermal fraction of $\sim$ 13\%.  This can be compared to the H$\alpha$ mixture method at the same resolution and frequency, which gives $\sim$ 13\% as well, with an estimated uncertainty on this fraction of 20\%, assuming that the optical and infrared calibrations are reliable for these edge-on galaxies.		

For NGC~5775, our fitting procedure shows that a fit of Type C (non-zero $\beta_{\rm NT}$) gives the best results (Table~\ref{tab:spectral_fits_27points_N5775}). A key comparison with H$\alpha$ data is best seen in Figure~\ref{fig:Halpha_ITH_N5775}, where thermal emission from the radio-only method follows the H$\alpha$ emission in the disk.  Over the detected region of $\sim$ 49 beams, our radio-only thermal fraction is
44\% with a 17\% uncertainty on this value (Table~\ref{tab:imaging_results_N5775}). This is slightly higher than the value estimated from H$\alpha$ results of 32\%, depending on the adopted value of $\alpha_{NT}$ for extrapolating to the same frequency. Our restricted map region may also be favouring higher thermal emission compared to the global H$\alpha$-related value.

}

\subsection{The Apparent Vertical Outflow Feature in NGC~3044}
\label{sec:vertical_feature}

Figure~\ref{fig:ITH_fractions_15arcsec} visualizes the main difference between the H$\alpha$-derived thermal fraction and the radio-derived thermal fraction.  
A surprising result is the appearance of an apparent vertical feature centered approximately at the nucleus of the galaxy. This feature is {\it not} observed in either the H$\alpha$ map or the thermal fraction map of \cite{var18}. It is also not obvious in the radio-only modeled $I_{\rm NT}$ or $I_{\rm TH}$ maps (Figures~\ref{fig:INT_result}, \ref{fig:ITH_result}), which both show decreases in intensity with distance from the center. 
The fact that the thermal map shows barely a hint of a vertical extension emphasizes how subtle this feature is. 
It really 
becomes obvious only in the thermal fraction map (Figure~\ref{fig:ITH_fraction_result}).
%NGC~3044 is not exactly edge-on, with an inclination of $i\,=\,85\degr$ (Table~\ref{tab:N3044params}).  Could the apparent vertical feature actually represent a structure that is in the plane? A 12 arcsec beam, as in Figure~\ref{fig:ITH_fraction_result}, corresponds to a projected size of 1.18 kpc. If it were to represent emission along the disk of the galaxy, instead of height, the distance along the plane would be 13.5 kpc at the galaxy's center. However, at the northern peak of the feature (2.6 kpc in projection, see Sect.~\ref{sec:thermaloutflow}) the corresponding in-plane distance is 30 kpc which is well outside of any disk emission. 
%Moreover, its morphology bears no resemblance to an inclined disk. We also know that globally, disk emission is dominated by nonthermal emission and we would not expect to see a thermal fraction increasing with the radius in the disk.  We therefore interpret this apparent vertical feature as being actually vertical in the galaxy. 
There is a maximum in this vertical feature 22 arcsec (2.2 kpc in projection) to the north of the nucleus, close to the northern limit of the map (Figure~\ref{fig:ITH_fraction_result}), at
position: RA$\,=\,09^{\rm h}\,53^{\rm m}\,41\rasec 2$, 
DEC$\,=\,01^{\rm d}\,35^\prime\,10\decsec 2$.

To investigate this further, we show the radio-only thermal fraction map as contours over a colour image of the H$\alpha$ mixture method $I_{\rm TH}$ image in Figure~\ref{fig:Vargas_radio_comparison} (Bottom). In this case, for the colour map, we show a wider field since the H$\alpha$ map extends farther than the radio-only map. The galaxy center is marked with a magenta `+' and the northern peak in the vertical feature is marked with a yellow `+'. 
We now see that the peak in the vertical thermal  fraction map is at the center of a loop (colour plus white contours) in the  dust-corrected H$\alpha$ map. At this position, a vertical outflow will have escaped from the disk and the H$\alpha$ loop likely represents a thermal shell.

Note that there is another extended H$\alpha$ feature to the east (left) of the vertical feature that extends to $\sim$ 1 arcmin from the midplane towards the north, or almost 6 kpc in projection (white contours, hereafter the `eastern extension').  The thermal fraction map also shows an extension toward this feature up to the point at which the thermal fraction map is cut off (see black contours).

Apparently, there is a vertical outflow from the nuclear regions of this galaxy which has a high thermal fraction ($\sim\,65$ \% from Figure~\ref{fig:ITH_fraction_result}), assuming that  our method is valid at this high latitude position. 
At this peak position, from the matching 15 arcsec resolution maps, $I_{\rm TH}$ (radio-only) = $244\,\upmu$Jy beam$^{-1}$ whereas
$I_{\rm TH}$ (H$\alpha$ mixture method) = $6.9\,\upmu$Jy beam$^{-1}$. 
 Consequently, the radio-only thermal map is high compared to the H$\alpha$ mixture method at this location. 
A search for possible background sources at this location using the Aladin Sky Atlas\footnote{https://aladin.u-strasbg.fr/} as well as our previous high resolution images \citep[e.g.][]{irw22} showed no other sources at this position.

{Assuming a temperature of $T_e\,=\,10^4$ K, in a 15 arcsec beam, thermal emission of 244 $\mu$Jy beam$^{-1}$ corresponds to an emission measure of ${\cal EM} \,= \,466$ cm$^{-6}$ pc. This is well above the detectable rms of
${\cal EM}\,=\,9.31$ pc cm$^{-6}$ of \cite{var18} and therefore should have been detected in the mixture method maps unless there is strong dust obscuration at the same position. It may be, then, that there is still a contribution from nonthermal emission at this high latitude position. 

We can investigate this further. As indicated in Sect.~\ref{sec:initial_spectrum}, a fit of Type B ($\beta_{\rm NT}\, = \,0$ with $I_{\rm TH}$, as was modelled here) and a fit of Type D ($\beta_{\rm NT}\,\ne\,0$ with $I_{\rm TH}\,=\,0$, as was rejected because of the lack of a thermal component) cannot really be distinguished mathematically.  As $z$ increases and the position becomes far from the disk, we know that the nonthermal fraction should increase and so a fit of Type D should not be rejected. Therefore, we carried out the same analysis as presented in Table~\ref{tab:spectral_fits_27points}, but for the position of the peak in the vertical feature, 2.2 kpc above the plane. Again, we find that the two fits cannot be distinguished from the rms.  For a Type B fit, we find $\alpha_{\rm NT}\,=\,-2.09\,\pm\,2.35$ with a thermal fraction, $Fr\,=\,0.65\,\pm\,0.45$, as shown in Figures~\ref{fig:alpha_NT_result} and \ref{fig:ITH_fraction_result}
For a Type D fit, we find that $\alpha\,=\,-0.89\,\pm\,0.32$ and $\beta\,=\,0.98\,\pm\,2.2$.  In other words, no thermal component with possibly a flatter spectral index matches the data just as well. In this case, positive $\beta$ is what provides the flattening at high frequency.
%The facts that there is abundant detected dust \citep{var18} as well as polarization \citep[e.g.][]{coo23} in the halo of NGC~3044 argues for the presence of 

There is, in any event, evidence for a nuclear outflow in NGC~3044, possibly a mixture of thermal and nonthermal emission, similar to the galaxy, M~82 
\citep[e.g.][ and references therein]{lop20}.}
%, although at a less powerful level. 
The halo of M~82 shows significant polarization \citep{reu94} even though the outflow represents a starburst wind, indicating that synchrotron emitting plasma is part of the outflow.  For NGC~3044, we also observe significant polarization in the halo and largely vertically directed fields \citep[][see their Fig. 3]{kra20}. The star formation rate of M~82 is $\sim$ 10 M$_\odot$ yr$^{-1}$ \citep{deg01}, about 6 times greater than NGC~3044 (Table~\ref{tab:N3044params}). It is possible that NGC~3044 is a less powerful analogue to M~82 or at a different evolutionary stage. {The results of this analysis argue that our thermal/nonthermal analysis is best used in the disk of the galaxy where we are sure that a thermal component is present and significant.}

%We searched through the literature to see whether any analogues to the vertical feature are observed at other wavebands. At present, we do not see any other evidence, beyond the H$\alpha$ map, but this is not surprising, given that it is undetectable in our own $I_{\rm TH}$ map (Figure~\ref{fig:ITH_result}.  Infra-red spectral line observations are recommended as a follow-up, or radio recombination lines.
 %A line of sight distance equal to the beam FWHM is 1.476 kpc, 

\subsection{Vertical Extension into the Halo of NGC~3044}
\label{sec:vertical}

%{\bf MICHAEL, any change to this section? Probably the vertical extension is not just from an increasing thermal fraction but from nonthermal emission too, as outlined in the previous section
%\textcolor{red}{Michael: I think from my side we do not need to change things. However, we note in this section that we use a 2\% calibration error for the fitting. I think this need to be updated. If I remember correctly, you performed the spectral fitting for the box averages, right? }
%}

{In this section, we attempt to extend our analysis to higher $z$-heights. We know that, in the plane, the data can best be explained by including a thermal component with no non-thermal curvature (Sect.~\ref{sec:initial_spectrum}). However, from Sect.~\ref{sec:vertical_feature} we found that, by a projected height of 2.2 kpc, the vertical feature could be explained either by a high thermal fraction of $\sim\,65$\% or by no thermal fraction and a curved non-thermal spectral index $\beta\,\sim\,1$. Our data cannot distinguish between these possibilities for high $z$, but as a test and for consistency with our in-plane analysis, we continue with our analysis to high $z$ allowing for a thermal fraction.}

To improve the S/N, we have taken vertical slices through the galaxy and averaged the specific intensities over a broader region.  The galaxy was first rotated so that the major axis is horizontal. {We then used  the routine \texttt{BoxModels} that is implemented \texttt{NOD3} \citep{nod3}\footnote{\url{https://gitlab.mpifr-bonn.mpg.de/peter/NOD3}}. For the box integration, we use a similar setup, as motivated in \citet{ste23}. To account for potentially different feedback processes in the central area of NGC~3044 and its outskirts, we divide the galaxy into three broad strips, each of width, $61\arcsec$, to cover the entire galaxy disk diameter of $3\farcm 07$ (Table~\ref{tab:N3044params}). As box-height, we choose the FWHM of the beam ($12\arcsec$) to properly sample the intensity profile of the galaxy. The reported $z$-heights were corrected for the galaxy's inclination of $85\degr$ (Table~\ref{tab:N3044params}).}

\begin{figure*}[hbt!]
    \centering    
    \includegraphics[width=7truein,height=4.0truein]{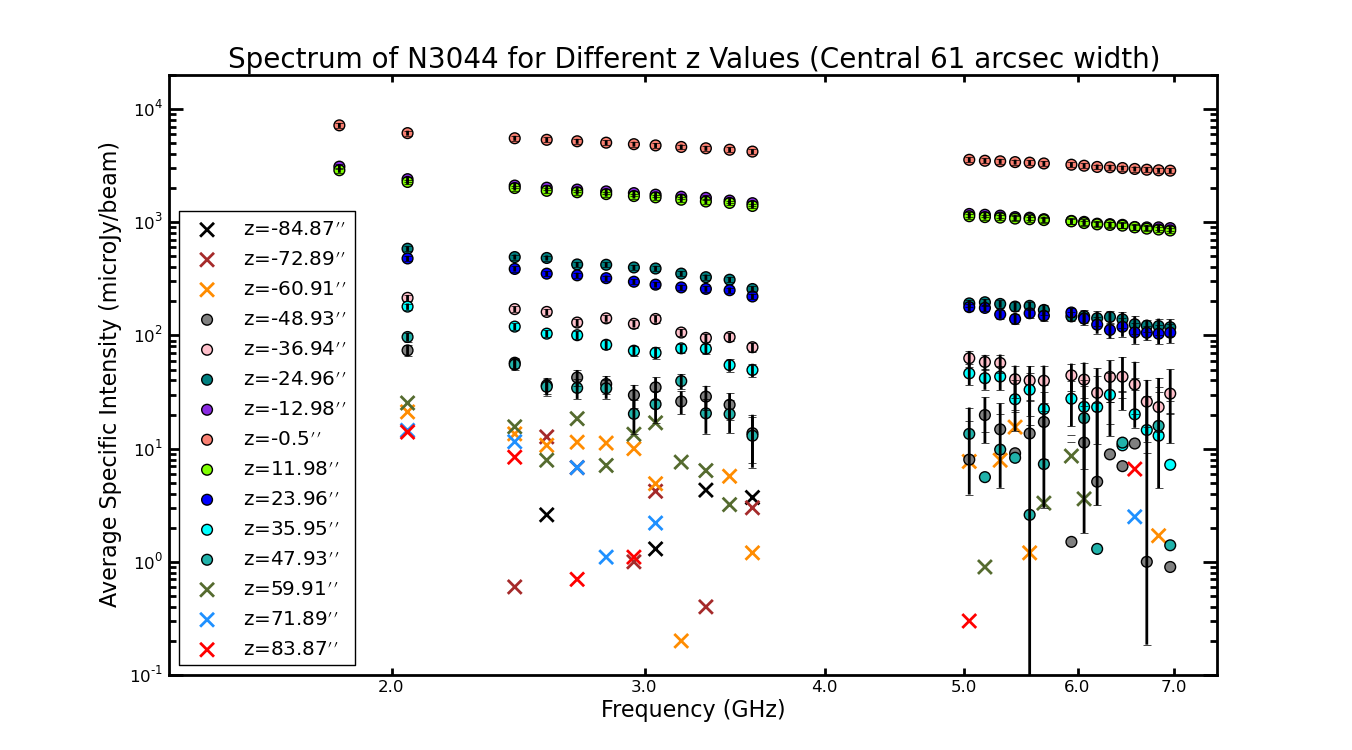}    
    \caption{Spectra over 27 frequencies for the box integrated slices from -84 arcsec (south) to +85 arcsec (north) in the central slice of width, 61 arcsec.
 Circles with error bars show the $z$-heights that were modelled and the unmodelled points are shown as X's without error bars. }
    \label{fig:box_int_spectra}
\end{figure*}

We calculate the uncertainty for the mean flux in each box by including the background noise (measured on empty regions in each narrow-band spectral map) corrected for the number of beams in each box, as well as a 2\% calibration error. %(cf. Sect.~\ref{sec:initial_spectrum}).
{In the following analysis, we focus on the box integration results from the central strip only, to further investigate the possible thermal outflow in the northern galaxy halo. }
In Figure~\ref{fig:box_int_spectra} we plot the spectra of the averages for each $z$-height from $-84\farcs 87$ to $+83\farcs 87$ through the central slice. As the height increases, the average values decrease, the error bars increase, and many points become negative. Heights for which thermal/nonthermal modelling was successful are shown as filled circles with error bars  and those for which modelling was unsuccessful are shown as X's. For most modelled heights, the error bars are smaller than the plot markers.

The modelling was executed in the same way as outlined in Sect.~\ref{sec:initial_spectrum} and Sect.~\ref{sec:cube_fits}, using Eq.~\eqref{eqn:fittedeqn}. All parameters
%($I_{\rm NT}, ~I_{\rm TH}, \alpha_{\rm NT},~ {\rm and}~ I_{\rm TH}/I_{\rm tot}$) 
were determined as before but, because of our interest in the outflow, we show only the thermal fraction, 
$I_{\rm TH}/I_{\rm tot}$ in Figure~\ref{fig:box_int_thermalfrac}.  The other modelled results for the box averages can be found in Appendix~\ref{app:zslice}. Altogether, 9 $z$-heights were successfully modelled. 
However, the flanking points, at $z\,=\,-49\arcsec$ and $z\,=\,+48\arcsec$ are shown in black because several points in the spectra became negative and had to be omitted.

Reliable results from $z\,=\,-37\arcsec$ to $z\,=\,+36\arcsec$ suggest that the thermal fraction shows a continuing trend to high values on the north side of the galaxy, consistent with Figure~\ref{fig:ITH_fraction_result}. These averages extend the thermal/nonthermal separation another $15\arcsec$ into the halo beyond the northern peak pointed out in  
Sect.~\ref{sec:vertical_feature} and visible in Figure~\ref{fig:ITH_fraction_result}. That is, a high thermal fraction can be traced to $z\,\sim\,3.5$ kpc. {If the thermal fraction does {\it not} increase with $z$ height, as suggested by this analysis, then we require an increasingly positive value of $\beta$ with height in order to flatten the spectrum at high frequency. }

\begin{figure*}[hbt!]
    \centering    
    \includegraphics[width=7truein,height=4.0truein]{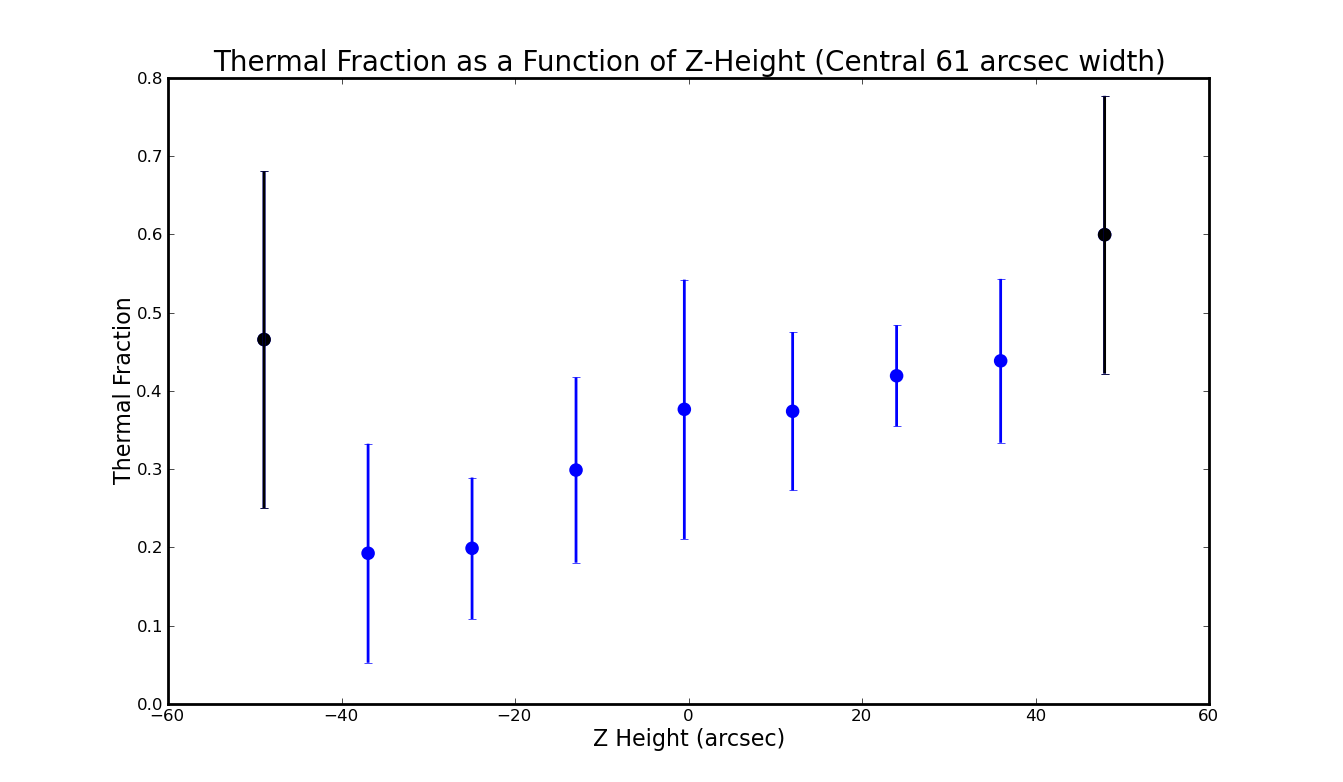}    
    \caption{Thermal fraction as a function of $z$-height from spectra shown in Figure~\ref{fig:box_int_spectra}. Positive is north of the (rotated) major axis and negative is south. The two black points that flank the ends were modelled with fewer frequency points and are less reliable. }
    \label{fig:box_int_thermalfrac}
\end{figure*}

\subsection{The L-Band Change in Spectral Slope}

A surprising result in NGC~3044 in Figures~\ref{fig:spectra_all} and \ref{fig:spectra_27points} is a flattening %(top plots in the figures) or turnover (bottom plots in the figures) 
in the spectrum at L-band ($\nu\,\sim\,1.5$ GHz). At these frequencies, the emission is largely nonthermal. {Both galaxies (Figures~\ref{fig:spectra_27points} and \ref{fig:spectra_27points_N5775}) show a change in slope, though it is more dramatic for NGC~3044.}
 %{\bf L-band data were combined from three different arrays (B, C, and D arrays) which had been brought to the same weighting scheme. In the event that this combination might have inadvertently caused the turnover, we separated each array and repeated the imaging.  results... We also used the combined arrays and re-measured the fluxes over all flux in the galaxy, rather than just the flux enclosed in the grey contour of Figure~\ref{fig:N3044_fig1}.xxx}
Preliminary results for the galaxy, NGC~2683, also in the CHANG-ES sample, show a similar spectral flattening in L-band. 
 
 \label{sec:turnover}
 The spectral flattening does {\it not} continue to lower frequencies for NGC~3044. The LOFAR Multifrequency Snapshot Survey (MSSS) reports a global flux density at 150 MHz of 0.36 $\pm$ 0.13 Jy \citep{chy18}, while the global flux in L-band at 1.575 GHz is 
 104.2 $\pm$ 2.1 mJy  \citep{wie15}.  
 A simple spectral index between these two points is 
 $\alpha_{0.150-1.575~{\rm GHz}}\,=\,-0.53\,\pm\,0.15$.
 Consequently, at frequencies lower than L-band, the spectrum shows the usual rise expected for synchrotron emission.  { %In fact, this low frequency spectral index actually marginally agrees with the weighted spectral index, $\overline{\alpha_w}\,=\,-0.682\,\pm\,0.016$, that we find by putting all wide-band data together (L, S, and C-bands, Table~\ref{tab:widebandresults}).
 It is only by breaking up the wide band into narrow channels that the spectral shape is delineated. The nonthermal spectral index is seen to be steeper above L-band once the thermal component is taken into account (Tables~\ref{tab:spectral_fits_27points} and \ref{tab:spectral_fits_27points_N5775}).}
 %https://astro.subhashbose.com/tools/error-propagation-calculator
 %The global spectral index between these two frequencies is $\alpha_{low}\,=\,-0.53\,\pm\,0.16$, which agrees with
 %$\alpha_{low}$\,=\,-0.56 $\pm\,0.01$. barely..

% In this section, we consider the possible reasons.   
%The global spectral index between 1.3 GHz and 5 GHz is  $\alpha_{high}$\,=\,-0.73 $\pm\,0.03$.

{This change of slope at L-band appears to be more universal than in the few galaxies that we have examined.} 
Recent work by \citet{chy18}, who looked at the  behaviour of spectra from 150 MHz to 5 GHz in more than 100 galaxies (including NGC~3044) also find a change of slope at the L-band when plotting the global flux (i.e. the unresolved flux) of these galaxies.  Like our results for NGC~3044, they
find a change in slope from flatter at low frequencies (50 MHz to 1.5 GHz, referred to as $\alpha_{low}$) to steeper at high frequencies (1.3 GHz to 5 GHz, called $\alpha_{high}$).  For NGC~3044, they find
$\alpha_{low}\,=\,-0.56\,\pm\, 0.01$, which agrees with $\alpha_{0.150-1.575~{GHz}}$ listed above,  and
$\alpha_{high}\,=\,-0.73\,\pm\, 0.03$ which compares well with 
 our wide-band value of $\overline{\alpha_w}\,=\,-0.682\,\pm\,0.016$ over a slightly modified frequency range of 1.25 to 7.02 GHz (Table~\ref{tab:widebandresults}). 
 Apparently, the change of slope in L-band is a `universal' phenomenon, applying to the total flux of many galaxies.  
%For NGC~3044, we have now also shown that the change in slope can be seen in a spatially resolved case.

{The reason for this spectral break is not yet clear. } 
%We can immediately rule out synchrotron self-absorption (SSE) as the cause of the L-band flattening/turnover because this effect is seen globally in the galaxy whereas SSE is only observed in very compact regions. For an example of SSE in the compact core of a CHANG-ES galaxy, see \citet{irw15}.  SSE should also continue to depress the emission at even lower frequencies, which is not observed.
Low frequency thermal absorption is a possibility. 
 A previous attempt at explaining a flattening in the spectrum of M~82 below $\sim$ 1 GHz, followed by an increase at lower frequencies (similar to NGC~3044) can be found in \citet{lac13}.  He assumed that absorbing HII regions form a kind of `pudding' within a spatially unresolved starbursting region.  For NGC~3044, we will take a simpler `feasibility' approach of adopting a thermal foreground screen that produces absorption of background synchrotron emission. 
 The optical depth of a thermal component, expressed in the radio regime, is 
 \begin{equation}
\tau_\nu({\rm TH})\,=\,8.24\,\times\,10^{-2}\,\left[{\rm \frac{T_e}{K}} \right]^{-1.35}
\left[{\frac{\nu}{\rm GHz}} \right]^{-2.1}\,\left[\frac{\cal EM}{{\rm cm}^{-6}\,{\rm pc}}\right]\label{eqn:tau}
 \end{equation}
where $\rm T_e$ is the electron temperature, and $\cal EM$ is the emission measure given by Eq.~\eqref{eqn:EM},
 with $n_{\rm e}$, the electron density, and $l$ the line of sight distance. We see that the optical depth should increase with decreasing frequency and potentially produce a flattening or turnover at low frequencies.

We can test for this possibility because we have measured values of $I_{\rm TH}$, unaffected by dust, from our fitting. At the center of NGC~3044, 
%$I_{\rm TH}\,=\,4.27$ 
$I_{\rm TH}\,=\,2.19$
mJy beam$^{-1}$ for a 12 arcsec beam at 4.13 GHz (Table~\ref{tab:spectral_fits_27points}), corresponding to a brightness temperature of 
%${\rm T_B}\,=\,3.07$ K. 
${\rm T_B}\,=\,1.57$ K.
This is firmly in the optically thin limit, in which case ${\rm T_B}\,=\,\tau_\nu\,{\rm T_e}$ and, for ${\rm T_e}\,=\,10^4$ K, we find $\tau_{4.13~{\rm GHz}}\,=\,1.57\,\times\,10^{-4}$, and hence 
%${\cal EM}\,=\,1.8\,\times\,10^4$ pc cm$^{-6}$ 
${\cal EM}\,=\,9.5\,\times\,10^3$ pc cm$^{-6}$
by Eq.~\eqref{eqn:tau}.  Using this ${\cal EM}$, the optical depth at $\nu\,=\,1.5$ GHz is then 
%\tau_{\nu=1.5~{\rm GHz}}\,=\,2.5\,\times\,10^{-3}$.
$\tau_{\nu=1.5~{\rm GHz}}\,=\,1.3\,\times\,10^{-3}$.
In our simple screen geometry, a turnover of the largely nonthermal spectrum at $\nu\,\sim\,1.5$ GHz, due to thermal absorption is 
\begin{equation}
I_\nu~ {\rm (absorbed)} \,=\, I_{\rm NT}\,e^{\tau_\nu({\rm TH})}.
\label{eqn:thermabs}
\end{equation}  Eq.~\eqref{eqn:thermabs} with such a low optical depth indicates that a depression in the spectrum at 1.5 GHz due to thermal absorption should be less than one \%.
%no more than 0.25\%.  
%Even accounting for a 36\% uncertainty in $I_{\rm TH}$, 
This is insufficient to produce the observed flattening of the spectrum. \citet{chy18} come to the same conclusion for their large sample of galaxies. They also find that highly inclined galaxies do not tend to have flattened spectra compared to more face-on galaxies, which would be expected if thermal absorption is important.

\cite{chy18} also present a number of possibilities that could account for the change of slope, including cosmic ray propagation and energy losses, synchrotron and inverse Compton losses accompanied by advective transport in a wind, or a recent cutoff in cosmic ray (CR) propagation. %Detailed modelling is likely required to sort out the relative importance of these various processes.

{We speculate that a change in slope at L-band may, in fact, be an even broader-scale phenomenon than in spiral galaxies alone.}  
An interesting comparison is with galaxy clusters \citep[see][for a review]{van19} in which steep spectrum ($\alpha_{NT}\,\ltabouteq\,-1$) sources appear to be associated with the intra-cluster medium (ICM) rather than with individual galaxies  \citep[e.g.][]{bru14}.  `Spectral breaks' around 1 to 1.5 GHz are clearly observed in Coma and other galaxy clusters, with flatter slopes at lower frequencies and steeper slopes at higher frequencies
 \citep{cas12, raj21}.

%What is producing the apparently universal  change of slope in L-band?  
As for clusters of galaxies, although radio structures in clusters are not fully understood, an energy cutoff has been proposed for some features, implying
 a maximum CR energy of a few GeV \citep{van19}.  An alternate proposal for galaxy clusters is the acceleration of cosmic ray electrons by magnetohydrodynamic (MHD) turbulence \citep{bru16,zuh13,don18}.  This idea is not new. 
 For example, \citet{eil84} showed that turbulence can accelerate particles sufficiently to produce synchrotron emission in the interstellar medium with spectral indices in the range, $-1.0 \,\ltabouteq\,\alpha\,\ltabouteq\,-0.5$. Moreover, \citet{hen82} showed that there should be a spectral break at a frequency for which the synchrotron lifetime is about equal to the turbulent `turnover time'.  
 %Acceleration and loss rates are then equal which should occur when the electron Lorentz factor is $\gamma_{\rm e}\,\sim\,2\,\times\,10^4$ \citep{eil84}. The frequency is then (cgs units)
 %\begin{equation}
 %    \nu\,=\,\frac{\gamma_{\rm e}^2\,{\rm e}\,B}{2\,\pi\,{\rm m_e}\,c}
 %\end{equation}
 %where e is the electronic charge, $B$ is the magnetic field strength, $\rm m_e$ is the electron mass and $c$ is the speed of light.  The magnetic field strength in NGC~3044 is $B\,=\,14~\upmu$G, so a predicted turnover due to turbulence is then
  %which is a low efficiency process and more likely to lead to steeper spectral indices than in typical galaxies. In galaxies like NGC~3044, a variety of cosmic ray sources are available but turbulent acceleration could be a contributor. 
  Turbulence is, in fact, {\it required} in order to produce the global magnetic dynamo in NGC~3044 and other CHANG-ES galaxies in the models of \citep[e.g.][]{hen21}.   Detailed modeling is likely required to sort out the relative importance of these various processes.

\clearpage
 
\section{Conclusions}\label{sec:conclusions}

{For two galaxies, NGC~3044 and NGC~5775}, we have combined VLA L-band, S-band, and C-band observations whose frequency coverage is almost contiguous from 1.25 to 7.02 GHz. For NGC~3044, a single map and corresponding spectral index map were initially made from all data by fitting a simple spectral index, $I_\nu\,=\,I_{\nu_0}\,\nu^\alpha$ (Figure~\ref{fig:widebandmap}). The average spectral index, weighted by intensity, is $\alpha_w\,=\,-0.682\,\pm\,0.016$.

The goal of this paper was to carry out  a spatially resolved thermal/nonthermal separation {in these galaxies} using radio data only. We have shown that this is indeed possible, provided the calibration and signal-to-noise are sufficient, and (as much as possible) the data are contiguous  over a broad radio band.  S-band, newly added to the CHANG-ES program, is crucial. This approach eliminates the need for separate H$\alpha$ and IR observations which are currently the standard method of obtaining information on the thermal component. It also eliminates uncertainties related to dust, which can be especially problematic in an edge-on galaxy. {The penetrating power of radio emission allows for a view clear through the galaxy that is not possible in other standard wavebands.}

In our initial examples, the broad frequency band was broken up into narrow channels of width, $\Delta\,\nu\,=\,0.13$ GHz within which maps were made with a common (12 arcsec) spatial resolution. The spectrum was then plotted for the central point (longest line of sight in an edge-on galaxy) as well as for a larger region corresponding to 33 beams (Figure~\ref{fig:spectra_all} or \ref{fig:spectra_27points}) {for NGC3044, and 47 beams (Figure~\ref{fig:spectra_27points_N5775}) NGC~5775.} 

{There is a change in slope for both galaxies in L-band, such that the spectrum flattens compared with the higher frequencies.  This flattening is more dramatic for NGC~3044 and is not well fit by functions of the form given in Eq.~\ref{eqn:sum3}. Therefore, we have excluded low frequency points from the fitting, where the change of slope occurs.}
%The spectrum flattens (central point) or turns over (33 beam region) in L-band ($\nu\,\sim\,1.5$ GHz).  
%cannot be explained by thermal absorption. Such a change in slope appears to be a common phenomenon in star-forming galaxies. %and even in some radio structures in clusters of galaxies (Sect.~\ref{sec:vertical_feature}).  

%Excluding the first 3 L-band points, each spectrum in the data cube  
%These data were fit with a variety of functions both all frequency points, as well as excluding the low frequency points.  
 {For NGC~3044, we cannot distinguish between fits that do or do not include curvature in the nonthermal spectrum. The simpler function described Eq.~\ref{eqn:fittedeqn}, i.e. a single nonthermal slope plus a thermal component, was therefore fitted, point-by-point in a cube of data. For NGC~5775, the best result had the form, Eq.~\ref{eqn:fittedeqnbeta}, i.e. a nonthermal slope with curvature plus a thermal component; this fit was carried out for the cube-fitting.
}
%That is, the spectrum consists of nonthermal plus thermal emission, but the nonthermal emission shows no detectable curvature. 

The resulting images are shown in Figures~\ref{fig:INT_result} to \ref{fig:ITH_fraction_result} {for NGC~3044  and in Figures~\ref{fig:INT_result_N5775} to \ref{fig:ITHfrac_result_N5775} for NGC~5775} for a frequency of $\nu_0\,=\,4.13$ GHz.
For NGC~3044, smoothed to 15 arcsec resolution (Table~\ref{tab:imaging_results_15arcsec}) we find a median thermal fraction of 13\% with an estimated uncertainty of 50\% {on this fraction} over a 26 beam region. The thermal fraction reaches up to 65\% in some regions. These results are comparable to the H$\alpha$ mixture method result. At our highest common resolution of 12 arcsec, the median thermal fraction is 25\% with an estimated uncertainty of 50\% {on this value} in measured regions, but this reduces to $\sim 10$ \% when extrapolated to a region that is comparable to the nonthermal emisson
(Table~\ref{tab:imaging_results}). 
{For the higher SFR galaxy, NGC~5775 over a 44-beam region at 12 arcsec resolution, we find a median thermal fraction of 44\% with an estimated uncertainty on this value of 17\% (Table~\ref{tab:imaging_results_N5775})}.

The thermal structure of NGC~3044 found from the radio-only method matches the H$\alpha$ map is most regions except for a  vertical region at the center (e.g. Figure~\ref{fig:Vargas_radio_comparison}). This is a new result, likely revealed because of the ability of the radio data to penetrate through the dust of an edge-on galaxy. The vertical feature 
%This is higher than the estimated 20\% using H$\alpha$ data \citep{var18} but the latter approach cannot account for uncertainties related to some assumptions regarding dust obscuration in an edge-on galaxy.
 is suggestive of outflow similar to M~82.   There is a corresponding H$\alpha$ loop on the northern side of the vertical feature (Sect.~\ref{sec:vertical_feature}) but no other related features have yet been observed. The thermal/nonthermal modelling using Eq.~\ref{eqn:fittedeqn} was extended into the halo regions by taking averages at different $z$-heights over a swath of width, $\sim\,1$ arcmin. The result indicates that the high thermal fraction continues into the northern halo to at least $z\,\sim\,3.5$ kpc (Sect.~\ref{sec:vertical}), { {\it provided} the nonthermal spectrum has no curvature as $z$ increases away  from the disk. For NGC~5775, the radio-only thermal map matches the H$\alpha$ map extremely well (see Figure~\ref{fig:Halpha_ITH_N5775}).}

 This method of radio-only thermal/nonthermal separation has the potential to reveal new features in galaxies that are obscured by dust, especially in the {\it disks} of edge-on star-forming galaxies.  {Mapping the separated emission enables an analysis of local changes in cosmic ray transport and star forming regions.  } The recommended frequency range is from approximately 1.3 to 8 GHz with fully contiguous data, if possible. NGC~3044, with a flux density of $S_{4.12 GHz}\,=\,53.4 \pm\,1.2$, is a minimal case in the sense that galaxies of this order or higher total flux density are recommended targets, ensuring there is no lost flux because of large angular size and the calibration is of high quality. {A future improvement to the code would be to include physically motivated functions for the nonthermal emission.} The CASA-compatible code used in the paper for cube-fitting is available at {\tt https://projects.canfar.net/changes}.

\section*{Acknowledgements}{The first author wishes to thank the Natural Sciences and Engineering Research Council of Canada for  Discovery Grant \# RGPIN-2021-02456. 
We thank Marie-Lou Gendron-Marsolais for assistance with Python scripting and also Ren{\'e} Walterbos for assistance with the H$\alpha$ image of NGC~5775. Thanks also to R. N. Henriksen for useful discussions.
This research has used the Karl G. Jansky Very Large Array operated by the National Radio Astronomy Observatory (NRAO). NRAO is a facility of the National Science Foundation operated under a cooperative agreement by Associated Universities, Inc. M.S. and R.-J. D. acknowledge funding from the German Science Foundation DFG, within the Collaborative Research Center SFB1491 “Cosmic Interacting Matters – From Source to Signal. TW acknowledges financial support from the grant CEX2021-001131-S funded by MICIU/AEI/ 10.13039/501100011033, from the coordination of the participation in SKA-SPAIN, funded by the Ministry of Science, Innovation and Universities (MICIU).
}
%%%%%%%%%%%%%%%%%%%%%%%%%%%%%%%%%%%%%%%%%%%%%%%%%%

%%%%%%%%%%%%%%%%%%%%%%%%%%%%%%%%%%%%%%%%%%%%%%%%%%
\section*{Data Availability}
Maps in FITS format are available for L-band and C-band data as well as H$\alpha$ maps are available at
{\tt https://projects.canfar.net/changes}. 


\vspace{5mm}
\facilities{VLA
}


\software{{\tt astropy} \citep{ast13}, CASA \citep{mcm07}.
The code, {\tt fitcurves\_CUBE.py}, discussed in Sect.~\ref{sec:cube_fits}, is available at {\tt https://projects.canfar.net/changes/publications}.
}



\clearpage
\appendix

\section{Full-Band Imaging {of NGC~3044}}\label{app:fullband}
\renewcommand{\thetable}{\thesection.\arabic{table}}
\setcounter{table}{0}

Table~\ref{tab:tclean_params_wide} lists {\tt tclean} inputs that were used to make the wide-band map shown in Figure~\ref{fig:widebandmap}. Parameters that are not listed were left at their default values. The input data are listed in Table~\ref{tab:observing}. Note that there are two small frequency gaps in the observations, one within the L-band and one between the S-band and C-band.

A variety of input parameters were tried in an attempt to find the best map. This was determined by minimizing the map noise which includes thermal noise plus residual sidelobes from the cleaning process.  An important input is the number of Taylor terms that is being fit ({\tt nterms}). With our best result using {\tt nterms} = 2, this effectively puts a straight line through the data in log space {(i.e. $log I_\nu\,=\,\alpha\,log\nu$ plus a constant}).  We show in Sect.~\ref{sec:initial_spectrum} that there is, indeed, curvature in the data in which case the rms noise in a map that fits a straight line spectrum will be larger than if there were no curvature.  A spectral index error map is an output of {\tt tclean} and is calculated following \citet{rau11} (their Eq. 39).

The correction for the primary beam (PB) was carried out using the CASA task, {\tt widebandpbcor}.  This task takes into account the fact that the PB varies with frequency. The spectral index map is also corrected for the PB frequency dependence at this time. For the spectral index, a 5
$\sigma$ cutoff is used, where $\sigma$ represents the measured rms noise in the map prior to PB-correction. The spectral index error map is also corrected for the PB at this time. The resulting maps are shown in Figure~\ref{fig:widebandmap}.

\begin{table}
%\begin{center}
\caption{{\tt tclean} Inputs for the Wide-Band Map}\label{tab:tclean_params_wide}
\begin{center}
    \begin{tabular}{l | c}
    \hline
    Parameter & Value \\
    \hline
 imsize$^{\rm a}$  &  2700\\
 cells (arcsec)$^{\rm b}$ & 1.5 arcsec\\
 wprojplanes$^{\rm c}$ & 128 \\
 scales$^{\rm d}$ & [0,10,20]\\
 weighting$^{\rm e}$ & briggs, robust = 2\\
 gridder$^{\rm f}$ & widefield \\
 deconvolver$^{\rm g}$ & mtmfs \\
 threshold (mJy beam$^{-1}$)$^{\rm h}$ & 0.020 \\
 spw$^{\rm i}$ & all\\
 boxes$^{\rm j}$ & multiple regions specified\\
 nterms$^{\rm k}$ & 2 \\
 \hline
    \end{tabular}
\end{center}
$^{\rm a}$ Number of pixels per side of the image.\\
$^{\rm b}$ Pixel (cell) size.\\
$^{\rm c}$ Number of W-values in computing gridding convolution functions for W-projection.\\
$^{\rm d}$ Spatial scales used in deconvolution (in number of pixels), as described in \citet{rau11}.\\
$^{\rm e}$ Weighting scheme, from \citet{bri95} as implemented in CASA.\\
$^{\rm f}$ Gridding option for resampling visibilities onto a regular grid.\\
$^{\rm g}$ Deconvolution option, mtmfs, does a multi-term, multi-scale, multi-frequency synthesis, as described in \citet{rau11}.\\
$^{\rm h}$ Lower limit that determines when to stop the clean.\\
$^{\rm i}$ Spectral windows that are included.\\
$^{\rm j}$ Clean boxes within which cleaning is carried out. These are consistent for all arrays/frequencies.\\
$^{\rm k}$ Number of Taylor terms used. \\
\end{table}

\section{Narrow-Band Imaging {of NGC~3044}}\label{app:narrowband}
\renewcommand{\thetable}{\thesection.\arabic{table}}
\setcounter{table}{0}

As described in Sect.~\ref{sec:initial_spectrum}, a series of maps were produced, frequency-by-frequency, so that the initial spectrum could be examined.   The CASA algorithm, {\tt tclean}, was used to make and clean\footnote{Cleaning involves deconvolving the point spread function (PSF, also known as the 'dirty beam'), from the map (also known as the 'dirty map') and then reconvolving with a 'clean beam' which is a gaussian fit to the PSF without the sidelobes.} the maps. 

The process is challenging because the effective bandwidths of each map must be small enough that any curvature in the spectral index should be captured, yet the effective bandwidths must be wide enough that a reasonable signal-to-noise (S/N) results and {\tt tclean} does not diverge. Unlike Figure~\ref{fig:widebandmap}, these narrow frequency maps have a higher noise because of the restrictive narrow frequency range. 
Moreover, a wide field must be mapped because of the presence of a number of distant outlier sources. Various input parameters to the CASA {\rm tclean} algorithm were attempted in order to obtain convergence to reasonable results.  

The adopted parameters are given in Table~\ref{tab:tclean_params}.
The most important criterion for ensuring convergence was specifying boxes around all sources so that only real sources were cleaned and regions between sources were not cleaned. An alternative to specifying boxes over such large fields is to place outlier sources in their own fields {in which case the phase center is adjusted for the center of each field}; however, trials revealed that this did not improve the results. Without restricting the clean regions with boxes, the maps diverged in almost all cases.

The central frequencies of each map, frequency ranges, and spectral windows used are given in Table~\ref{tab:im_params}.    Several spws had to be omitted because they were either completely flagged during calibration because of RFI (spws 17 and 18 in S-band) or the results were poor, meaning an unrealistic drop off in flux and/or very high rms (spws 29 to 31  at the upper end of S-band). There are also two small frequency ranges that were not observed: a small gap in frequency at L-band from 1.5 to 1.65 GHz was omitted because of radio frequency interference (RFI), and a gap from 4 to 5 GHz between S-band and C-band which was not observed.  
The result was 30 maps over the entire frequency band from L-band, through S-band, and C-band.  {Once the data sets of Table~\ref{tab:observing} in a given band  were combined, several more attempts of self-calibration did not improve the maps (although some individual data sets had already been self-calibrated, Sect.~\ref{sec:radioobs}),
and neither did applying a uv-taper}.

We use {\tt nterms} = 1, meaning that no Taylor expansion is applied to each map (the spectrum is taken to be flat within each narrow band).  A region of steep spectral index, $\alpha\,=-1$, would result in a variation of less than 1\% at the lowest frequency and less than 5\% at the highest frequency, between the center of the band and its edge.

Because of the changing frequency, map-to-map, each map had a slightly different spatial resolution.  All maps were then smoothed to 12 arcsec which is just larger than the lowest resolution. 

Because the 
galaxy itself subtends only $\sim\,4$ 
arcmin (see Fig.~\ref{fig:N3044_fig1}), primary beam (PB) corrections are small (see also Appendix~\ref{app:calibration}).
For example, the PB full width at half maximum (FWHM) ranges from 32.3 arcmin at the lowest frequency (1.31 GHz) to 5.99 arcmin at the highest frequency (6.96 GHz) \citep{per16}. 
Nevertheless, we carry out the PB corrections in order to accurately measure correct fluxes. For this, we
used the CASA task, {\tt impbcor}, using
the PB that corresponds to center frequency of each map (Table~\ref{tab:im_params}). Note that, 
over one-half of the frequency width of any given map (0.128/2 GHz), the FWHM varies by only $\sim$ 1 to 2\% \citep{per16}. Consequently, the wide-band task, {\tt widebandpbcor} was not necessary for these narrow-band images.

The spectra shown in Sect.~\ref{sec:initial_spectrum} were made from the common spatial resolution, PB-corrected maps.

The rms noise was measured from the PB-corrected maps, so the rms increases with distance from the center.  Four boxes were adopted that were within a few arcmin of the galaxy emission for these rms measurements.  The same rms boxes were used for each map and are listed in Table~\ref{tab:im_params}.
However, the final error bars %shown in Figures~\ref{fig:spectra_all} and \ref{fig:spectra_27points} 
include an additional 2\% or 5\% calibration error (see Appendix~\ref{app:calibration}) which was added to the rms value in quadrature.  Note that the final error is dominated by the calibration error, not the rms.

\begin{table}
%\begin{center}
\caption{{\tt tclean} Inputs for Narrow-Band Maps}\label{tab:tclean_params}
\begin{center}
    \begin{tabular}{l | c}
    \hline
    Parameter & Value \\
    \hline
 imsize$^{\rm a}$  &  4000\\
 cells (arcsec)$^{\rm b}$ & 1.0 arcsec\\
 wprojplanes$^{\rm c}$ & 128 \\
 scales$^{\rm d}$ & [0,10,20]\\
 weighting$^{\rm e}$ & briggs, robust = 0\\
 gridder$^{\rm f}$ & widefield \\
 deconvolver$^{\rm g}$ & mtmfs \\
 threshold (mJy beam$^{-1}$)$^{\rm h}$ & 0.050 for L-band, 0.025 for S and C-bands \\
 spw$^{\rm i}$ & as in Table~\ref{tab:im_params}\\
 boxes$^{\rm j}$ & multiple regions specified\\
 nterms$^{\rm k}$ & 1 \\
 \hline
    \end{tabular}
\end{center}
$^{\rm a}$ Number of pixels per side of the image.\\
$^{\rm b}$ Pixel (cell) size.\\
$^{\rm c}$ Number of W-values in computing gridding convolution functions for W-projection.\\
$^{\rm d}$ Spatial scales used in deconvolution (in number of pixels), as described in \citet{rau11}.\\
$^{\rm e}$ Weighting scheme, from \citet{bri95} as implemented in CASA.\\
$^{\rm f}$ Gridding option for resampling visibilities onto a regular grid.\\
$^{\rm g}$ Deconvolution option, mtmfs, does a multi-term, multi-scale, multi-frequency synthesis, as described in \citet{rau11}.\\
$^{\rm h}$ Lower limit that determines when to stop the clean.\\
$^{\rm i}$ Spectral windows that are included.\\
$^{\rm j}$ Clean boxes within which cleaning is carried out.\\
$^{\rm k}$ Number of Taylor terms used. \\
\end{table}

\clearpage
{
\section{Calibration Errors}\label{app:calibration}
\renewcommand{\thetable}{\thesection.\arabic{table}}
\setcounter{table}{0}

As indicated in Sect.~\ref{sec:initial_spectrum}, calibration uncertainties dominate over rms variations that are normally measured in blank regions of a map. By `calibration uncertainties', we mean any uncertainties that are not reflected in rms map measurements. In this section, we probe more deeply into these uncertainties.

The primary calibrator, 3C~286, is well-known to have a stable flux at the frequencies used.
Over the time period of the NGC~3044 observations (Table~\ref{tab:observing}), the flux density calibrations were improved from the original `Perley-Taylor-99' scale\footnote{\tt http://www.vla.nrao.edu/astro/calib/manual/baars.html} to the `Perley-Butler-17' scale \citep{per17}. Our measurements of these two scales over the spw frequencies listed in Table~\ref{tab:im_params}, give an average difference of $(1.56\,\pm\,0.90)$\%, where the error represents the standard deviation, spw-to-spw. Bootstrap errors, that is, transference from the primary to the secondary calibrators, are also typically 1 to 2\%\footnote{\tt https://www.vla.nrao.edu/astro/calib/manual/boot.html}. Consequently, known errors related to the adopted calibrators are estimated to be $\sim\,2$\% \citep{wie15}.

Uncertainties related to the Primary Beam (PB) correction are minor over the enclosed 33-beam region that we have concentrated on for NGC~3044, as shown by the grey curve of Figure~\ref{fig:N3044_fig1}. This region spans 3 arcmin in diameter along the major axis (out to 1.5 arcmin from the pointing center) and 1.5 arcmin along the minor axis direction (0.75 arcmin from the pointing center).  The modelled region (e.g. Fig.~\ref{fig:INT_result}) has similar dimensions.
This emission is well within the full-width-half-maximum (FWHM) of the PB at all frequencies. For example, at the lowest frequency, $\nu\,=\,1.31$ GHz, the PB falls to only 0.5\% of its peak value by a radius of 1.5 arcmin. At the highest frequency, $\nu\,=\,6.96$ GHz, the PB falls to 15\% of the peak. Therefore, the actual correction for the PB is always less than 15\%. The
accuracy of the PB at all frequencies within the FWHM is known to be $\sim\,0.5$\%\footnote{https://library.nrao.edu/public/memos/evla/EVLAM\_195.pdf} so the accuracy within the more restricted region that we concentrated on, is better than this. For NGC~5775, the elliptical region shown in Figure~\ref{fig:N5775_fig10} is similar, with a semimajor axis of 1.6 arcmin. Consequently, our corrections for the PB do not introduce any worrisome error.

%HERE are refs to the two scales:
%Perley and Taylor (1999.2); plus Reynolds (1934-638; 7/94) Details can be found at http://www.vla.nrao.edu/astro/calib/manual/baars.html. wherein is found:
%valid between 300 MHz and 50 GHz
%Next is Perley-Butler
%https://ui.adsabs.harvard.edu/abs/2017ApJS..230....7P/abstract

Variations, spw to spw, depend on the accuracy of the bandpass calibration which uses the primary calibrator, 3C~286. In some cases, bad RFI or inaccuracies near the band edges were present.  For cases such as these, poor images resulted (cf. Appendix~\ref{app:narrowband}) and these were simply omitted from the fitting procedure.

Of the remaining data, flux density variations can occur due to the sensitivities of the arrays to various spatial scales.
For example, the largest angular scale (LAS) detectable is given in Table~\ref{tab:observing}.  The smallest LAS for any combined-array data is at C-band which is 4.0 arcmin.  Since the modelled regions for both NGC~3044 and NGC~5775 are smaller than this, the LAS is not expected to place any limits on the detectable flux.

There is, moreover, a way to estimate the final calibration error, using the observations themselves. This is because, for NGC~3044, we have observed with multiple arrays at different times and sometimes also have two set of independent observations using the same array and frequency (Table~\ref{tab:observing}). For example, independent observations at S-band C array, were taken approximately 1 month apart, as were observations at C-band C array.
This allows us to investigate the differences in flux between these different observing sessions. 

We therefore, remade all narrow-band maps, spw-by-spw, for each observing session independently. We then smooth to equivalent beam sizes and correct for the frequency-dependent primary beam. The results are given in Table~\ref{tab:calibration} in which we specify how different the peak values and flux densities (within the 33-beam region shown in Fig.~\ref{fig:N3044_fig1}) are, between the different observations or combinations of observations. %The results should incorporate, not only the calibration errors, as described above, but also the rms map error, between individual observations. 

The largest differences occur at L-band. 
For example, a comparison between B and C array results (16 arcsec resolution) shows $\approx$ 21\% different in flux density and 5\% difference in the peak value.  The difference in flux density, however, 
can be attributed to the grossly different spatial scales (cf. the LAS) that are detected in these two arrays. 
The comparison between C and D arrays (50 arcsec resolution) shows an $\approx\,15$\% difference.  Although the specified LAS is equivalent for these two arrays, the surface brightness sensitivity and image fidelity to extended structure is considerably inferior at C array compared with D array\footnote{\tt https://science.nrao.edu/facilities/vla/docs/manuals/oss/performance/resolution}. Consequently, the C array to D array difference may also be due mainly to differences between spatial scale sensitivity. Note also that a 50 arcsec beam is significantly larger than the 12 arcsec beam used in our final narrow-band images.
In S-band and C-band, the differences are smaller, of order a few percent or less.

This exercise has allowed us to understand the variations expected in flux density from observation to observation. The comparison takes into account calibration errors, as defined above and, since the measurements are directly on the maps, they also include rms map errors. However, the final images are made from the {\it combined} data sets, e.g. B array, C array and D array at L-band.  Consequently, the calibration error on the combined data will be smaller than any individual values in Table~\ref{tab:calibration}. Given the intrinsic calibration errors listed above as well as this comparison between individual data set, we
therefore estimate a final calibration uncertainty of 5\% for L-band and 2\% for S-band and C-band. 

}

\begin{table*}
\begin{center}
\caption{Calibration Errors via Comparison between Individual Observations of NGC~3044\label{tab:calibration}}
\begin{tabular}{lcccccc}
\hline
Band$^{\rm a}$ & & Beam size$^{\rm b}$ & LAS$^{\rm a}$
&\% Diff Flux Density$^{\rm c}$ & \% Diff Peak Value$^{\rm d}$ & \\ 
~~~~~ [Array] & & (arcsec) & (arcmin) &  & &\\
\hline\hline
L-band & &  &  & &   \\  
~~~~~[B] with [C] & &16 &2.0, 16.2 & $20.6\,\pm\,6.4$&  $4.9\,\pm\,3.0$&\\   
%~~~~~ [B+C]$^{\rm f}$ with [B]$^{\rm f}$& & 16 & & $20.2\,\pm\,5.7$&  $5.1\,\pm\,2.3$&\\   
%~~~~~ [B+C]$^{\rm f}$ with [C] & & 16 & & $0.71\,\pm\,0.38$& $1.1\,\pm\,1.6$& \\   
%~~~~~ [B+C+D] and [C] & &16 & &$8.8\,\pm\,4.1$ & $4.6\,\pm\,3.9$\\
%~~~~~ [B+C+D] and [B] & &16 & &$22.6\,\pm\,16.3$ & $7.3\,\pm\,7.3$\\
\multicolumn{7}{c}{\dotfill}\\
~~~~~ [C] with [D] & &50 &16.2 & $14.0\,\pm\,4.8$ & $15.7\,\pm\,6.0$ &\\
%~~~~~ [C+D]$^{\rm g}$ with [C]$^{\rm g}$& & 50 & 16.2& $10.6\,\pm\,3.7$ & $10.3\,\pm\,4.0$ &\\   
%~~~~~[C+D]$^{\rm g}$ with [D]& & 50 & 16.2& $4.0\,\pm\,1.4$& $5.2\,\pm\,1.9$& \\  
%~~~~~[B+C+D]$^{\rm g}$ and [D]& & 50 & 16.2& $5.2\,\pm\,2.3$& $6.7\,\pm\,2.7$& \\  
%~~~~~[B+C+D]$^{\rm g}$ and [C]& & 50 & 16.2& $8.9\,\pm\,3.6$& $8.8\,\pm\,3.7$& \\  
\hline
S-band &  &  & & &  \\
~~~~~[Ca] with [Cb] &  & 10& 8.2 & $0.99 \,\pm\,0.57$ & $0.37 \,\pm\, 0.27$& \\
\hline
C-band  &  &  &  & &  \\ 
~~~~~[Ca] with [Cb] & & 4.5 & 4.0 & $2.7\,\pm\,2.7$ & $0.94\,\pm\,0.63$&\\
\multicolumn{7}{c}{\dotfill}\\
~~~~~[Ca+Cb]$^{\rm h}$ with [D]$^{\rm h}$ & & 14 &  4.0&  $3.5\, \pm\,2.2$& $3.6\, \pm\,1.7$& \\
\hline
\end{tabular}
\end{center}
$^{\rm a}$ Bands and arrays are as in Table~\ref{tab:observing}. Indication of which data sets are being compared; for example, "[B+C] with [B]" means that the combined B and C array  images are being compared with B-array images.\\
$^{\rm b}$ Synthesized beam size for every spw of the given observing set. Spws are listed in Table~\ref{tab:im_params}. \\
$^{\rm c}$ Percentage difference of the flux density as measured within the region shown in Figure~\ref{fig:N3044_fig1} between the data set or sets of the first bracket and the data set of second bracket, averaged over the spws in those data sets. The error bar represents the standard deviation, spw-to-spw. All values are measured from the PB-corrected images.\\
$^{\rm d}$ Percentage difference of the peak specific intensity 
between the data set or sets of the first bracket and the data set of second bracket, averaged over the spws in those data sets.
The error bar represents the standard deviation, spw-to-spw. All values are measured from the PB-corrected images.\\
$^{\rm f}$ A 10 k$\lambda$ uv taper was used.\\
$^{\rm g}$ A 2 k$\lambda$ uv taper was used.\\
$^{\rm h}$ A common uvrange of 1 to 22.5  k$\lambda$ was used.\\
\\ 
%https://science.nrao.edu/facilities/vla/docs/manuals/oss/performance/resolution
\end{table*}
%judith used listobs to get these numbers
%https://science.nrao.edu/facilities/vla/docs/manuals/oss/referencemanual-all-pages
% 

\clearpage

\renewcommand\thefigure{\thesection.\arabic{figure}} 
\setcounter{figure}{0}
\setcounter{table}{0}

\section{Curve-Fitting Details}
\label{app:curve_fitting}
%\subsection{Fitting Methods}
%\label{sec:fitting}

{We will refer to the initial tests for the central point and larger flux regions (e.g. Sect.~\ref{sec:initial_spectrum} and Sect.~\ref{sec:narrowband_N5775}) as the `spectral-fits'; and we will refer to the point-by-point cube fitting (Sect.~\ref{sec:cube_fits} and Sect.~\ref{sec:cube_fitting_N5775}), as the `cube-fits'. Python codes were written to carry out fits for both of these cases.

For the fitting, the python code, {\tt curve\_fit} was imported from {\tt scipy\_optimize}. 
The results were robust to input guesses. For example, input guesses that deviated from the final result by more than a factor of 10 still converged to the same result. In all cases, each point was weighted by its error during the fitting process. {\tt Curve\_fit} allows for bounds to be placed on the parameters; for the spectral-fits, we restricted $I_{\rm NT}$ and  $I_{\rm TH}$ to be positive, but put no restrictions on the other parameters. This was also the case for the cube-fits of NGC~3044.  For the cube-fits of NGC~5775, since there was an additional free parameter ($\beta_{\rm NT}$, Eq.~\ref{eqn:fittedeqnbeta}), we added the restrictions: $-3.0\,\leq\,\alpha_{\rm NT}\,\leq\,1.0$ and $-5.0\,\leq\,\beta_{\rm NT}\,\leq\,5.0$. For the cube-fits of NGC~3044, we also applied a cut-off wherever the relative error in $I_{\rm TH}$, $I_{\rm NT}$ or $\frac{I_{\rm TH}}{I_{\rm TOT}}$ exceeded a value of 2.  This value was varied to investigate how much of the resulting map was blanked.
In practice, the errors increased the most when $I_{\rm TH}$ became very low.  Whether or not such blanking is applied, the error maps, which are shown for each modelled parameter 
(Figures~\ref{fig:INT_result} through \ref{fig:ITH_fraction_result}) can be used to assess the accuracy of the modelled result, position-by-position. In general, it is quite clear where the uncertainties become large.
For the cube-fits of NGC~5775, we opted to blank the output if $I_{\rm TH}$ became indistinguishable from zero, but not to do additional blanking where relative errors exceeded 2. This provided for a smoother output map and, again, the error maps can be examined to assess the accuracy of the modelled maps. Tabular results for these maps (Table~\ref{tab:imaging_results_N5775}) now include values with large uncertainties; however, tests showed that median values with and without the high relative error blanking differ from each other by $\ltabouteq~10$\%. 

The {\tt curve\_fit} algorithm allows for three different methods of doing the fitting: `trf' 
(Trust Region Reflective, the default), `dogbox', or `lm' (Levenberg-Marquardt). All three were tested.  For the spectral-fits, we found consistent results for all three.  However, note that lm does not accept bounds so the bounds were removed.  For the cube-fits, lm generally did not converge.  
Since we do have good boundaries, we used either trf or dogbox, whose results were statistically indistinguishable.
%https://docs.scipy.org/doc/scipy/reference/generated/scipy.optimize.least_squares.html#scipy.optimize.least_squares
For information on these algorithms, see
{\tt https://www.mathworks.com/help/optim/ug/equation-solving-algorithms.html}.
We also used the package, {\tt lmfit} (Non-Linear Least-Square Minimzation and Curve-Fitting for Python) \citep{new16}, to do several additional checks, again finding consistent results.}

\begin{figure*}
\centering\includegraphics[width=0.75\linewidth]{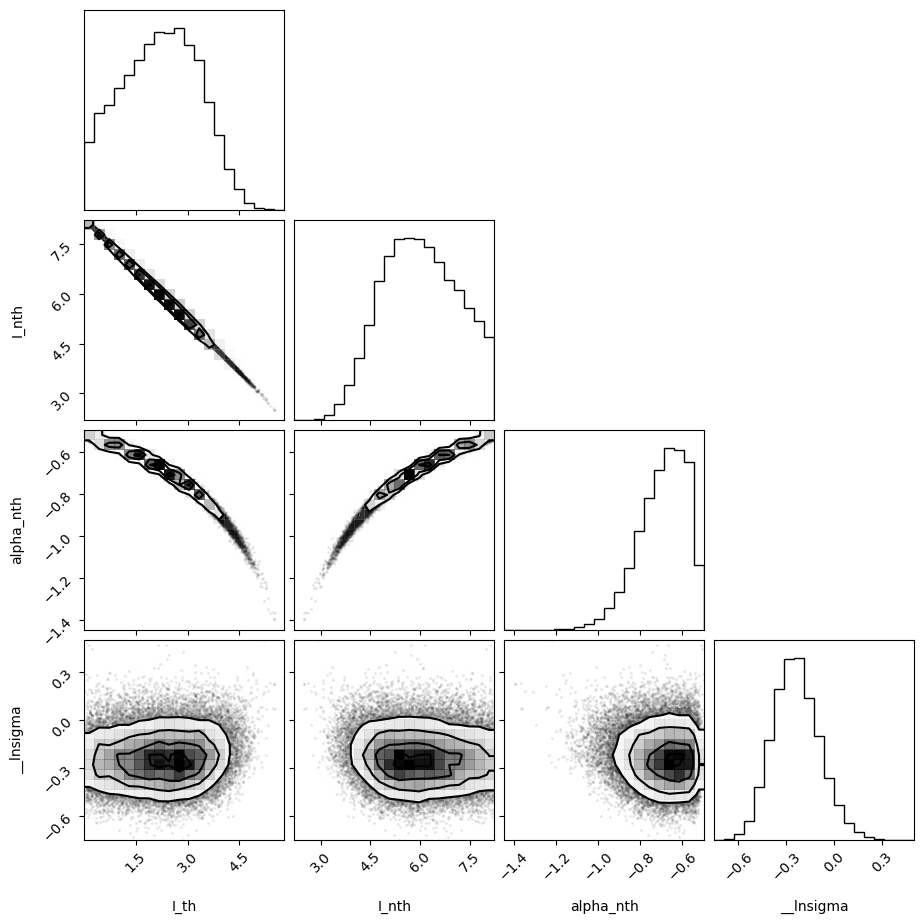}
    \caption{Corner Plot of the MCMC fitting routine}
    \label{fig:app_MCMC}
\end{figure*}

{{In addition to the aforementioned minimization algorithms that rely on local gradient information in the Jacobian to infer the covariance matrix for deriving the parameter uncertainties, we also test a random sampling approach (Markov Chain Monte Carlo Sampler, MCMC) to explore the parameter space. Here, we rely on the EMCEE implementation of \texttt{lmfit} (\url{https://lmfit.github.io/lmfit-py/index.html}). MCMC based fitting produces more robust uncertainty estimates, especially in multimodal distribution. However, it is computationally much more expensive. Therefore, to further test the reliability of the classic minimization algorithms, we refit the peak position spectrum of NGC~3044 (a `spectral fit') with the MCMC approach. The fitting results are presented in Table \ref{tab:app_MCMC} and Fig. \ref{fig:app_MCMC}. For the peak position spectrum, the MCMC fitting routine predicts even lower uncertainties than the classical fitting approach, however we do expect to find comparable uncertainties for low S/N regions. As noted before, running the MCMC is computationally much more expensive (factor $\sim 4000$), and due to our test results, we rely on the classical fitting routines, as described above, in this work.}}

\begin{table}
    \centering
    \caption{Comparison of fitting routines.}
    \label{tab:app_MCMC}
    \begin{tabular}{lrrr}
    \hline
         Minimization Approach & $I_{\mathrm{NT}}$ & $\alpha_{\mathrm{NT}}$ & 
         $I_{\mathrm{TH}}$   \\
            \hline\hline     
         lm & $6.1 \pm 1.5 (25.05\%)$ & $ -0.67 \pm 0.14 (20.69\%)$ & $2.2 \pm 1.5 (67.69\%)$\\
         MCMC & $5.99 \pm 1.23 (20.51\%)$ & $-0.68 \pm 0.12 (17.19\%)$ &  $ 2.25 \pm 1.20 (53.24\%)$\\
        \hline
    \end{tabular}
\end{table}

%\subsection{Correlations Between Parameters}
%\label{sec:correlations}
{The straightforward, classical approach, then, solves for the three free parameters, $I_{\rm NT}$, $\alpha_{\rm NT}$, and $I_{\rm TH}$, as well as the fourth,  $\beta_{\rm NT}$ from Eqns.~\ref{eqn:fittedeqn} or \ref{eqn:fittedeqnbeta}.  It is clear that $I_{\rm NT}$ and $I_{\rm TH}$ are {\it anti-correlated}; as the thermal fraction increases, the nonthermal fraction decreases.
This anti-correlation is implicit in these two fitted equations.  We also explored whether the observed data could be reproduced without a thermal component or with both a thermal component as well as nonthermal curvature (Tables~\ref{tab:spectral_fits_27points} and \ref{tab:spectral_fits_27points_N5775}). We know, by physical arguments, that the thermal component must be present, at least in the disk.  The only other issue is whether an additional curvature term ($\beta_{\rm NT} \ne 0$) is required. For NGC~3044, this could not be determined, but for NGC~5775, the fits clearly improved when a curvature term was included. }

\section{Spectral Fitting Results using all Narrow-band Frequencies {of NGC~3044}}
\label{app:spectra_allfreqs}

\renewcommand{\thetable}{\thesection.\arabic{table}}
\setcounter{table}{0}
\setcounter{figure}{0}

\begin{figure*}[hbt!]
\centering    \includegraphics[width=5.5truein,height=3.4truein]
{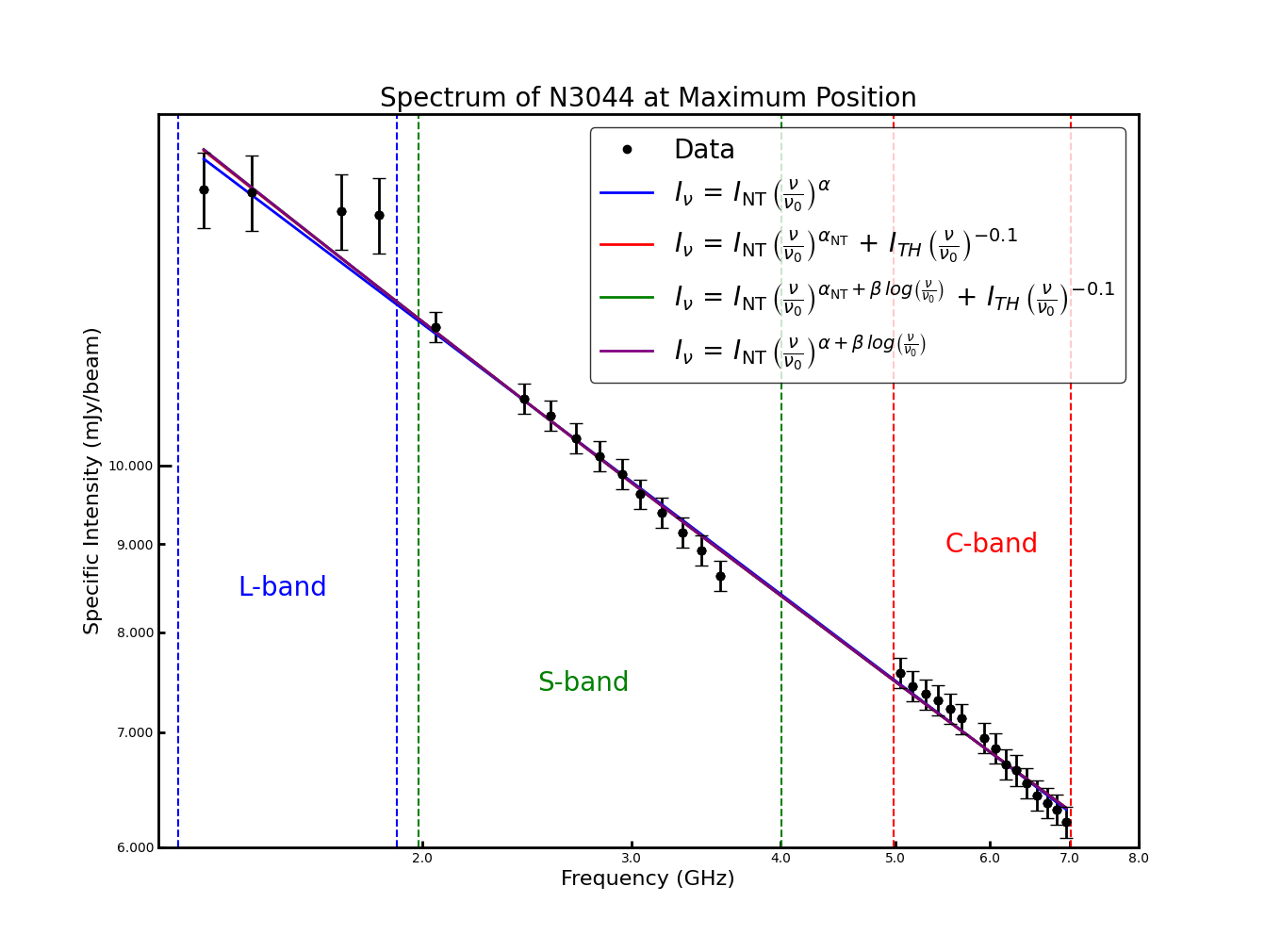}
\vskip -0.2truein
\includegraphics[width=5.5truein,height=3.4truein]{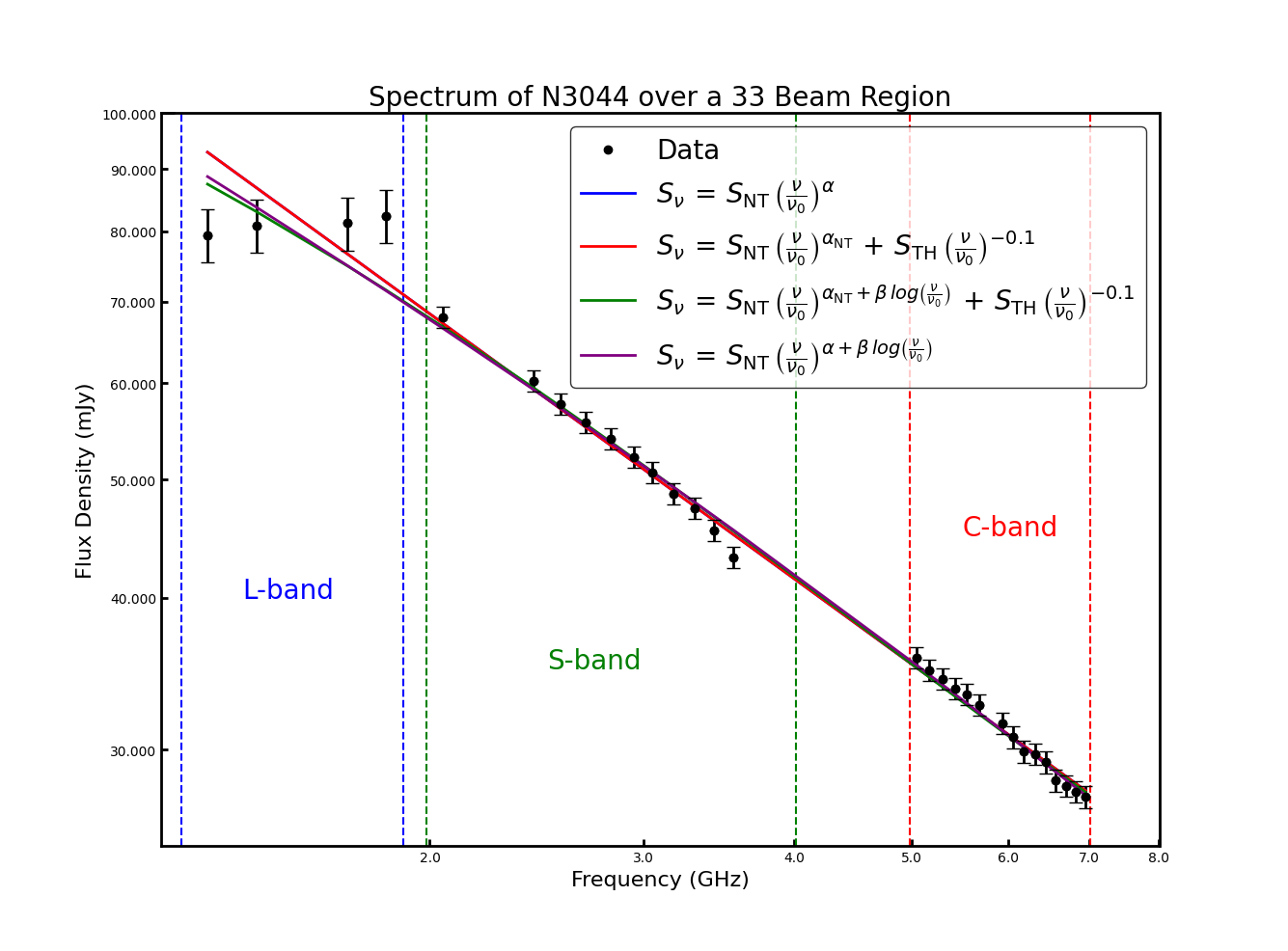}
\caption{Spectrum of NGC~3044 from narrow-band imaging using all frequency points.  Top: Values for the central maximum point only. Error bars add in quadrature: a 2\% or 5\% calibration error (Appendix~\ref{app:calibration}) and the rms from
Table~\ref{tab:im_params}. Bottom: flux density over a 33 beam region that is common to all maps and is encircled by the grey dashed curve in Figure~\ref{fig:N3044_fig1}. Error bars add in quadrature a 2\% or 5\% calibration error (Appendix~\ref{app:calibration}) with the rms from
Table~\ref{tab:im_params} multiplied by the square root of the number of beams. For each plot, the various frequency bands are marked, and the fitted equations are given in the legend.  Fitted parameters are given in Table~\ref{tab:spectral_fits_all}.  }
    \label{fig:spectra_all}
    \end{figure*}

Here we carry out spectral fitting of all 30 frequencies using the narrow-band maps, as described in Sect.~\ref{sec:initial_spectrum}. Figure~\ref{fig:spectra_all} shows the fits and Table~\ref{tab:spectral_fits_all} provides the corresponding numerical values.

\begin{table*}[ht]
\fontsize{7}{11}\selectfont
\begin{center}
\caption{Spectral Fits of NGC~3044 -- 12 arcsec resolution -- 30 points}\label{tab:spectral_fits_all}
    \begin{tabular}{lcccccc}
    \hline
    {\it Fitted Expression}$^{\rm a}$&&&&&&\\
    \hline
{\bf Maximum Position}$^{\rm b}$& $I_{\rm NT}$ $^{\rm c}$& $\alpha$ or $\alpha_{NT}$ $^{\rm c}$& $\beta$ or $\beta_{\rm NT}$ $^{\rm c}$&$I_{\rm TH}$$^{\rm c}$ & 
$\frac{I_{\rm TH}}
{I_{\rm tot}}$$^{\rm d}$ 
& rms scatter$^{\rm e}$
\\
 & (mJy beam$^{-1}$) & & & (mJy beam$^{-1}$) & & (mJy beam$^{-1}$)\\
\hline
\hline
{\bf A:} $I_\nu\,=\,I_{\rm NT}\,\left(\frac{\nu}{\nu_0} \right)^\alpha$&
$8.28\,\pm\,0.03 $& $-0.521\,\pm\,0.010$ &  & & & 0.34\\
{\bf B:} $I_\nu\,=\,I_{\rm NT}\,\left(\frac{\nu}{\nu_0} \right)^{\alpha_{\rm NT}}\,+\, I_{\rm TH}\,\left(\frac{\nu}{\nu_0} \right)^{-0.1}$ &
$7.20\,\pm\,1.96$ & $-0.582\,\pm\,0.126$& &$1.07\,\pm\,1.92$ & $0.13\,\pm\,0.21$ & 0.33\\
{\bf C:} $I_\nu\,=\,I_{\rm NT}\,\left(\frac{\nu}{\nu_0} \right)^{\alpha_{\rm NT}\,+\,\beta_{\rm NT}\,log(\frac{\nu}{\nu_0})}\,+\, I_{\rm TH}\,\left(\frac{\nu}{\nu_0} \right)^{-0.1}$ &
$8.26\,\pm\,66$ & $-0.52\,\pm\,3.37$ & $0.030\,\pm\,1.30$ &$0\,\pm\,66$ & $0\,\pm\,8$ & 0.33\\
{\bf D:} $I_\nu\,=\,I_{\rm NT}\,\left(\frac{\nu}{\nu_0} \right)^{\alpha\,+\,\beta\,log(\frac{\nu}{\nu_0})}$ & $8.26\,\pm\,0.05$&  $-0.520\,\pm\,0.010$
& $0.030\pm\,0.058$& & & 0.33\\
\hline
{\bf Region (33 beams)}$^{\rm b}$& $S_{\rm NT}$$^{\rm c}$& $\alpha$ or $\alpha_{NT}$ $^{\rm c}$& $\beta$ or $\beta_{\rm NT}$ $^{\rm c}$ &$S_{\rm TH}$$^{\rm c}$ & $\frac{S_{\rm TH}}{S_{\rm tot}}$$^{\rm d}$ & rms$^{\rm e}$
\\
& (mJy) & & & (mJy) & & (mJy) \\
\hline\hline
{\bf A:} $S_\nu\,=\,S_{\rm NT}\,\left(\frac{\nu}{\nu_0} \right)^\alpha$&
$40.50\,\pm\,0.16 $& $-0.723\,\pm\,0.009$ &  & & & 3.4\\
{\bf B:} $S_\nu\,=\,S_{\rm NT}\,\left(\frac{\nu}{\nu_0} \right)^{\alpha_{\rm NT}}\,+\, S_{\rm TH}\,\left(\frac{\nu}{\nu_0} \right)^{-0.1}$ &
$40.50\,\pm\,5.93$ & $-0.723\,\pm\,0.086$& &$0.00\,\pm\,5.74$ & $0.00\,\pm\,0.14$ & 3.3\\
{\bf C:} $S_\nu\,=\,S_{\rm NT}\,\left(\frac{\nu}{\nu_0} \right)^{\alpha_{\rm NT}\,+\,\beta_{\rm NT}\,log(\frac{\nu}{\nu_0})}\,+\, S_{\rm TH}\,\left(\frac{\nu}{\nu_0} \right)^{-0.1}$ &
$27.00\,\pm\,8.87$ & $-1.069\,\pm\,0.333$ & $-0.427\,\pm\,0.375$ &$13.68\,\pm\,8.66$ & $0.34\,\pm\,0.16$ & 2.8\\
{\bf D:} $S_\nu\,=\,S_{\rm NT}\,\left(\frac{\nu}{\nu_0} \right)^{\alpha\,+\,\beta\,log(\frac{\nu}{\nu_0})}$&
$40.85\,\pm\,0.24$& $-0.732\,\pm\,0.010$&
$-0.112\,\pm\,0.057$ &  & & 3.0\\
\hline
    \end{tabular}
\end{center}
$^{\rm a}$ The type of fit is given by the mathematical expressions shown.  For a single point (Maximum Position), the values are expressed as specific intensities, $I_\nu$, and are in mJy beam$^{-1}$. For a larger region (33 beams), the values are flux densities, $S_\nu$, in mJy. The beam is circular with a FWHM of 12 arcsec. Values are computed for the central frequency of $\nu_0\,=\,4.13$ GHz.\\
$^{\rm b}$ Position at which the fit was carried out. Maximum Position: center of the map at which the emission is a maximum.  Region (33 beams): Spatially integrated region within the areas enclosed by the grey curve in Figure~\ref{fig:N3044_fig1}. \\
$^{\rm c}$ Parameters of the fit with their standard deviations. \\
$^{\rm d}$ Thermal fraction at $\nu_0$.  The denominator is the sum of the thermal and nonthermal emission.\\
$^{\rm e}$ Root-mean-square scatter between the data and fitted curve.\\
\end{table*}

\clearpage

\section{Vertical Extension Results}\label{app:zslice}
\renewcommand{\thetable}{\Alph{section}\arabic{table}}
\setcounter{table}{0}

As indicated in Sect.~\ref{sec:vertical}, three vertical slices were made from the rotated galaxy covering $z$-heights from -84.87 to +83.87 arcsec.  
The output from the fitting using 
 Eq.~\eqref{eqn:fittedeqn} is given in Table~\ref{tab:z-heights}. Data are taken from averages within the boxes and
 results are only shown for fits that used all 27 frequencies.
  Strips are 61 arcsec wide and values are averages within this width with a height of 12 arcsec (the beam size). The reference frequency is $\nu_0\,=\,4.13$ GHz. The uncertainties can be significant in the weak halo region.

  %testing \curveshape here

\begin{table}[htb!]
%\begin{center}
\caption{Model Results for the three Vertical Strips\label{tab:z-heights}}
\begin{center}
    \begin{tabular}{lcccccc}
    \hline
    Strip Identifier & $z^{\rm a}$ & ${I_{\nu_{0_{\rm NT}}}}^{\rm b}$  & ${\alpha_{\rm NT}}^{\rm c}$ &${I_{\rm TH}}^{\rm d}$ & ${I_{\rm TH}/I_{\rm tot}}^{\rm e}$ & rms$^{\rm f}$\\
     & (arcsec) & ($\upmu$Jy beam$^{-1}$) & & ($\upmu$Jy beam$^{-1}$) & & ($\upmu$Jy beam$^{-1}$)\\
    \hline
 Left (Eastern) Slice  & $-12.98$ & $192\,\pm\,47$ & $-1.19\,\pm\,0.23$ & $19\,\pm\,45$ & $0.09\,\pm\,0.19$ &
12.2
\\
 Left (Eastern) Slice  & $-0.50$ & 
$472	\,\pm\,140$	&$-0.92	\,\pm\,0.22$	&$0\,\pm\,134$	&$0	\,\pm\,0.28$	&$11.1$
 \\
 Left (Eastern) Slice  & $11.98$ & 
$215\,\pm\,48$ &$-1.21	\,\pm\,0.21$ &
$0.72\,\pm\,45$	&$0.003	\,\pm\,0.207$	&$8.66$\\
 Left (Eastern) Slice  & $23.96$ & 
$61 \,\pm\,18$	&$-1.55	\,\pm\,0.31$&$	0\,\pm\,17$&	$0	\,\pm\,0.28$&	9.21\\
 Central Slice & $-36.94$ &
$57\,\pm\,12$&	$-1.89	\,\pm\,0.26$ &	$14	\,\pm\,12$&	$0.19	\,\pm\,0.14$ &	$11.6$\\
Central Slice & $-24.96$ &
$194\,\pm\,29$	&$-1.61	\,\pm\,0.18$	&$48	\,\pm\,26$	&$0.20	\,\pm\,0.09$	&$30.5$\\
Central Slice & $-12.98$ &
$946	\,\pm\,225$	&$-1.14	\,\pm\,0.22$&	$403	\,\pm\,207$	&$0.30	\,\pm\,0.12$	&$64.8$\\
Central Slice & $-0.50$ &
$2452	\,\pm\,931$	&$-0.94	\,\pm\,0.29$	&$1479	\,\pm\,880$&	$0.38	\,\pm\,0.17$&	$88.4$\\
Central Slice & $11.98$ &
$805	\,\pm\,191$	&$-1.20	\,\pm\,0.23$	&$481	\,\pm\,174$	&$0.37	\,\pm\,0.10$	&$55.7$\\
Central Slice & $23.96$ &
$114	\,\pm\,19$	& $-1.85	\,\pm\,0.21$ &	$82	\,\pm\,17$	&$0.42	\,\pm\,0.06$&	$19.1$\\
Central Slice & $35.95$ &
$25	\,\pm\,6$	& $-2.59	\,\pm\,0.33$	&$19	\,\pm\,7$	&$0.44	\,\pm\,0.10$	&$8.32$\\
Right (Western) Slice  & $-24.96$ & 
$53	\,\pm\,13$&$-1.79	\,\pm\,0.29	$&$0	\,\pm\,13	$&$0	\,\pm\,0.24	$&$9.80$\\
Right (Western) Slice  & $-12.98$ & 
$224\,\pm\,51	$&$-1.17	\,\pm\,0.21	$&$0	\,\pm\,48	$&$0	\,\pm\,0.21	$&$10.3$\\
Right (Western) Slice  & $-0.5$ & 
$371	\,\pm\,107	$&$-1.11	\,\pm\,0.26	$&$217	\,\pm\,99	$&$0.37	\,\pm\,0.13	$&$12.2$\\
Right (Western) Slice  & $11.98$ & 
$179\,\pm\,	49$	&$-1.13	\,\pm\,0.25	$&$9	\,\pm\,47	$&$0.05	\,\pm\,0.24	$&$9.02$\\
 \hline
    \end{tabular}
\end{center}
$^{\rm a}$ $z$-height of the measurement. Positive is north of the (rotated) major axis and negative is south.\\
$^{\rm b}$ Specific intensity of the nonthermal emission. \\
$^{\rm c}$ Nonthermal spectral index.\\
$^{\rm d}$ Specific intensity of the thermal emission.\\
$^{\rm e}$ Thermal fraction.\\
$^{\rm f}$ rms of the fitted curve.\\
\end{table}

\bibliography{sample631}{}
\bibliographystyle{aasjournal}



\end{document}